\documentclass[reqno]{amsart}

\usepackage{epsfig,amsmath, amsfonts,amsthm,amssymb}
\usepackage{dsfont}

\usepackage[english]{babel}

\usepackage[left=1.75cm, top=2.7cm, bottom=2.50cm,right=1.75cm]{geometry}

\usepackage{verbatim}
\usepackage{setspace}
\usepackage{mathrsfs}
\usepackage{bm}
\usepackage{color}
\usepackage{epsfig}
\usepackage{epstopdf}
\usepackage{graphicx}
\usepackage{graphics}

\usepackage{url}

\usepackage{bm}

\usepackage{colortbl}

 \newtheorem{thm}{Theorem}[section]
 \newtheorem{cor}[thm]{Corollary}
 \newtheorem{lem}[thm]{Lemma}

 \theoremstyle{definition}

 \newtheorem{rem}[thm]{Remark}
 \numberwithin{equation}{section}
\newcommand{\be}{\begin{equation}}
\newcommand{\ee}{\end{equation}}
\newcommand{\bq}{\begin{eqnarray}}
\newcommand{\eq}{\end{eqnarray}}

\newcommand{\half}{\frac{1}{2}}

    \def\ed{{\,\stackrel{\frak {D}}{=}\,}}
    \def\ld{\;{\stackrel{\frak {D}}{\longrightarrow}}\;}

    \def\ld{{\stackrel{\frak {D}}{\longrightarrow}}}

    \def\calF{{\mathcal F}}

    \def\calR{{\mathcal R}}

    \def\bbr{{\mathbb R}}
    \def\bbe{{\mathbb E}}
    \def\bbp{{\mathbb P}}
    
    \def\bbn{{\mathbb N}}

    \def\ed{\;{\stackrel{\frak {D}}{=}}\;}
    
    \def\ld{\;{\stackrel{\frak {D}}{\longrightarrow}}\;}

    \def\b{{b}}

    \def\q{{q}}

    \def\w{{w}}

    \def\I{{I}}
    \def\J{{J}}

    \def\N{{N}}

    \def\U{{U}}
    \def\V{{V}}
    \def\W{{W}}
    \def\X{{X}}
    
    \def\Z{{Z}}

  \definecolor{Red}{rgb}{0,0,0}
    
    \definecolor{DRed}{rgb}{0,0,0}
    
    \definecolor{Green}{rgb}{0,0,0}
    
    \definecolor{Blue}{rgb}{0,0,0}
    \newcommand{\Blue}{\color{Blue}}
    \definecolor{PaleGrey}{rgb}{0,0,0}

\title[{Skew asymptotics under  stochastic volatility and L\'evy jumps}]{Short-term asymptotics for the implied volatility skew under a stochastic volatility model with L\'evy jumps}

\author{{Jos\'e E. Figueroa-L\'opez}}
\address{
Department of Mathematics\\
Washington University in St. Louis, St. Louis, MO 63130, USA}
\email{{\tt figueroa@math.wustl.edu}}
\thanks{{The first author's research was supported in part by the NSF Grant DMS-1149692. {The authors gratefully acknowledge two anonymous reviewers and the editor for providing constructive and insightful comments, which improved significantly the quality of the manuscript. The authors would also like to thank Christian Houdr\'e and Frederi Viens for useful suggestions.}}}
\author{Sveinn \'Olafsson}
\address{Department of Statistics and Applied Probability, University of California, Santa Barbara, CA 93106, USA}
\email{{\tt olafsson@pstat.ucsb.edu}}

\begin{document}
\maketitle

\begin{abstract}
The implied volatility {skew} has received relatively little attention in the literature on short-term asymptotics for financial models with jumps, despite its importance in model selection and calibration. We rectify this by providing high-order asymptotic expansions for the at-the-money implied volatility skew, under a rich class of stochastic volatility models with independent stable-like jumps of infinite variation. The case of a pure-jump stable-like L\'evy model is also considered under the minimal possible conditions for the resulting expansion to be well defined.\ Unlike recent results for ``near-the-money'' option prices and implied volatility, the results herein aid in understanding how the implied volatility smile near expiry is affected by {important features of the continuous component}, such as the leverage and vol-of-vol parameters. As intermediary results we obtain high-order expansions for at-the-money digital call option prices, which furthermore allow us to infer analogous results for the delta of at-the-money options.\ Simulation results indicate that our asymptotic expansions give good fits for options with maturities up to one month, underpinning their relevance in practical applications, and an analysis of the implied volatility skew in recent S\&P500 options data shows it to be consistent with the infinite variation jump component of our models.

\vspace{0.2 cm}
\noindent\textbf{AMS 2000 subject classifications}: 60G51, 60F99, 91G20, 91G60.

\vspace{0.1 cm}
\noindent\textbf{Keywords and phrases}: Exponential L\'{e}vy models; stochastic volatility models; short-term asymptotics; ATM implied volatility slope; implied volatility skew; ATM digital call option prices.

\end{abstract}

\section{Introduction}

\subsection{{Motivation}}\label{motivation}
Since the emergence of the Black-Scholes option pricing model, there have been two main themes in the evolution of financial models: jumps and stochastic volatility. The features of both model {components} have their merits, 
and combining jumps in returns and stochastic volatility {is a viable way} to calibrate the implied volatility surface across strikes and maturities. However, the increased generality comes at a cost as models that combine jumps and stochastic volatility are often highly complicated to calibrate and implement. An important line of research has therefore been to examine various extreme regions of the volatility surface and attempt to understand how different model features and their associated parameters affect the behavior of option prices (see \cite{AlosLeonVives}, \cite{Andersen}, \cite{GerGul:2014}, \cite{FigForde:2012}, \cite{LopGonHou:2014}, \cite{LopOla:2014}, \cite{GaoLee:2013}, \cite{TankMij}, \cite{Tankov:2010}, and references therein).

In particular, a number of recent papers have shed light on the short-term asymptotic behavior of option prices and implied volatility, and revealed that those quantities exhibit markedly different behavior from one model setting to the next. 
For instance, while there typically exists a limiting implied volatility smile in continuous models, the smile exhibits explosive behavior in models with jumps. This phenomenon is due to the much slower decay of out-of-the-money (OTM) option prices in the presence of jumps, which causes the implied volatility to blow up off-the-money, while converging to the spot volatility at-the-money (ATM) (see, e.g.,\ \cite{FigForde:2012}, \cite{LopOla:2014}, and \cite{Tankov:2010}). This feature of jump models is actually desirable as it enables them to reproduce the pronounced smiles and skews observed  at short maturities.

\begin{figure}
	\centering
	\includegraphics[width=7cm,height=5.0cm]{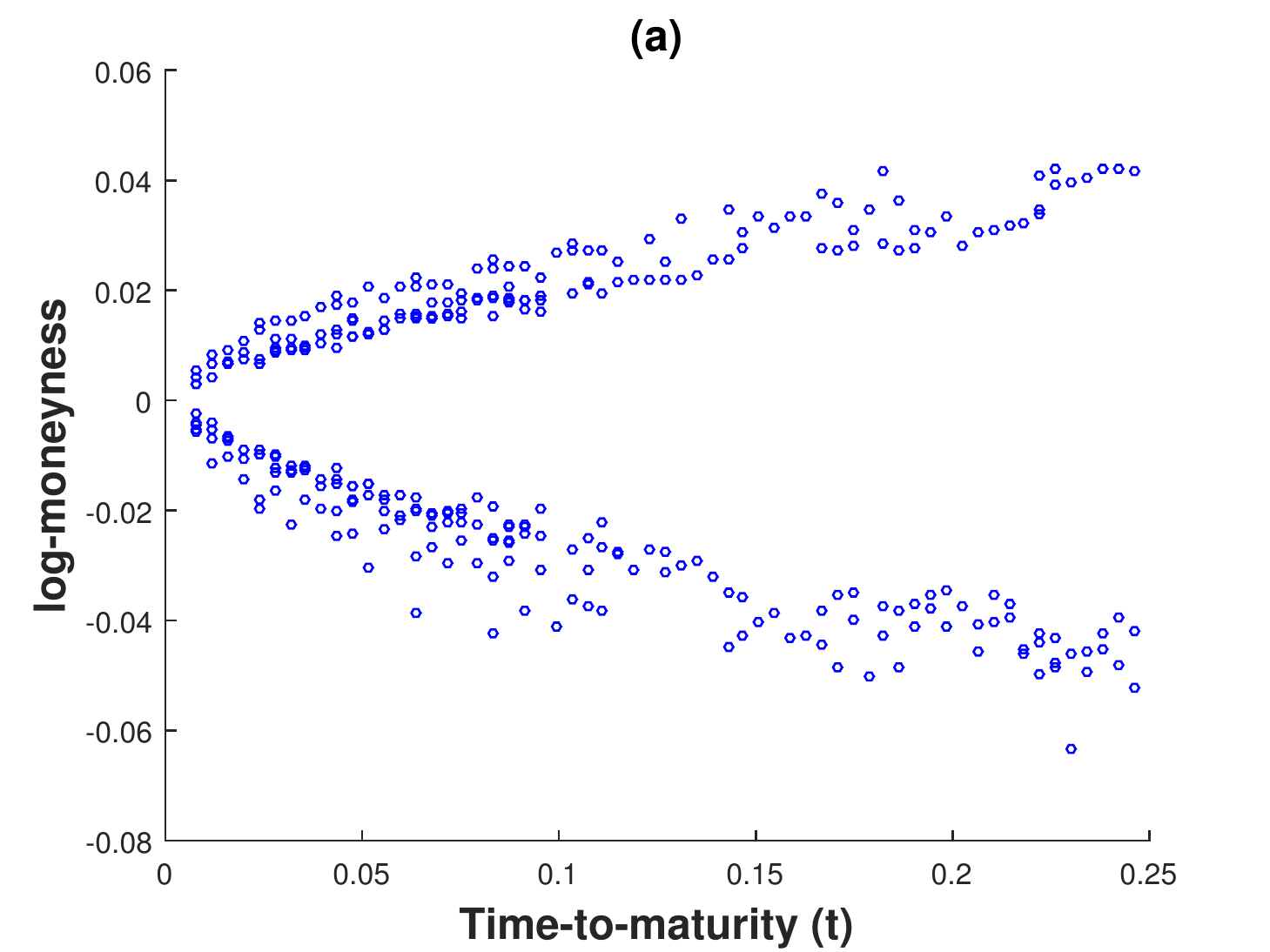}\hspace{1cm}
	\includegraphics[width=7cm,height=5.0cm]{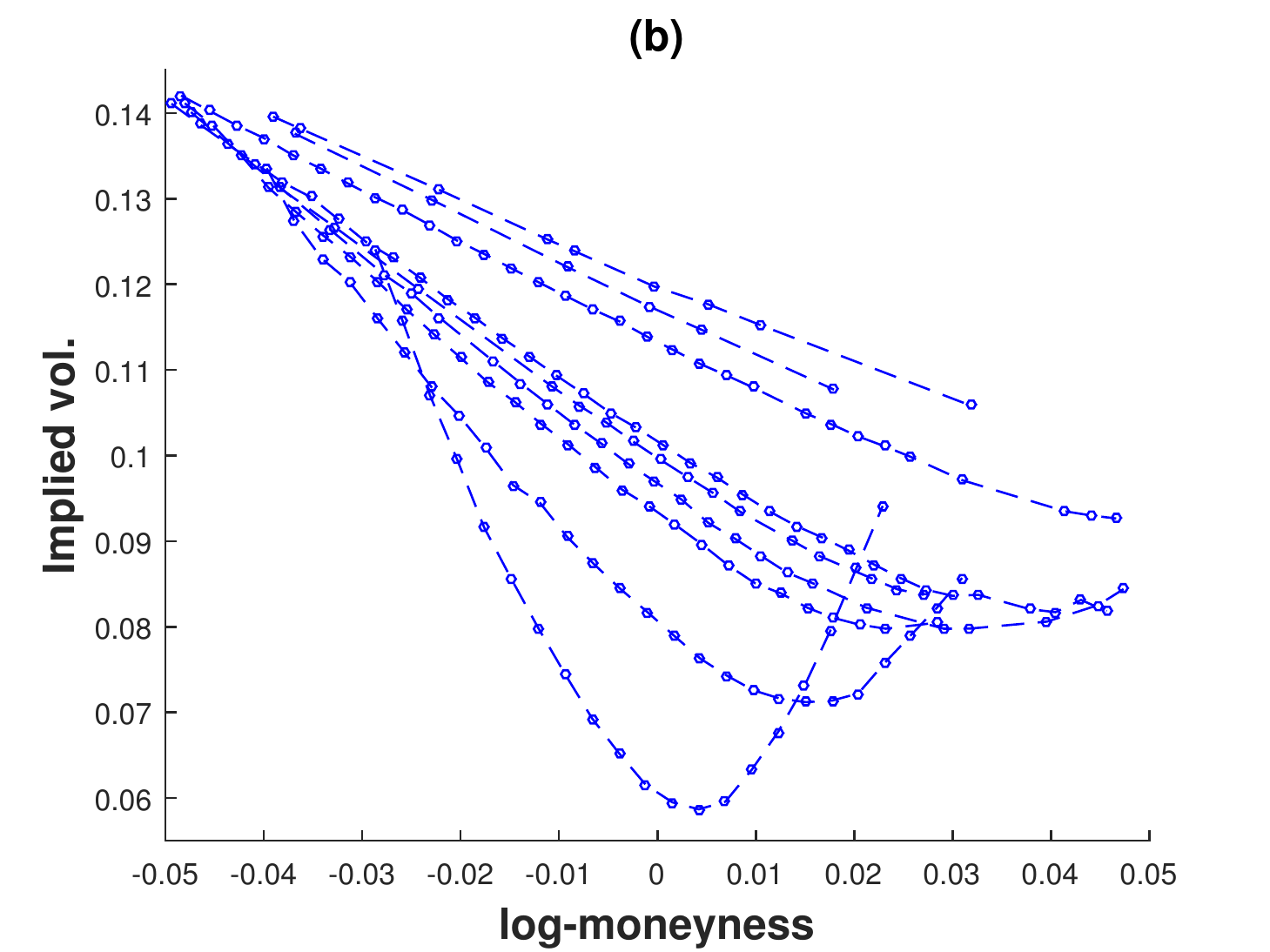}
    \caption{\small {\bf(a)} The log-moneyness ($\kappa=\ln(K/F)$, where $F$ is the forward price) of outstanding 25-delta call options ($\kappa>0$) and put options ($\kappa<0$) in Jan 2014. {\bf(b)} The implied volatility smiles on Jan 15, 2014, corresponding to maturities ranging from 0.012 (3 days) to 0.25 (3 months).}\label{NearTheMoney}
\end{figure}

{Another {stylized} empirical fact is that as time-to-maturity decreases, the liquid strike prices become increasingly concentrated around the ATM strike (see, e.g., \cite{TankMij}, and Figure \ref{NearTheMoney} herein).
ATM options are therefore of particular importance when it comes to short-term {asymptotics. The} leading order term of ATM option prices has been derived for various models containing jumps (\cite{MuhNut:2009}, \cite{Tankov:2010}), {but, unfortunately,} those approximations are known to require unrealistically small maturities to attain satisfactory accuracy, {and provide limited information about key model parameters. These drawbacks have} motivated the search for higher order asymptotics and, in the recent paper \cite{LopOla:2014}}, such
expansions are derived for a general class of stochastic volatility models with L\'evy jumps, in an asymptotic regime where time-to-maturity \emph{and} log-moneyness become small, which, as explained above, is of particular practical importance.
Still, certain shortcomings prevail, in particular{ when it comes to understanding the effect of the stochastic volatility component on} the short-term volatility smile, and its interaction with the jump component. For instance, the asymptotic expansions developed in \cite{LopOla:2014} depend on the volatility process only through the spot volatility.

In the present work, we alleviate the issues mentioned in the previous paragraph by analyzing the short-term behavior of the ATM implied volatility skew\footnote{Practitioners commonly use the terms ``skew'' and ``implied volatility skew'' for the ATM slope of the implied volatility curve for a given expiration date (see, e.g., \cite{Mixon}). We will use the terms interchangeably.}, which turns out to depend {on} key parameters of the underlying volatility process such as the leverage and vol-of-vol parameters, and {gives} a more accurate picture of the short-term behavior of the{ implied volatility smile.
{The ATM skew has received comparatively little attention in the literature, {despite the fact that it} is actively monitored in practice by traders and analysts {(cf.\ \cite{Mixon})}, stemming in part from its rich informational content, and considerable empirical support has in fact been provided for its significance in predicting future equity returns, and as an indicator of the risk of large negative jumps (see, e.g., \cite{Xing}, \cite{Yan}, for individual stock options, {and} \cite{Bates}, \cite{Pan}, for index options).

The ATM skew is also highly relevant in model selection and when calibrating models to observed option prices (cf.\ \cite[Ch.\ 5]{GatheralSV}), especially in FX markets, where it is standard to effectively quote directly on smile slope and convexity (cf.\ \cite[Sec.\ 2]{Andersen}). Moreover, it is also widely {believed} that the implied volatility skew and convexity reflect the skewness and kurtosis in the underlying risk-neutral 
distribution {and, thus,} a number of researchers have attempted to relate the smile skew and convexity to moments of the risk-neutral distribution (see, e.g., \cite{Bakshi}, \cite{Bouchaud}, \cite{Fajardo}, \cite{Zhang}). 

Finally, another important reason to study the skew is {its connection with the delta of options, which is of paramount importance in the trading and hedging of options. More specifically,} the same key quantities needed to derive the asymptotic behavior of the skew, can be used to derive short-term asymptotics for the delta of options (see Section \ref{Overview} {below}).

\subsection{Literature Review}

The literature on the short-term implied volatility skew in the presence of jumps is somewhat limited, but in recent years the leading order term has been obtained, albeit under somewhat restrictive assumptions. 
For models of jump-diffusive nature, several sources show that the skew converges to a nonzero value as time-to-maturity tends to zero (see, e.g., \cite{AlosLeonVives}, \cite{GerGul:2014}, \cite{Yan}). In such models, jumps are infrequent and can be interpreted as the occurrences of rare events, but substantial empirical evidence supports models with {infinite jump activity}. In particular, 
S\&P500 index options are used in {\cite{MedSca07}} to reach the conclusion that adding Poisson jumps to a stochastic volatility model is not sufficient to account for the implied volatility skew at short maturities. 

For models with infinite jump activity, a recent result \cite{GerGul:2014} shows that the skew is of order $t^{-1/2}$ for bounded variation L\'evy processes, as well as a few specific infinite variation cases with Blumenthal-Getoor index\footnote{For a L\'evy process $X$ with L\'evy measure $\nu$, the Blumenthal-Getoor index is defined as $\inf\{p\geq 0:\int_{|x|\leq 1}|x|^p\nu(dx)<\infty\}.$} equal to one, such as the Normal Inverse Gaussian (NIG) and Meixner processes. This growth rate is in fact the fastest possible one in the absence of arbitrage (cf.\ \cite[Sec.\ 3]{RLee}), but in \cite{GerGul:2014} the authors also show that for certain L\'evy models with a Brownian component, the skew can explode at a rate slower than $-1/2$. This can be viewed as a special case of Corollary \ref{corXW} in Section \ref{SectionPureJump} herein, and the same can be said about Proposition $8.5$ in a recent survey paper \cite{Andersen}, which can be 
interpreted as a skew approximation for tempered stable L\'evy processes as defined in \cite{CT04}, but under some extremely restrictive assumptions on the model parameters.

A somewhat different approach is adopted in \cite{Bouchaud}, which belongs to the stream of literature attempting to relate {features of the implied volatility smile to properties of the} risk-neutral distribution of the underlying. Concretely, the authors derive, partly by heuristic arguments, the following expansion for near-the-money implied volatility $\tilde\sigma(\tilde\kappa,t)$, parameterized in terms of the scaled moneyness $\tilde\kappa=(K-S_0)/(S_0\sigma\sqrt{t})$, where {$K$ and $t$ are the option's strike and time-to-maturity, $S_{0}$ is the spot price of the underlying, and $\sigma$ is a measure of the ``overall volatility level":} 
\begin{align}\label{BouchaudFormula}
\tilde\sigma(\tilde\kappa,t) = \sigma\big(\alpha_t + \beta_t\tilde\kappa + \gamma_t\tilde\kappa^2 + O(\tilde\kappa^3)\big).
\end{align}
{Explicit expressions for the coefficients $\alpha_{t}$, $\beta_{t}$, and $\gamma_{t}$ are also proposed (see Eq.~(2) therein). 
In particular, the form of the skew coefficient $\beta_t$} suggests, at least heuristically, the following expression for the ATM skew,
\begin{align}\label{BouchaudForSkew}
	\left. \frac{\partial\tilde\sigma\left({\tilde\kappa},t\right)}{\partial\tilde\kappa}\right|_{\tilde\kappa=0}
	 = {\sigma}\sqrt{\frac{2\pi}{t}}\bigg(\frac{1}{2}-\bbp\left(S_{t}\geq{}0\right)\bigg) + o(t^{-\half}),\quad t\to 0, 
\end{align}
which agrees with formula (\ref{ATMslope}) below, but can at best be used to obtain the leading order term of the skew. More importantly, a mathematically sound justification is needed when passing from the asymptotic expansion (\ref{BouchaudFormula}) for $\tilde{\sigma}(\tilde\kappa,t)$ to an analogous asymptotic expansion for its derivative of the form (\ref{BouchaudForSkew}). In a certain sense, the results herein therefore formalize and extend the heuristic approach for the skew in \cite{Bouchaud}.
Another important contribution of the {approach} in \cite{Bouchaud} is that it attempts to explain the general shape of volatility smiles and, in particular, how the smile skew, $\beta_t$, is related to the skewness of the distribution of the underlying. {To this end, the authors} use S\&P500 index time series to argue that there cannot be a simple relation between the skew of the smile and the skewness.

\subsection{Overview of New Results}\label{Overview}

It is important to stress that most of the aforementioned results still suffer from the same shortcomings as the corresponding results for ATM option prices, {in that} their domain of validity is extremely small. This, in part, stems from the 
focus being on obtaining the leading order term, which can be done in some generality, but in return tends to {only} depend on the most general model parameters. 
A significant contribution of the present work is therefore to provide accurate higher order expansions for the implied volatility skew, under a class of models that goes beyond the homogeneous L\'evy framework by combining {stochastic volatility and jumps with high activity}. Empirical evidence generally supports the need for such models, not only to calibrate the implied volatility surface, but also to generate realistic future dynamics of implied volatility, in order to give reasonable prices for exotic derivatives (see, e.g., \cite{Bergomi}). 

{Throughout,} we assume that the {risk-free interest rate $r$ and the} dividend yield $\delta$ {of the underlying} are $0$, and that the price process $S:=(S_t)_{t\geq 0}$ of the underlying asset is a $\bbp$-martingale. We denote the implied volatility {of an option} by $\hat\sigma(\kappa,t)$, where $\kappa:=\log(K/S_0)$ is the log-moneyness and $t$ is the time-to-maturity. {For simplicity, the ATM implied volatility $\hat\sigma(0,t)$ is denoted by $\hat{\sigma}(t)$}. 
Let us start by recalling some basic relationships
that are fundamental to our approach (see Section \ref{TSP} for details). First, under some mild conditions the ATM skew satisfies,
\begin{align}\label{ATMslope}
\left.\frac{\partial\hat\sigma\left(\kappa,t\right)}{\partial\kappa}\right|_{\kappa=0}
 = \sqrt{\frac{2\pi}{t}}\bigg(\half-\bbp\left(S_t\geq S_{0}\right)-\frac{\hat\sigma(t)\sqrt{t}}{2\sqrt{2\pi}}+O\Big(\big(\hat\sigma(t)\sqrt{t}\big)^3\Big)\bigg)\bigg(1+\frac{\left(\hat\sigma(t)\sqrt{t}\right)^2}{8}+O\Big(\big(\hat\sigma(t)\sqrt{t}\big)^4\Big)\bigg),
\end{align}
which enables us to separate the problem of studying the asymptotic behavior of the skew into finding the asymptotics of two quantities that are important in their own right: (i) the ATM implied volatility $\hat\sigma(t)$, and (ii) the ATM digital call option price $\bbp\left(S_t\geq S_{0}\right)$. Interestingly enough, there is also a close connection between $\bbp\left(S_t\geq S_{0}\right)$ and the delta of ATM options, i.e.\ the sensitivity of the ATM option price {$C(S_{0},t)$}, {with respect to} the spot {price} of the underlying, $S_0$. Concretely, we show that 
\begin{align}\label{Delta}
\Delta(t) := \frac{\partial {C\left(S_{0},t\right)}}{\partial S_{0}}=\frac{1}{S_0}{C(S_{0},t)} + \bbp(S_t\geq S_0),
\end{align} 
so the asymptotic results for the ATM option price {$C(S_{0},t)$} (cf.~\cite{LopOla:2014}), together with the present results for the transition probability $\bbp(S_t\geq S_0)$, can be used to obtain short-term asymptotic expansions for the delta of ATM call options.

Both the ATM implied volatility, $\hat\sigma(t)$, and the corresponding option price, $C( S_{0},t)$, have received considerable attention in the literature, and it is well documented that their short-term behavior is strongly tied to various pathwise properties of the log-returns process. For example, including a continuous component can significantly change the properties of pure-jump models, and the type of jump component can also have a drastic effect. 
The same is true when it comes to the ATM volatility skew, and we proceed to explain the different novelties of our work by separately analyzing the two cases of interest (pure-jump and mixed), and then elaborating on the accuracy and applicability of our results in model selection and calibration. We also briefly consider the OTM volatility skew, which, much like the OTM volatility (cf.\ \cite{FigForde:2012}), can be analyzed in much more generality than the ATM skew (see Remark \ref{OTMSkewH}).

\subsubsection{Pure-jump exponential L\'evy model}
In this paper, we consider {tempered} \emph{stable-like} L\'evy processes, as introduced in \cite{LopGonHou:2014} and \cite{LopOla:2014}, with and without an independent continuous component. {In this section we briefly describe our results related to the latter case.} More concretely, we consider the model
\begin{equation}\label{ELM0}
	S_t:=S_0e^{X_t}, 
\end{equation}
where $X$ stands for a pure-jump L\'evy process with a L\'evy measure of the form 
\begin{align}\label{tmpstblnu}
\nu(dx)=C\Big(\frac{x}{|x|}\Big)|x|^{-Y-1}\bar\q(x)dx,
\end{align}
for some constants $C(1),C(-1)\in[0,\infty)$ such that $C(1)+C(-1)>0$, $Y\in(0,2)$, 
and a {bounded} function $\bar{q}:\bbr\backslash\{0\}\to[0,\infty)$ such that $\bar{q}(x)\to{}1$, as $x\to{}0$. This framework includes most of the infinite activity L\'evy models used in practice, and its short-term behavior depends strongly on the \emph{index of jump activity} $Y$, which coincides with the Blumenthal-Getoor (BG) index of the process. In what follows we impose the condition $Y\in(1,2)$, which implies that  $X$ has infinite variation.
This restriction is supported by recent econometric studies of high-frequency financial data (see Remark 2.2 in \cite{LopGonHou:2014}), and we will also argue in Section \ref{Examples}, using recent S\&P500 options data, that it is the most relevant case since it gives the flexibility needed to calibrate the short-term implied volatility skew observed in practice.                          

For models of the form (\ref{ELM0}), a second order short-term expansion for the ATM implied volatility, $\hat\sigma(t)$, is given in Theorem $3.1$ of \cite{LopOla:2014}, under a minimal integrability condition on $\bar\q$ around the origin. The key to studying the implied volatility skew is therefore the transition probability appearing in (\ref{ATMslope}), for which we have $\bbp\left(S_t\geq S_0\right)=\bbp\left(X_t\geq 0\right)$, but while a lot is known about $\bbp\left(X_t\geq x_0\right)$ for nonzero $x_0$ (cf.\ \cite{LopHou:2009}), much less has been said about $\bbp\left(X_t\geq 0\right)$ for processes with infinite jump activity. {The leading order term for bounded variation L\'evy models, as well as certain models with BG-index one (e.g., NIG, Meixner) is obtained in \cite{GerGul:2014}}, while, for tempered stable-like processes, 
\begin{align}
\bbp\left(X_t\geq 0\right) 
\longrightarrow \widetilde\bbp\left(Z_1\geq 0\right),\quad t\to 0,\label{SLOTST}
\end{align}
where $Z_1$ is a strictly $Y$-stable random variable under $\widetilde\bbp$. This limit 
is a consequence of the fact that $t^{-{1}/{Y}}X_{t}$ converges {in distribution} to $Z_1$, as $t\to{}0$ (cf.\ \cite{rosenbaum.tankov.10}), and cannot be extended to higher order terms. However, procedures similar to the ones used to derive near-the-money option price expansions in \cite{LopGonHou:2014} and \cite{LopOla:2014} {will} allow us to get a closer look at the convergence. 
Concretely, the following novel higher order asymptotic expansion is obtained,
\begin{align}\label{probX0}
\bbp\left(X_t\geq 0\right) - \widetilde\bbp\left(Z_1\geq 0\right)= \sum_{k=1}^n d_{k}t^{k\left(1-\frac{1}{Y}\right)}+ e\,t^{\frac{1}{Y}}+f\,t+o(t),\quad t\to 0, 
\end{align}
where $n:=\max\{k\geq 3:k\left(1-{1}/{Y}\right)\leq 1\}$. It is important to point out that the leading order term is $d_{1}t^{1-{1}/{Y}}$ for all $Y\in(1,2)$. This can be compared to the expansion for ATM option prices given in Theorem $3.1$ of \cite{LopOla:2014}, where the first and second order terms are of order $t^{{1}/{Y}}$ and $t$, i.e.\ the convergence here of digital option prices is slower.

{{With the expansion (\ref{probX0}) at our disposal, we can use (\ref{ATMslope}) to} deduce an expansion for the ATM implied volatility skew, which turns out to exhibit explosive behavior in short time, but unlike jump-diffusion models where the skew is always bounded, and finite variation models where the skew is of order $t^{-1/2}$ (cf.\ \cite{GerGul:2014}), the order of the skew here actually depends on the index of jump activity, $Y$, and ranges between $0$ and $-1/2$. This provides an important model selection and calibration tool, and in Section \ref{Examples} we will show that our result is in line with the short-term skew in S\&P500 index options, while the same {cannot} be said about models where the dynamics are driven by a purely continuous model, nor a model with a finite variation jump component.
Furthermore, important qualitative properties such as the sign of the skew can easily be recovered from the model parameters, and used to create simple parameter restrictions that can be used for calibration and model selection purposes (see Remark \ref{remark1} for further details).} 

\subsubsection{Exponential L\'evy model with stochastic volatility}

Empirical work has generally supported the need for both jumps to reflect shorter maturity option prices, and stochastic volatility to calibrate the longer maturities where the smile effect of jump processes is limited {(see, e.g., \cite[Ch.5]{GatheralSV} and \cite{Konikov})}.\ In order to incorporate a continuous component of diffusive type into the price dynamics, we consider the model 
\begin{align}
S_t:=S_0e^{X_t+V_t},
\end{align} 
where $(V_{t})_{t\geq{}0}$ is a stochastic volatility process of the form 
\begin{align}\label{modelVY}
\begin{split}
		dV_t & = \mu(Y_t) dt+\sigma(Y_t) \big(\rho dW_t^1 + \sqrt{1-\rho^2}dW_t^2\big),\qquad V_0=0, \\ 
		dY_t & = \alpha(Y_t)dt+\gamma(Y_t)dW_t^1,\qquad Y_0=y_0, 
\end{split}
\end{align}
and $(W_t^1)_{t\geq 0}$ and $(W_t^2)_{t\geq 0}$ are independent standard Brownian motions, independent of the pure-jump L\'evy process $X$. This framework includes the most commonly used stochastic volatility models, such as the mean-reverting Heston and Stein-Stein models, and we remark that in such models it is {generally believed} that the leverage parameter $\rho$ is responsible for generating asymmetric volatility smiles. A question of interest is therefore to what extent, and in what way, the leverage parameter contributes to the short-term skew in models with jumps. {As in the pure-jump case, 
the key to studying the ATM skew is the transition probability $\bbp(S_t\geq S_0)$, and for models of the form $S_t=S_0e^{X_t+V_t}$ we have $\bbp(X_t+V_t\geq 0)\to 1/2$, as $t\to 0$. The literature is quite sparse beyond that, but in Section \ref{SectionContComp} we derive the following higher order expansion,}
\begin{align}\label{probXW0}
\bbp\left(X_t+V_t\geq 0\right)&= \half + \sum_{k=1}^n d_{k}\,t^{k\left(1-\frac{Y}{2}\right)} + e\,t^{\half} + f\,t^{\frac{3-Y}{2}} + o(t^{\frac{3-Y}{2}}),\quad t\to 0,
\end{align}
where $n:=\max\left\{k\geq 3:k\left(1-Y/2\right)\leq (3-Y)/2\right\}$. Comparing this result to the short-term expansion for ATM call option prices given in Theorem $4.1$ of \cite{LopOla:2014}, where the first and second order terms are of order $t^{1/2}$ and $t^{{(3-Y)}/{2}}$, reveals that the convergence of ATM digital option prices is slower, as it was in the pure-jump case, unless $C(1)=C(-1)$, in which case the summation term vanishes. 

Piecing together the results above now gives 
the following expansion for the ATM implied volatility skew,
\begin{align}\label{volSlope0cont}
\left.\frac{\partial\hat\sigma\left(\kappa,t\right)}{\partial\kappa}\right|_{\kappa=0} 
=-\sqrt{2\pi}\,\sum_{k=1}^{n}d_k\,t^{\left(1-\frac{Y}{2}\right)k-\half} -  \frac{c}{\sigma(y_0)}+\Big(\sqrt{2\pi} f+\half\bar\sigma_1\Big)\,t^{1-\frac{Y}{2}}+o(t^{1-\frac{Y}{2}}),\quad t\to 0,
\end{align}
where $c:=\tilde\gamma-(\rho\sigma'(y_0)\gamma(y_0))/2$, and $\tilde\gamma$ is a constant that depends only on the parameters of $X$.
{As in the pure-jump case, the order of the skew ranges between 0 and $-1/2$ depending on the index of jump activity, $Y$, 
but it is {also} observed that for a fixed value of $Y$, the skew is less explosive than in the pure-jump case. {In particular,} in the symmetric case $C(1)=C(-1)$, the skew actually converges to a nonzero value, $-c/\sigma(y_0)$, as in jump-diffusion models. This is to be expected since including a continuous component has a limited effect on OTM volatility, while raising the limiting ATM volatility from zero to the spot volatility $\sigma(y_0)$, effectively flattening the smile.}  
The expression (\ref{volSlope0cont}) also offers a significant improvement over existing results in that it depends on both the correlation coefficient $\rho$ and the volatility of volatility, $\sigma'(y_0)\gamma(y_0)$. In particular, this is in sharp contrast to the expansions for option prices and implied volatility in \cite{LopOla:2014}, where the impact of replacing the Brownian component by a stochastic volatility process was merely to replace the volatility of the Brownian component, $\sigma$, by the spot volatility, $\sigma(y_0)$. 

Lastly, it is noteworthy that the skew-effects of the jump-component and the continuous component (i.e.\ the correlation coefficient $\rho$) turn out to be additive to the leading order, and the contribution to the implied volatility skew of a nonzero correlation $\rho$ between the asset price and volatility can be quantified as 
 \[
 	\half\frac{\rho\sigma'(y_{0})\gamma(y_{0})}{\sigma(y_{0})}(1+o(1)),\quad t\to{}0.
\]
Interestingly enough, this shows that to the leading order, the skew-effect of stochastic volatility is the same as in jump-diffusion models (see Ex.\ 7.1 in \cite{AlosLeonVives} for a comparable result for such finite activity models), but the effect of the jump-component is drastically different, and so is the interaction between the two model components.

\subsubsection{Accuracy of the asymptotic formulas/Empirical analysis}

As explained above, the asymptotic expansions obtained herein can be used to infer several important features of the implied volatility smile.  However, we also mentioned that a common drawback of short-term approximations is that their domain of validity can be small, which could potentially limit their usefulness for practical work. {To better assess this point,} in Section \ref{Examples} we test the accuracy of our expansions using Monte Carlo simulation. {Our} results indicate that for the important class of tempered stable processes, they give good approximations for options with maturities up to one month, underpinning their practical relevance in essentially every major options market. 
In the second part of Section \ref{Examples} we look at the short-term implied volatility skew in {recent} S\&P500 option {prices}, with a view toward model selection and calibration. In particular, we consistently observe that the skew exhibits a power law of order between 0 and $-1/2$, which is in line with the skew-behavior of the models studied in Sections \ref{SectionPureJump} and \ref{SectionContComp}, while contradicting the behavior of purely continuous models, as well as model with a finite variation jump-component. Furthermore, we provide a simple calibration procedure for the index of jump activity of the process, $Y$, which can be viewed as a new forward-looking tool to assess this fundamental parameter, complementing the popular rear-facing estimation methods based on high-frequency observations of the underlying asset's returns.

\subsection{Outline}

The rest of this paper is organized as follows. Section \ref{TSP} provides the probabilistic relationships on which we build our analysis, and introduces the class of tempered stable-like L\'evy processes. Section \ref{SectionPureJump} contains our results for the transition probability, volatility skew, and delta, under a pure-jump exponential L\'evy model. Section \ref{SectionContComp} 
presents the analogous results under a L\'evy jump model with stochastic volatility. Section \ref{Examples} contains numerical examples to assess the accuracy of the asymptotic expansions, as well as an empirical analysis of the short-term skew in recent} S\&P500 option prices. Section \ref{conclusions} summarizes our results {and draws some further conclusions}. Finally, proofs of intermediary results are collected in the appendix.

\section{{Notation and auxiliary results}}\label{TSP}

{Throughout, $X:=(X_t)_{t\geq 0}$ denotes} a pure-jump tempered stable-like L\'evy process, as introduced \cite{LopGonHou:2014} and \cite{LopOla:2014}, defined on a filtered probability space $(\Omega,\calF,(\calF_t)_{t\geq 0},\bbp)$ satisfying the usual conditions. That is, $X$ is a L\'evy process with triplet $(0,b,\nu)$ relative to the truncation function ${\bf 1}_{\{|x|\leq 1\}}$ (see Section 8 in \cite{Sato}), where the L\'evy measure $\nu$ is given by (\ref{tmpstblnu}), for some constants $C(1),C(-1)\in[0,\infty)$ such that $C(1)+C(-1)>0$, $Y\in(1,2)$, and a bounded function $\bar{q}:\bbr\backslash\{0\}\to[0,\infty)$ such that $\bar{q}(x)\to{}1$ as $x\to{}0$. 
Let us also introduce the following additional technical conditions on the tempering function $\bar\q$, conveniently selected to facilitate the proofs of some of the results that follow:
\begin{align}\label{NewAssumEq}
  {\rm (i)}\; \int_{|x|\leq{}1}\big|\bar{q}(x)-1-\alpha\Big(\frac{x}{|x|}\Big)x\big||x|^{-Y-1}dx<\infty; \quad
	{{\rm (ii)}}\;\limsup_{|x|\to\infty} \frac{|\ln \bar\q(x)|}{|x|}<\infty;\quad
	{{\rm (iii)}}&\;\inf_{|x|<\varepsilon} \bar\q(x)>0,\;\forall\varepsilon>0. 
\end{align}
Here, $\alpha(1)$ and $\alpha(-1)$ are real-valued constants. We emphasize that the main results of Sections \ref{SectionPureJump} and \ref{SectionContComp} only require condition (i) to be satisfied, which controls the behavior of the L\'evy density around the origin. In particular, a sufficient condition for all $1<Y<2$ is given by $\bar\q(x)=1+\alpha({x}/{|x|})x+O(x^2)$, as $x\to 0$. 
Next, define a measure transformation $\bbp\to\widetilde\bbp$, so that $X$ has L\'evy triplet $(0,\tilde\b,\tilde\nu)$ under the measure $\widetilde\bbp$, where 
\begin{align}\label{nutilde}
\tilde\nu(dx)=C\Big(\frac{x}{|x|}\Big)|x|^{-Y-1}dx,
\end{align}
is the $Y$-stable L\'evy measure, and $\tilde\b$ is given by 
\begin{align*}
\tilde\b:=b+\int_{|x|\leq 1}{x}(\tilde\nu-\nu)(dx)
&= b+C(1)\int_{0}^{1}(1-\bar\q(x))x^{-Y}dx-
C(-1)\int_{-1}^{0}(1-\bar\q(x))|x|^{-Y}dx.
\end{align*} 
In particular, the centered process $(Z_t)_{t\geq 0}$, defined by
\begin{align}\label{Z}
Z_t:=X_t-\tilde\gamma t,
\end{align} 
is a strictly $Y$-stable process under $\widetilde\bbp$, and 
\begin{equation}\label{tildegamma}
\tilde\gamma := \widetilde{\bbe}(X_1) =  b+\frac{C(1)-C(-1)}{Y-1}+C(1)\int_0^1x^{-Y}(1-\bar\q(x))dx-C(-1)\int_{-1}^0|x|^{-Y}(1-\bar\q(x))dx.
\end{equation}
As is well known, necessary and sufficient {conditions} for $S_t:=S_0e^{X_t}$ to be a martingale are given by 
\begin{equation}\label{CndFrMrtX}
{\rm (i)} \;\;\int_{\bbr_0}e^{x}\nu(dx)<\infty,\quad\qquad 
{\rm (ii)}\;\;	b=-\int_{\bbr_0}(e^{x}-1-x{\bf 1}_{|x|\leq{}1})\nu(dx),
\end{equation}
and in that case, $\tilde\gamma$ can also be written as
\begin{equation*}
\tilde\gamma=-\int_0^{\infty}(e^{x}\bar{q}(x)-\bar{q}(x)-x)C\Big(\frac{x}{|x|}\Big)|x|^{-Y-1}dx.
\end{equation*}
By virtue of Theorem $33.1$ in \cite{Sato}, a necessary and sufficient condition for the measure transformation $\bbp\to\widetilde\bbp$ to be well defined is given by
\begin{align}\label{ChangeMeasureCond}
	\int_{\bbr_{0}}\big(e^{\varphi(x)/2}-1\big)^{2}\nu(dx)<\infty,
\end{align}
where, hereafter,  {$\varphi(x):=-\ln\bar\q(x)$.}
In what follows it will be useful to write the log-density process $U_{t}:={\Blue\ln}\frac{d \widetilde{\bbp}|_{\calF_{t}}}{d \bbp|_{\calF_{t}}}$ as 
\begin{align}\label{Uplusminuseta}
	U_{t}=\widetilde{U}_{t}+\eta t:=\int_{0}^{t}\int_{\bbr_{0}} \varphi(x)\bar{N}(ds,dx)
	+t\int_{\bbr_{0}}\big(e^{-\varphi(x)}-1+\varphi(x)\big)\tilde{\nu}(dx),
\end{align}
which follows from Theorem 33.2 in \cite{Sato}, and is valid provided that
\begin{align}\label{eta}
\int_{\bbr_{0}}\big|e^{-\varphi(x)}-1+\varphi(x)\big|\tilde{\nu}(dx)<\infty.
\end{align}
\noindent We shall also make use of the following decomposition
\begin{align}
Z_t=\int_0^t\int x\bar\N(ds,dx) = \int_0^t\int_0^{\infty} x\bar\N(ds,dx) + \int_0^t\int_{-\infty}^0 x\bar\N(ds,dx) =:\Z_t^{(p)} + \Z_t^{(n)},\label{ZDecomp}
\end{align}
where, under $\widetilde\bbp$, $\Z_t^{(p)}$ and $\Z_t^{(p)}$ are strictly $Y$-stable random variables with respective L\'evy measures
\begin{align}\label{nutildePM}
\tilde\nu^{(p)}(dx):=C(1)|x|^{-Y-1}{\bf 1}_{\{x>0\}}dx,\quad  \tilde\nu^{(n)}(dx):=C(-1)|x|^{-Y-1}{\bf 1}_{\{x<0\}}dx.
\end{align}  
Finally, for future reference, we denote by $L_{Z}$ the infinitesimal generator of the process $(Z_t)_{t\geq 0}$, which, for a function $g\in C^{2}_{b}$, is given by
\begin{equation}\label{DfnLZ}
	(L_Zg)(x)=\int_{\bbr_0}(g(u+x)-g(x)-ug'(x))C\Big(\frac{u}{|u|}\Big)|u|^{-Y-1}du.
\end{equation}

We conclude this section by collecting a couple of lemmas that will be needed in the sequel. The first one shows that the conditions in (\ref{NewAssumEq}) are
sufficient to justify the measure transformation $\bbp\to\widetilde\bbp$, 
as well as the representation (\ref{Uplusminuseta}).
It is a simple adaptation of Lemma A.1 in \cite{LopOla:2014}, and the proof therefore omitted (see further details in \cite{Olafsson}):
\begin{lem}\label{FirstLem}
Under (\ref{NewAssumEq}), both (\ref{ChangeMeasureCond}) and (\ref{eta}) hold true.
\end{lem}

The next lemma provides key probabilistic relationships for the skew and the delta, that are fundamental to our approach. 
Concretely, let $R_{t}:=\ln\left(S_{t}/S_{0}\right)$ be the log-return of the underlying, $C(S_{0},t,K)=\bbe\left(S_{0}e^{R_{t}}-K\right)^{+}$ be the price of an option with strike price $K$, time-to-maturity $t$, and spot price $S_{0}$, and $C^{BS}(S_{0},t,K;\sigma)$ be the price of the corresponding option 
under a Black-Scholes model with volatility $\sigma$. Then, the implied volatility $\hat{\sigma}(\kappa,t)$ and delta $\Delta(\kappa,t)$, 
parameterized in terms of the 
log-moneyness $\kappa:=\log(K/S_0)$, 
are respectively defined so that 
\[
	C^{BS}\left(S_{0},t,S_{0}e^{\kappa};\hat\sigma(\kappa,t)\right)=C\left(S_{0},t,S_{0}e^{\kappa}\right),\qquad 
	\Delta(\kappa,t)=\left.\frac{\partial C(S_{0},t,K)}{\partial S_{0}}\right|_{K=S_{0}e^{\kappa}}.
\]
Finally, denote by $\Phi$ and $\phi$ the standard Gaussian cumulative distribution and probability density functions. 
\begin{lem}\label{lem22}
Suppose that $S_t$ admits a density function. Then, 
\begin{align}\label{slope1}
{\rm (i)}\;\frac{\partial\hat\sigma(\kappa,t)}{\partial \kappa} = 
-\frac{{e^{\kappa}}\bbp\left(S_t\geq S_0e^{\kappa}\right)-{e^{\kappa}}\Phi\left(-\frac{\kappa+\half\hat\sigma^2(\kappa,t)t}{\hat\sigma(\kappa,t)\sqrt{t}}\right)}{\sqrt{t}\phi\left(\frac{-\kappa+\half\hat\sigma^2(\kappa,t)t}{\hat\sigma(\kappa,t)\sqrt{t}}\right)},\quad 
{\rm (ii)}\;\Delta(\kappa,t) = \frac{1}{S_0}{C(S_0,t,S_0e^{\kappa})} + e^{\kappa}\bbp(S_t\geq S_0e^{\kappa}).
\end{align} 
\end{lem}
Formula (\ref{slope1}-i) for the skew is well known in the literature (cf.\ \cite[Ch.\ 5]{GatheralSV}) and is a simple consequence of the implicit function theorem, together with the identity 
\begin{equation}\label{WKF}
	\frac{\partial C(S_{0},t,K)}{\partial K}=-\bbp\left(S_t\geq K\right), 
\end{equation}
which holds true, e.g., under the assumption stated in the previous lemma. Formula (\ref{ATMslope}) {for the ATM skew} is then obtained by using that $\hat\sigma(0,t)\sqrt{t}\to 0$ as $t\to 0$ (which holds under the framework considered herein \cite{LopOla:2014}, and, in fact, holds in a much more general model setting \cite{Roper:2009}), and the standard approximations $\Phi(x)=1/2+{x}/{\sqrt{2\pi}}+O(x^3)$ and $1/(\sqrt{2\pi}\phi(x))=1+x^{2}/2+O(x^{4})$, as $x\to{}0$. 
Formula (\ref{slope1}-ii) for the delta follows from the chain rule and (\ref{WKF}), and formula (\ref{Delta}) {for the ATM delta} then immediately follows by taking $\kappa=0$.

\section{Pure-jump L\'evy model}\label{SectionPureJump}
In this section, we study the short-time asymptotic behavior of the ATM implied volatility skew under the exponential L\'evy model $S_t:=S_0e^{X_t}$, where $X:=(X_{t})_{t\geq{}0}$ is a pure-jump tempered stable-like process as described in the previous section. 
For this model, a second order expansion for the ATM implied volatility is given in Theorem $3.1$ of \cite{LopOla:2014}, under a minimal integrability condition on $\bar\q$ around the origin. Specifically, it is shown that 
\begin{align}\label{sigImpX}
\hat\sigma(t)=\sqrt{2\pi}\sigma_1t^{\frac{1}{Y}-\half}+\sqrt{2\pi}\sigma_2 t^{\half}+o(t^{\half}),\quad t\to 0,
\end{align} 
with $\sigma_1:=\widetilde\bbe(Z_1^+)$, where, under $\widetilde\bbp$, $(Z_t)_{t\geq{}0}$ is a strictly stable process with L\'evy measure
$\tilde\nu(dx)=C({x}/{|x|})|x|^{-Y-1}dx$, and 
\begin{align}\label{sigImpX2}
\begin{split}
\sigma_2&:=\widetilde{\bbp}(Z_{1}< 0)\,C(1)\int_{0}^{\infty}\big(e^{x}\bar{q}(x)-\bar{q}(x)-x\big)x^{-Y-1}dx\\
 &\quad-\widetilde{\bbp}(Z_{1}\geq 0)C(-1)\int_{-\infty}^{0}\big(e^{x}\bar{q}(x)-\bar{q}(x)-x\big)|x|^{-Y-1}dx.
 \end{split}
\end{align}
As explained in the introduction, the ATM skew is then related to the probability of the process $X$ being positive. The following theorem gives an asymptotic expansion, in small time, for such a probability, which sometimes is termed the positivity parameter of a process (cf.\ \cite[p.\ 218]{Bertoin}). The expansion is explicit up to a term of order $O(t^{1/Y})$ (see the subsequent Remark \ref{ExplExprCoe}). Below, $\tilde\gamma$ is as in Eq.\ (\ref{tildegamma}), $Z_1^{(p)}$ and $Z_1^{(n)}$ are as in Eq.~(\ref{ZDecomp}), and, finally, $f_{Z_{1}^{(p)}}$, $f_{Z_{1}^{(n)}}$, and $f_Z$, are the probability density functions of $Z_1^{(p)}$, $Z_1^{(n)}$, and $Z_1=Z_1^{(p)}+Z_1^{(n)}$, respectively.

\begin{thm}\label{thmX}
Let $X$ be a tempered stable-like L\'evy process with a L\'evy measure as described in (\ref{tmpstblnu}). Furthermore, assume that the condition 
\begin{align}\label{origin}
\int_{|x|\leq{}1}\big|\bar{q}(x)-1-\alpha\Big(\frac{x}{|x|}\Big)x\big||x|^{-Y-1}dx<\infty, 
\end{align}
is satisfied for some constants $\alpha(1),\alpha(-1)\in \bbr$. Then,
\begin{align}\label{probX}
\bbp(X_t\geq 0)-\widetilde\bbp(Z_1\geq 0)
=\sum_{k=1}^nd_{k}t^{k\left(1-\frac{1}{Y}\right)}+e\,t^{\frac{1}{Y}}+f\,t+o(t),\quad t\to 0, 
\end{align}
where $n:=\max\{k\geq 3:k(1-{1}/{Y})\leq 1\}$, and 
\begin{align}
d_{k} &:= \frac{(-1)^{k-1}}{k!}\tilde\gamma^{k}f^{(k-1)}_Z(0),\quad 1\leq k\leq{}n,\label{d}\\
e &:= \alpha(1)\widetilde\bbe\big(Z_1^{(p)}{\bf 1}_{\{Z_1^{(p)} + Z_1^{(n)}\geq 0\}}\big)
+ \alpha(-1)\widetilde\bbe\big(Z_1^{(n)}{\bf 1}_{\{Z_1^{(p)} + Z_1^{(n)}\geq 0\}}\big),\label{e}\\
f &:= \tilde\gamma{(\alpha(1)-\alpha(-1))}\widetilde\bbe\big(Z_1^{(p)}f_{Z_1^{(n)}}\big(-Z_1^{(p)}\big)\big)\label{d1X}
\nonumber\\
&\quad + \widetilde\bbp(Z_1\leq 0) C(1)\int_{0}^{\infty}(\bar\q(x)-1-\alpha(1)x)x^{-Y-1}dx \\
&\quad -\widetilde\bbp(Z_1> 0)C(-1)\int_{-\infty}^0 (\bar\q(x)-1-\alpha(-1)x)|x|^{-Y-1}dx.\nonumber
\end{align}
\end{thm}
\begin{rem}\label{ExmplTSP}
	The processes covered by Theorem \ref{thmX} include stable processes, where $\bar\q(x)\equiv 1$, and tempered stable processes as defined in \cite{CT04}, where $\bar{q}(x)=e^{-\alpha(1)x}{\bf 1}_{\{x>0\}}+e^{\alpha(-1)x}{\bf 1}_{\{x<0\}}$, with $\alpha(1),\alpha(-1)>0$. 
	They are of particular importance for practical applications, and will be studied numerically in Section \ref{Examples}. It is also important to note that condition (\ref{origin}) is the minimal condition needed for the expansion to be valid. That is, if (\ref{origin}) does not hold, the coefficient $f$ is not well defined.
\end{rem}

If the L\'evy triplet $(0,b,\nu)$ of $X$ satisfies the martingale condition (\ref{CndFrMrtX}), then the previous result can be interpreted as an asymptotic expansion for ATM digital call option prices. Together with (\ref{ATMslope}) and (\ref{sigImpX})-(\ref{sigImpX2}), it then gives an asymptotic expansion for the ATM implied volatility skew. Moreover, together with (\ref{Delta}) and Theorem 3.1 of \cite{LopOla:2014}, it also gives an asymptotic expansion for the delta of ATM call options:
\begin{cor}\label{corX}
Let $X$ be a tempered stable-like L\'evy process as in Theorem \ref{thmX}, with $b$ and $\nu$ satisfying (\ref{CndFrMrtX}), so that $S_{t}:=S_0e^{X_{t}}$ is a martingale. Then,
\begin{enumerate}
\item
The ATM implied volatility skew satisfies
\begin{align}\label{slopeX}
\bigg.\frac{\partial\hat\sigma(\kappa,t)}{\partial\kappa}\bigg|_{\kappa=0} 
&= \sqrt{\frac{2\pi}{t}}\bigg(\half-\widetilde\bbp(Z_1\geq 0) 
-\sum_{k=1}^nd_{k}t^{k\left(1-\frac{1}{Y}\right)}
-\Big(e+\frac{\sigma_1}{2}\Big)t^{\frac{1}{Y}}
-\Big(f+\frac{\sigma_2}{2}\Big)t+o(t)\bigg),\quad t\to 0,
\end{align}
with $\sigma_1:=\widetilde\bbe\left(Z_1^+\right)$ and $\sigma_2$ as in (\ref{sigImpX2}).
\item
The delta of an ATM call option satisfies
\begin{align}
\Delta(t)& = \widetilde\bbp(Z_1\geq 0)
+\sum_{k=1}^nd_{k}t^{k\left(1-\frac{1}{Y}\right)} +(\sigma_1+e)\,t^{\frac{1}{Y}}+(\sigma_2+f)\,t+o(t),\quad t\to 0.
\end{align}
\end{enumerate}
\end{cor}

\begin{rem}\label{remark1}
A few comments are in order:\hfill

\noindent{\rm (a)}
	It is important to point out that the leading order term of (\ref{probX}) is $d_{1}t^{1-{1}/{Y}}$ for all $Y\in(1,2)$. Furthermore, 
	the coefficients of (\ref{probX}) can be ranked as 
	\[
		d_{1}t^{1-\frac{1}{Y}}\succ{}\dots\succ{}d_{m}t^{m\left(1-\frac{1}{Y}\right)}\succeq {e t^{\frac{1}{Y}}}\succ{}d_{m+1}t^{(m+1)\left(1-\frac{1}{Y}\right)}\succ\dots \succ d_{n}t^{n\left(1-\frac{1}{Y}\right)}\succeq {f\, t,}\qquad t\to{}0,
	\]
	where $m=\max\{k\geq{}2: k(1-1/Y)\leq 1/Y\}$ and, as usual, $h(t)\succ g(t)$ (resp.\ $h(t)\succeq g(t)$) if $g(t)=o(h(t))$ (resp. $g(t)=O(h(t))$), as $t\to 0$. This can be compared to the expansion for ATM option prices given in Theorem $3.1$ of \cite{LopOla:2014}, where the first and second order terms are of order $t^{{1}/{Y}}$ and $t$, respectively.

\noindent{\rm (b)}
	It is informative to note that the summation term in (\ref{probX}) comes from expanding the probability of a stable process with drift being positive. Specifically, we have
	\begin{align*}
	\widetilde\bbp(Z_t + \tilde\gamma t\geq 0) - \widetilde\bbp(Z_1\geq 0) 
	=\sum_{k=1}^n d_{k}t^{k\left(1-\frac{1}{Y}\right)} + O\big(t^{(n+1)\left(1-\frac{1}{Y}\right)}\big),\quad t\to 0.
	\end{align*}
	The other terms, $e$ and $f$, arise as a result of the discrepancy between $X$ and a stable process. In particular, this implies that the probability of $X$ being positive at time $t$ can be approximated, for small $t$, by the analogous probability for a stable process, up to an error term of order $O(t^{{1}/{Y}})$. Similarly, the same probability can be approximated by that of a tempered stable process (as defined in Remark \ref{ExmplTSP}), up to an error term of order $O(t)$.

\noindent{\rm (c)}
	As mentioned in the introduction, the implied volatility of OTM options explodes as $t\to 0$, while for ATM options it converges to the volatility of the continuous component. Here we see that the ATM implied volatility slope also blows up as $t\to 0$, with a sign that can easily be recovered from the model parameters. Indeed, when $Z_1$ is symmetric (i.e. $C(1)=C(-1)$), it is of order $t^{1/2-{1}/{Y}}$, with the same sign as the parameter $\tilde\gamma$, i.e. the center of $X$ under $\widetilde\bbp$, but, when $C(1)>C(-1)$ (resp. $C(1)<C(-1)$), it is of order $t^{-1/2}$ with a negative (resp. positive) sign. 
\end{rem}

\begin{rem}\label{ExplExprCoe}
There exist explicit expressions for $\widetilde{\bbe}(Z_{1}^{+})$ and $\widetilde{\bbp}(Z_{1}\geq 0)$ (see \cite{LopGonHou:2014} and references therein):
\begin{align*}
\widetilde{\bbe}\left(Z_{1}^{+}\right)&=\frac{A^{\frac{1}{Y}}}{\pi}\Gamma(-Y)^{\frac{1}{Y}}\Big|\cos\Big(\frac{\pi Y}{2}\Big)\Big|^{\frac{1}{Y}}\cos\Big(\frac{1}{Y}\arctan\Big(\frac{B}{A}\tan\left(\frac{Y\pi}{2}\right)\Big)\Big)\Gamma\left(1-\frac{1}{Y}\right)\Big(1+\left(\frac{B}{A}\right)^{2}\tan^{2}\left(\frac{\pi Y}{2}\right)\Big)^{\frac{1}{2Y}},\\
\widetilde{\bbp}(Z_{1}\geq{}0)&=\frac{1}{2}+\frac{1}{\pi Y}\arctan\Big(\frac{B}{A}
	\tan\Big(\frac{Y\pi}{2}\Big)\Big),
\end{align*}
where $A:=C(1)+C(-1)$ and $B:=C(1)-C(-1)$. The derivatives $f_{Z}^{(k-1)}(0)$ can also be explicitly computed from the polynomial expansion for the stable density (see, e.g., Eq.~(4.2.9) in \cite{Zolotarev}). Indeed, it follows that
\[
	f_{Z}^{(k-1)}(0)=(-1)^{k-1}\frac{\Gamma\Big(\frac{k}{Y}+1\Big)}{k \pi}\sin({\rho k\pi})\Big(\frac{c_0}{c}\Big)^{\frac{k}{Y}},
\]
where 
\begin{align*}
	\rho = \frac{\delta+Y}{2Y}, \qquad
	\delta=\frac{2}{\pi}\arctan\Big(\beta\tan\left(\frac{Y\pi}{2}\right)\Big), \qquad
	c_{0}=\cos\Big(\arctan\Big(\beta\tan\Big(\frac{\pi Y}{2}\Big)\Big)\Big),
\end{align*}
and $\beta=({C(1)-C(-1)})/({C(1)+C(-1)})$ and $c=-\Gamma(-{Y})\cos({\pi{Y}}/{2})(C(1)+C(-1))$ are the skewness and scale parameters of $Z_1$.
\end{rem}

{\smallskip
{\noindent\textbf{Proof of Theorem \ref{thmX}.}}}

\noindent
\textbf{Step 1:} Let $X$ be a tempered stable-like process as in the statement of the theorem. In this step, we will show that (\ref{probX}) holds under the additional assumptions that the $\bar\q$-function of $X$ satisfies (\ref{NewAssumEq}-ii) and (\ref{NewAssumEq}-iii), so that 
Lemma \ref{FirstLem} is valid. Throughout, we use the notation introduced in the previous section. Let us start by noting {that
\begin{align}\label{Decomp}
\bbp(X_t\geq 0) - \widetilde\bbp(Z_1\geq 0)
= \widetilde\bbe\big({\bf 1}_{\{Z_1 \geq -\tilde\gamma t^{1-\frac{1}{Y}}\}} -{\bf 1}_{\{Z_1\geq 0\}}\big)
+ \widetilde\bbe\big(\big(e^{-U_t}-1\big){\bf 1}_{\{Z_t \geq -\tilde\gamma t\}}\big)
=: 
I_1(t)+I_2(t),
\end{align}
and} we look at each of the two terms separately. For the first one, we have 
\begin{align}\label{DecompI1}
I_1(t) = \widetilde\bbp(Z_1 \geq -\tilde\gamma t^{1-\frac{1}{Y}})-\widetilde\bbp\left(Z_1\geq 0\right)
=\int_{-\tilde\gamma t^{1-\frac{1}{Y}}}^{0}f_Z(z)dz, 
\end{align}
and, since $f_Z$ is a smooth function (see e.g.\ \cite{Sato}, Prop. $28.3$), {we can use its Maclaurin series expansion to show that
\begin{align}\label{AsyI11st}
I_1(t) 
&= \sum_{n=1}^{N}\frac{(-1)^{n+1}\tilde\gamma^{n}}{n!}f_Z^{(n-1)}(0)t^{n\left(1-\frac{1}{Y}\right)} + O\big(t^{(N+1)\left(1-\frac{1}{Y}\right)}\big),\qquad t\to 0.
\end{align}
For $I_{2}$, we further decompose it as
\begin{align}\label{DecompI2}
I_2(t) 
&=  \widetilde\bbe\big(\big(e^{-\widetilde\U_t}-1\big){\bf 1}_{\{Z_t \geq -\tilde\gamma t\}}\big) 
+ (e^{-\eta t}-1)\widetilde\bbe\big(\big(e^{-\widetilde\U_t}-1\big){\bf 1}_{\{Z_t \geq -\tilde\gamma t\}}\big)
+\big(e^{-\eta t}-1\big)\widetilde\bbe\big({\bf 1}_{\{Z_t \geq -\tilde\gamma t\}}\big)\nonumber\\
&=: I_2^1(t)+I_2^2(t)+I_2^3(t),
\end{align}
where} it is clear that
\begin{align}
I_2^2(t) = o(t),\qquad
I_2^3(t) =
 -\eta\,\widetilde\bbp\left(Z_1\geq 0\right)t+o(t),\qquad t\to 0.
\end{align}
We use Fubini's theorem on the first term to write 
\begin{align}\label{Decomp121}
I_2^1(t) &= \widetilde\bbe\big(\big(e^{-\widetilde\U_t}-1+\widetilde\U_t\big){\bf 1}_{\left\{Z_t \geq -\tilde\gamma t\right\}}\big)-\widetilde\bbe\big(\widetilde\U_t{\bf 1}_{\left\{Z_t \geq -\tilde\gamma t\right\}}\big)\nonumber\\
&= \int_{-\infty}^0(e^{-x}-1)\widetilde\bbp\big(Z_t \geq -\tilde\gamma t,\widetilde\U_t\leq x\big)dx
-\int_{0}^{\infty}(e^{-x}-1)\widetilde\bbp\big(Z_t \geq -\tilde\gamma t,\widetilde\U_t\geq x\big)dx
-\widetilde\bbe\big(\widetilde\U_t{\bf 1}_{\left\{Z_t \geq -\tilde\gamma t\right\}}\big)\nonumber\\
&=:J_2^{1}(t)+J_2^{2}(t)+J_2^{3}(t).
\end{align}
Analogous arguments to those in \cite[Eqs.~(A.11)-(A.14)]{LopOla:2014} can be used to apply the dominated convergence theorem and obtain:
\begin{align}\label{J21}
\lim_{t\to 0}\frac{1}{t}J_2^1(t) 
&= \int_{-\infty}^{0}(e^{-x}-1)\lim_{t\to 0}\frac{1}{t}\widetilde\bbp\big(Z_t \geq -\tilde\gamma t,\widetilde\U_t\leq x\big)dx
= \int_{-\infty}^{0}(e^{-x}-1)\int_{0}^{\infty}{\bf 1}_{\{\varphi(y)\leq x\}}\tilde{\nu}(dy)dx=:\vartheta_1,\\
\label{J22}
\lim_{t\to 0}\frac{1}{t}J_2^2(t)
&= -\int_{0}^{\infty}(e^{-x}-1)\lim_{t\to 0}\frac{1}{t}\widetilde\bbp\big(Z_t \geq -\tilde\gamma t,\widetilde\U_t\geq x\big)dx
= -\int_{0}^{\infty}\left(e^{-x}-1\right)\int_{0}^{\infty}{\bf 1}_{\{\varphi(y)\geq x\}}\tilde{\nu}(dy)dx=:\vartheta_2.
\end{align}
Finally, to deal with the third term of (\ref{Decomp121}), we decompose $\widetilde\U_t=\int_0^t\int\varphi(x)\bar\N(ds,dx)$ as
\begin{align}\label{UDecomp}
\widetilde\U_t=\int_0^t\int\big(\varphi(x)+\alpha\Big(\frac{x}{|x|}\Big)x\big)\bar\N(ds,dx)-\int_0^t\int \alpha\Big(\frac{x}{|x|}\Big)x\bar\N(ds,dx) =: \widetilde\U_t^{(1)} - \widetilde\U_t^{(2)},
\end{align}
so that
\begin{align}\label{J23decomp}
J_2^3(t)=-\widetilde\bbe\big(\widetilde\U_t^{(1)}{\bf 1}_{\left\{Z_t \geq -\tilde\gamma t\right\}}\big) 
+\widetilde\bbe\big(\widetilde\U_t^{(2)}{\bf 1}_{\left\{Z_t \geq -\tilde\gamma t\right\}}\big)
=:-J_2^{31}(t) + J_2^{32}(t).
\end{align}
First, for $J_2^{32}(t)$, note that
\begin{align}\nonumber
\widetilde\bbe\big(Z_t^{(p)}{\bf 1}_{\left\{Z_t\geq -\tilde\gamma t\right\}}\big)
&= \widetilde\bbe\big(Z_t^{(p)}{\bf 1}_{\left\{Z_t\geq 0\right\}}\big) + \widetilde\bbe\big(Z_t^{(p)}\big({\bf 1}_{\left\{Z_t^{(p)}+Z_t^{(n)}\geq -\tilde\gamma t\right\}}-{\bf 1}_{\left\{Z_t^{(p)}+Z_t^{(n)}\geq 0\right\}}\big)\big)\\
&= t^{\frac{1}{Y}}\widetilde\bbe\big(Z_1^{(p)}{\bf 1}_{\left\{Z_1\geq 0\right\}}\big)
+ t^{\frac{1}{Y}}\widetilde\bbe\Big(Z_1^{(p)}{\int_{-\tilde\gamma t^{1-\frac{1}{Y}}-Z_1^{(p)}}^{-Z_1^{(p)}}f_{Z_1^{(n)}}(z)dz}
\Big)\nonumber
\\
&= t^{\frac{1}{Y}}\widetilde\bbe\big(Z_1^{(p)}{\bf 1}_{\left\{Z_1\geq 0\right\}}\big)
+ \tilde\gamma t\,\widetilde\bbe\big(Z_1^{(p)}f_{Z_1^{(n)}}(-Z_1^{(p)})\big)+o(t),\quad t\to 0, \label{FLHG1}
\end{align}
since $\displaystyle\sup_{z\in\bbr}f_{Z_1^{(n)}}(z)<\infty$. Similarly,
\begin{align}
\widetilde\bbe\big(Z_t^{(n)}{\bf 1}_{\left\{Z_t\geq -\tilde\gamma t\right\}}\big)
&= t^{\frac{1}{Y}}\widetilde\bbe\big(Z_1^{(n)}{\bf 1}_{\left\{Z_1\geq 0\right\}}\big)
+ \tilde\gamma t\,\widetilde\bbe\big(Z_1^{(n)}f_{Z_1^{(p)}}(-Z_1^{(n)})\big)+o(t),\quad t\to 0.\label{FLHG2}
\end{align}
From (\ref{FLHG1})-(\ref{FLHG2}) and the fact that $\widetilde\bbe\big(Z_1^{(n)}f_{Z_1^{(p)}}(-Z_1^{(n)})\big)=-\widetilde\bbe\big(Z_1^{(p)}f_{Z_1^{(n)}}(-Z_1^{(p)})\big)$, we {get
\begin{align}\label{J23}
J_2^{32}(t)
&= t^{\frac{1}{Y}}\big(\alpha(1)\widetilde\bbe\big(Z_1^{(p)}{\bf 1}_{\left\{Z_1\geq 0\right\}}\big)
+\alpha(-1)\widetilde\bbe\left(Z_1^{(n)}{\bf 1}_{\left\{Z_1\geq 0\right\}}\right)\big)+ \tilde\gamma t(\alpha(1)-\alpha(-1))\widetilde\bbe\big(Z_1^{(p)}f_{Z_1^{(n)}}\big(-Z_1^{(p)}\big)\big)
+o(t).
\end{align}
For} $J_2^{31}$, we will show that
\begin{align}\label{Ut1}
J_2^{31}(t) = \widetilde\bbe\big(\widetilde\U_t^{(1)}{\bf 1}_{\left\{Z_t \geq -\tilde\gamma t\right\}}\big) 
= \vartheta t + o(t),\quad t\to 0,
\end{align}
where
\begin{align}\label{vartheta}
\vartheta :=C(1)\widetilde\bbp(Z_1\leq 0)\int_0^{\infty}(\alpha(1)x-\ln\bar\q(x))x^{-Y-1}dx
-C(-1)\widetilde\bbp(Z_1\geq 0)\int_{-\infty}^0(\alpha(-1)x-\ln\bar\q(x))|x|^{-Y-1}dx.
\end{align}
Combining (\ref{DecompI2})-(\ref{vartheta}) then gives an asymptotic expansion for $I_2(t)$, which together with (\ref{Decomp}) and (\ref{AsyI11st}) yields (\ref{probX}), after some standard simplifications. To complete the proof we therefore only need to show (\ref{Ut1}).
In order to do that, define $f(x):=\varphi(x)+\alpha({x/|x|})x$, and, for $\varepsilon>0$, further decompose $\widetilde\U_t^{(1)}$ as
\begin{align}\label{U1Decomp}
\widetilde\U_t^{(1)}=\int_0^t\int f(x)\bar\N(ds,dx)
=\int_0^t\int_{|x|\leq\varepsilon} f(x)\bar\N(ds,dx) + \int_0^t\int_{|x|>\varepsilon} f(x)\bar\N(ds,dx) 
=:\widetilde\U_{\varepsilon}^{(1,1)}(t) + \widetilde\U_{\varepsilon}^{(1,2)}(t),
\end{align}
and let
\begin{align}\label{J231}
J_2^{31}(t) = \widetilde\bbe\big(\widetilde\U_{\varepsilon}^{(1,1)}(t){\bf 1}_{\left\{Z_t \geq -\tilde\gamma t\right\}}\big)
+\widetilde\bbe\big(\widetilde\U_{\varepsilon}^{(1,2)}(t){\bf 1}_{\left\{Z_t \geq -\tilde\gamma t\right\}}\big):=\tilde\J_ {1,\varepsilon}(t)+\tilde\J_{2,\varepsilon}(t).
\end{align}
For future reference, recall that $\varphi(x)=-\ln\bar\q(x)$, and 
\begin{equation}\label{Bndfwrtnu}
	\int |f(x)|\tilde\nu(dx)
	\leq \int\big|\alpha\Big(\frac{x}{|x|}\Big)x+1-\bar{q}(x)\big|\tilde\nu(dx)
	+\int \big|\bar{q}(x)-1-\ln\bar{q}(x)\big|\tilde\nu(dx)  <\infty,
\end{equation}
in light of (\ref{origin}), the boundedness of $\bar\q$, and the fact that (\ref{NewAssumEq}) implies (\ref{eta}) as proved in Lemma \ref{FirstLem}.
Now note that $\widetilde\U_{\varepsilon}^{(1,2)}(t)$ is a compound Poisson process with drift; i.e., we can write
\begin{align*}
\widetilde\U_{\varepsilon}^{(1,2)}(t)=\beta^{(\varepsilon)} t+\sum_{i=1}^{N_t^{(\varepsilon)}}f(\xi_i^{(\varepsilon)}), 
\end{align*}
where $\beta^{(\varepsilon)}:= -\int_{|x|>\varepsilon}f(x)\tilde\nu(dx)$, $\big(N_t^{(\varepsilon)}\big)_{t\geq 0}$ is a counting process with intensity $\lambda^{(\varepsilon)}:=\int_{|x|>\varepsilon}\tilde\nu(dx)$, and $\big(\xi_i^{(\varepsilon)}\big)_{i\in\bbn}$ are i.i.d. random variables with probability measure $\tilde\nu(dx){\bf 1}_{\{|x|>\varepsilon\}}/\lambda^{(\varepsilon)}$. We can also write 
\begin{align*}
Z_t=\int_0^t\int x\bar\N(ds,dx) = \int_0^t\int_{|x|\leq\varepsilon}x\bar\N(ds,dx) + \sum_{i=1}^{N_t^{(\varepsilon)}}\xi_i^{(\varepsilon)} - t\int_{|x|>\varepsilon}x\tilde\nu(dx) =: \breve\Z_t^{(\varepsilon)} + \sum_{i=1}^{N^{(\varepsilon)}_t}\xi^{(\varepsilon)}_i + c^{(\varepsilon)}t,
\end{align*}
and, under $\widetilde{\bbp}$, $t^{-\frac{1}{Y}}\breve\Z_t^{(\varepsilon)}\ld  Z_1$ as $t\to 0$ 
(see \cite{rosenbaum.tankov.10}, Proposition 1). Then, by conditioning on $N_t^{(\varepsilon)}$, we have
\begin{align}\label{J312}
\tilde\J_{2,\varepsilon}(t)
&=e^{-\lambda^{(\varepsilon)} t}\beta^{(\varepsilon)} t\,\widetilde\bbp\big(\big.Z_t\geq -\tilde\gamma t\big|N^{(\varepsilon)}_t=0\big)
+\lambda^{(\varepsilon)} te^{-\lambda^{(\varepsilon)} t}\,\widetilde\bbe\big(\big.\big(\beta^{(\varepsilon)} t+f(\xi^{(\varepsilon)}_1)\big){\bf 1}_{\{Z_t\geq -\tilde\gamma t\}}\big|N_t^{(\varepsilon)}=1\big)+o(t)\nonumber\\
&=e^{-\lambda^{(\varepsilon)} t}\beta^{(\varepsilon)} t\,\widetilde\bbp\big(t^{-\frac{1}{Y}}\breve\Z_t^{(\varepsilon)}\geq -(\tilde\gamma+c^{(\varepsilon)}) t^{1-\frac{1}{Y}}\big)
+\lambda^{(\varepsilon)} te^{-\lambda^{(\varepsilon)} t}\,\widetilde\bbe\big(f(\xi_1^{(\varepsilon)}){\bf 1}_{\{ \breve\Z_t^{(\varepsilon)}+\xi^{(\varepsilon)}_1\geq -(\tilde\gamma+c^{(\varepsilon)}) t\}}\big)+o(t)\nonumber\\ 
&= \vartheta^{(\varepsilon)}t + o(t),\quad t\to 0,
\end{align}
where
\begin{align*} 
\vartheta^{(\varepsilon)}
&:=\beta^{(\varepsilon)}\,\widetilde\bbp(\Z_1\geq 0)
+\lambda^{(\varepsilon)}\,\widetilde\bbe\big(f(\xi^{(\varepsilon)}_1){\bf 1}_{\{\xi^{(\varepsilon)}_1>0\}}\big)\nonumber\\
&=C(1)\widetilde\bbp(\Z_1\leq 0)\int_{\varepsilon}^{\infty}\left(\alpha(1)x-\ln\bar\q(x)\right)x^{-Y-1}dx
-C(-1)\widetilde\bbp\left(\Z_1\geq 0\right)\int_{-\infty}^{-\varepsilon}\left(\alpha(-1)x-\ln\bar\q(x)\right)|x|^{-Y-1}dx,
\end{align*}
and the second equality follows from standard simplifications. Moreover, in light of (\ref{Bndfwrtnu}), 
\begin{align}\label{varthetadiff}
\vartheta-\vartheta^{(\varepsilon)} 
&\longrightarrow 0,\quad \varepsilon\to 0.
\end{align}
For $\widetilde\U_{\varepsilon}^{(1,1)}(t)$ we note {that, by (\ref{Bndfwrtnu})
and Theorem 10.15 in \cite{Kallenberg}, 
we have}
\begin{align}\label{SENH}
|\tilde\J_{1,\varepsilon}(t)| \leq \widetilde\bbe\big|\widetilde\U_{\varepsilon}^{(1,1)}(t)\big| \leq 2t\int_{|x|\leq\varepsilon}|f(x)|\tilde\nu(dx)=:K^{(\varepsilon)}t\to 0,\quad \text{ as } \varepsilon\to 0.
\end{align}
Finally, (\ref{Ut1}) follows since, by (\ref{J231}), (\ref{J312}) and (\ref{SENH}),
\begin{align*}
	-K^{(\varepsilon)}+\vartheta^{(\varepsilon)}\leq \liminf_{t\to{}0}\frac{\J^{31}_{2}(t)}{t}
	\leq \limsup_{t\to{}0}\frac{\J^{31}_{2}(t)}{t}\leq  K^{(\varepsilon)}+\vartheta^{(\varepsilon)},
\end{align*}
and the lower and upper bounds converge to $\vartheta$ as $\varepsilon\to{}0$ in view of (\ref{varthetadiff}) and (\ref{SENH}).

\noindent
{\textbf{Step 2:}  Now} assume that $X$ is a tempered stable-like process whose $\bar\q$-function satisfies (\ref{origin}), but not necessarily the additional conditions imposed in Step $1$: (\ref{NewAssumEq}-ii) and (\ref{NewAssumEq}-iii). We would like to approximate it by a process whose $\bar\q$-function satisfies those conditions, and for which the result (\ref{probX}) is therefore known by Step $1$. To do that, first note that since $\bar\q(x)\to 1$ as $x\to 0$, we can find $\varepsilon_{0}>0$ such that {$\inf_{|x|\leq\varepsilon_{0}}\bar\q(x)>0$}. Next, for each $\delta>0$, let $(\Omega^{(\delta)},\calF^{(\delta)},\bbp^{(\delta)})$ be an extension of the original probability space $(\Omega,\calF,\bbp)$, carrying a L\'evy process $R^{(\delta)}$, independent of the original process $X$, with L\'evy triplet $(0,\beta^{(\delta)},\nu_R^{(\delta)})$ given by
\begin{align}
	\nu_R^{(\delta)}(dx)  := C\Big(\frac{x}{|x|}\Big)e^{-\frac{|x|}{\delta}}{\bf 1}_{\{|x|\geq{}\varepsilon_0\}}|x|^{-Y-1}dx,\quad\quad
	\beta^{(\delta)} := \int_{|x|\leq 1}x\nu_R^{(\delta)}(dx).
\end{align} 
In particular, $R^{(\delta)}$ is a compound Poisson process and can be written as
\begin{align}
R_t^{(\delta)} = \sum_{i=1}^{N^{(\delta)}_t}\xi_i^{(\delta)},
\end{align}
where $(N^{(\delta)}_t)_{t\geq 0}$ is a Poisson process with intensity $\lambda^{(\delta)}:=\int_{|x|\geq\varepsilon_0}\nu_R^{(\delta)}(dx)$, and $(\xi_i^{(\delta)})_{i\in\bbn}$ are i.i.d.\ random variables with probability measure $\nu_R^{(\delta)}(dx)/\lambda^{(\delta)}$. Let us recall that, by the definition of a probability space extension (see \cite{Kallenberg}), the law of $X$ under $\bbp^{(\delta)}$ remains unchanged. Also, all expected values in the sequel will be taken with respect to the extended probability measure $\bbp^{(\delta)}$, so for simplicity we denote the expectation under $\bbp^{(\delta)}$ by $\bbe$. Next, we approximate the law of the process $X$ with that of the following process, again defined on the extended probability space $(\Omega^{(\delta)},\calF^{(\delta)},\bbp^{(\delta)})$:
\begin{align}
	{X}_{t}^{(\delta)}:=X_t+R_t^{(\delta)}.
\end{align}
{Then,} the L\'evy triplet $(0,b^{(\delta)},\nu^{(\delta)})$ of ${X}^{(\delta)}$ is given by
\begin{align}
	b^{(\delta)} := b + \beta^{(\delta)},\quad
	{\nu}^{(\delta)}(dx) :=C\Big(\frac{x}{|x|}\Big)|x|^{-Y-1} {\bar{q}^{(\delta)}(x)}dx:=C\Big(\frac{x}{|x|}\Big)|x|^{-Y-1}\big(\bar\q(x)+e^{-|x|/\delta}{\bf 1}_{|x|\geq{}\varepsilon_0}\big)dx,		
\end{align}
so it is clear that ${\bar{q}^{(\delta)}}$ satisfies the conditions in (\ref{NewAssumEq}).
Hence, the probability measure $\widetilde\bbp$ can be defined as described in {Section \ref{TSP}}, using the jump measure {of $X^{(\delta)}$}, and note that for $\tilde\gamma^{(\delta)}:=\widetilde\bbe(X_1^{(\delta)})$, we have, using the expression (\ref{tildegamma}),
\begin{align}
\tilde\gamma^{(\delta)} &=  b^{(\delta)}+\frac{C(1)-C(-1)}{Y-1}+C(1)\int_0^1x^{-Y}\big(1-\bar\q^{(\delta)}  (x)\big)dx-C(-1)\int_{-1}^0|x|^{-Y}\big(1-\bar\q^{(\delta)}(x)\big)dx 
= \tilde\gamma\nonumber,
\end{align}
after plugging in the above expressions for $b^{(\delta)}$ and $\bar\q^{(\delta)}$. Now, since $\bar\q^{(\delta)}$ satisfies the conditions in Step $1$, we know that (\ref{probX}) holds for $X^{(\delta)}$, i.e.\ that
\begin{align}\label{ScnOrderApM}
\bbp(X_t^{(\delta)}\geq 0) - \widetilde\bbp(Z_1\geq 0)=\sum_{k=1}^n d_{k}t^{k\left(1-\frac{1}{Y}\right)}+e\, t^{\frac{1}{Y}}+f^{(\delta)}t+o(t),\quad t\to 0, 
\end{align}
where $n:=\max\{k\geq 3:k\left(1-{1}/{Y}\right)\leq 1\}$, $d_k$ and $e$ are independent of $\delta$ and given by (\ref{d})-(\ref{e}), and
\begin{align}\label{d1delta}
\begin{split}
	f^{(\delta)} := \tilde\gamma(\alpha(1)-\alpha(-1))\widetilde\bbe\big(Z_1^{(p)}f_{Z_1^{(n)}}\big(-Z_1^{(p)}\big)\big)
&+ \widetilde\bbp(Z_1\leq 0) C(1)\int_{0}^{\infty}(\bar\q^{(\delta)}(x)-1-\alpha(1)x)x^{-Y-1}dx \\
& -\widetilde\bbp(Z_1> 0)C(-1)\int_{-\infty}^0 (\bar\q^{(\delta)}(x)-1-\alpha(-1)x)|x|^{-Y-1}dx.
\end{split}
\end{align}
Now, from the triangle inequality, it follows that
\begin{align}\label{tri}
\bbp(X_t^{(\delta)}\geq 0) - \big|\bbp(X_t\geq 0)-\bbp(X_t^{(\delta)}\geq 0)\big| 
\leq \bbp(X_t\geq 0) 
\leq \bbp(X_t^{(\delta)}\geq 0) + \big|\bbp(X_t\geq 0)-\bbp(X_t^{(\delta)}\geq 0)\big|,
\end{align}
and, by conditioning on the number of jumps of the process $R^{(\delta)}$, we have
\begin{align}\label{diffXY}
\mathcal{R}_{t}^{(\delta)}
:=\big|\bbp(X_t\geq 0)-\bbp(X_t^{(\delta)}\geq 0)\big|
=\lambda^{(\delta)}te^{\lambda^{(\delta)}t}\big|\bbp(X_t\geq 0)-\bbp(X_t+\xi_{1}^{(\delta)}\geq 0)\big|+o(t),\quad t\to 0,
\end{align}  
which, in particular,  implies that 
\begin{align}\label{LFRN}
\lim_{\delta\to 0}\lim_{t\to 0}t^{-1}\calR_t^{(\delta)}= 0, 
\end{align}
since $\lambda^{(\delta)}=\int_{|x|\geq\varepsilon_0}e^{-|x|/\delta}\tilde\nu(dx)\to 0$, as $\delta\to 0$.
By subtracting $\widetilde\bbp\left(Z_1\geq 0\right)+\sum_{k=1}^n d_{k}t^{k\left(1-{1}/{Y}\right)}+e\,t^{{1}/{Y}}$ from the inequalities in (\ref{tri}), applying the expansion (\ref{ScnOrderApM}), dividing by $t$, taking the limit as $t\to{}0$, and using (\ref{LFRN}), it is clear that 
\[
	\lim_{t\to{}0}\frac{1}{t}\Big(\bbp\left(X_t\geq 0\right)-\widetilde\bbp\left(Z_1\geq 0\right)-\sum_{k=1}^n  d_{k}t^{k\left(1-\frac{1}{Y}\right)}-e\,t^{\frac{1}{Y}}\Big)=\lim_{\delta\to{}0}f^{(\delta)}.
\]
Therefore, to conclude, it suffices to show that
$\displaystyle\lim_{\delta\to 0}f^{(\delta)}= f$, with $f$ as in (\ref{d1X}), 
which follows from (\ref{d1delta}) and the dominated convergence theorem.
\hfill\qed

\section{L\'evy jump model with stochastic volatility}\label{SectionContComp}
In this section we consider the case when an independent continuous component is added to the pure-jump L\'evy process $X$. Concretely, let $S_t:=S_0e^{X_t+V_t}$, with $X$ as in the previous section, while for the continuous component, $V$, we consider an independent stochastic volatility process of the form {(\ref{modelVY})},
defined on the same probability space as $X$, {where $(W_t^1)_{t\geq 0}$ and $(W_t^2)_{t\geq 0}$ are standard Brownian motions, relative to the filtration $(\calF_t)_{t\geq 0}$, $-1<\rho<1$, and $\alpha,\gamma,\mu$, and $\sigma$, are such that $V$ and $Y$ are well defined. Moreover, it is assumed that $\sigma_0:=\sigma(y_0)>0$, and that there exists a bounded open interval $I$, containing $y_0$, on which the function $\alpha$ is bounded, $\gamma$ and $\mu$ are Lipschitz continuous, and $\sigma$ is a $C^2$ function. 
In the sequel, $\phi_{\delta}$ (resp. $\phi$) denotes the probability density function of a $\mathcal{N}(0,\delta^2)$ (resp. $\mathcal{N}(0,1)$) random variable, while
\begin{align}\label{functional}
	\Psi(z):=\int_0^{z}\phi(x)dx, \;\, z\in \bbr\,, \quad\text{and }\quad {\xi:=\int_{0}^{\infty}\phi_{\sigma_0}(x)x^{1-Y}dx}.
\end{align}
Let us also recall that $L_{Z}$, defined in (\ref{DfnLZ}), denotes the infinitesimal generator of the strictly stable process $(Z_t)_{t\geq 0}$. 
The next theorem gives an asymptotic expansion for the probability of a tempered stable-like process being positive, in the presence of a continuous component satisfying the previously described conditions.
\begin{thm}\label{thmXW}
Let $X$ be a tempered stable-like process as in Theorem \ref{thmX} and $V$ a diffusion process as described above. 
Then,
\begin{align}\label{probXW}
\bbp\left(X_t+V_t\geq 0\right)&= \half + \sum_{k=1}^n d_{k}t^{k\left(1-\frac{Y}{2}\right)} + e\,t^{\half} + f\,t^{\frac{3-Y}{2}} + o(t^{\frac{3-Y}{2}}),\quad t\to 0,
\end{align}
where $n:=\max\big\{k\geq 3:k(1-{Y}/{2})\leq {(3-Y)}/{2}\big\}$, and
\begin{align}
d_k&:=\frac{\sigma(y_0)^{-kY}}{k!}L_{Z}^k\Psi(0),\quad 1\leq k\leq n,\label{dk}\\
e&:= \Big(\tilde\gamma+\mu(y_0)-\frac{\rho}{2}\sigma'(y_0)\gamma(y_0)\Big)\phi_{\sigma_0}(0),\label{d0c}\\
f&:=\Big(\frac{\alpha(1)C(1)+\alpha(-1)C(-1)}{Y-1}-\frac{C(1)+C(-1)}{\sigma^2(y_0)Y}\Big(\tilde\gamma+\mu(y_0)-\frac{\rho}{2}\sigma'(y_0)\gamma(y_0)(1+Y)\Big)\Big)\xi.\label{dn1c}
\end{align}
\end{thm}

\begin{rem}\label{remark2}
A few observations are in order:\hfill

\noindent{\rm (a)}
	Using the notation of Remark \ref{remark1}, the terms can be ordered with regard to their rate of convergence as 
	\[
		d_{1}t^{1-\frac{Y}{2}}\succ{}\dots\succ{}d_{m}t^{m\left(1-\frac{Y}{2}\right)}\succeq   e\,t^{\half}\succ{}d_{m+1}t^{(m+1)\left(1-\frac{Y}{2}\right)}\succ\dots \succ d_{n}t^{n\left(1-\frac{Y}{2}\right)}\succeq  f\,t^{\frac{3-Y}{2}},\qquad t\to{}0,
	\]
	where $m:=\max\{k: k(1-Y/2)\leq 1/2\}$. A comparison of this and the expansion for ATM option prices given in Theorem $4.2$ of \cite{LopOla:2014}, where the first and second order terms were of order $t^{1/2}$ and $t^{{(3-Y)}/{2}}$, reveals that the convergence here is slower, as in the pure-jump case, unless $C(1)=C(-1)$ (see (\ref{probXW2}) below).

\noindent{\rm (b)}
As stated in the proof, there is another useful characterization of the $d_{k}$-coefficients  in terms of a short-time expansion for a certain functional depending on (\ref{functional}). Concretely,
the coefficients $d_{1},\dots,d_{n}$ are such that
\[
	 \widetilde{\bbe}\left(\Psi\left(Z_{t}\right)\right)=\widetilde{\bbp}\left(Z_{t}+\sigma(y_{0})W_{1}^{1}\geq{}0\right)-\frac{1}{2}=\sum_{k=1}^{n}d_{k}t^{k}+O(t^{n+1}),\qquad t\to{}0.
\]
When $Z_1$ is symmetric (i.e., when $C(1)=C(-1)$), it follows that $\widetilde{\bbe}\big(\Psi\left(Z_{t}\right)\big)=0$, and all the $d_{k}$'s vanish. In that case, the expansion simplifies to
\begin{align}\label{probXW2}
\bbp\left(X_t+V_t\geq 0\right)&= \half + e\,t^{\half} + f\,t^{\frac{3-Y}{2}} + o(t^{\frac{3-Y}{2}}),\qquad t\to 0.
\end{align}

\noindent{\rm (c)}
Interestingly enough, the correlation coefficient $\rho$ appears in the expansion (\ref{probXW}). Moreover, so does $\sigma'(y_0)\gamma(y_0)$, i.e.\ the volatility of volatility. This is in sharp contrast to the expansions for near-the-money option prices and implied volatility, given in Theorem $4.2$ of \cite{LopOla:2014}, where the impact of replacing the Brownian component by a stochastic volatility process was merely to replace the volatility of the Brownian component, $\sigma$, by the spot volatility, $\sigma(y_0)$.

\noindent{\rm (d)}
Similarly, the leading order term $d_1$ depends on the jump-component via $L_Z$, and the parameters $\tilde\gamma$, $\alpha(1)$, and $\alpha(-1)$, also appear, containing information on the tempering function $\bar\q$, i.e.\ the L\'evy density away from the origin. This is again in contrast to what was observed in Theorem $4.2$ of \cite{LopOla:2014}, where the leading order term only incorporated information on the spot volatility, $\sigma(y_0)$, and the approximation was altogether independent of the $\bar\q$-function. In short, the ATM skew is more sensitive to various model parameters, which may be anticipated since it is a measure of the asymmetry in the volatility smile, i.e.\ the difference between the volatilities of OTM call and put options, while the ATM volatility is a measure of the overall level of volatility.

\noindent{\rm (e)}
Tempered stable-like processes are a natural extension of stable L\'evy processes. In the pure-jump case, the ``deviation" of $X$ from a stable process does not appear in terms of order lower than $t^{{1}/{Y}}$ (see Remark \ref{remark1}-b). Here, recalling that $X_t$ has the stable representation $X_t^{\text{stbl}}:=Z_t+\tilde\gamma t$ under $\widetilde\bbp$, 
Theorem \ref{thmXW} implies that, 
\[ \bbp\left(X_{t}+V_{t}\geq{}0\right)-\widetilde\bbp\left(X_{t}^{\text{stbl}}+V_t\geq{}0\right)
=\frac{\alpha(1)C(1)+\alpha(-1)C(-1)}{Y-1}\int_{0}^{\infty}\phi_{\sigma_0}(x)x^{1-Y}dx\,t^{\frac{3-Y}{2}}+o\big(t^{\frac{3-Y}{2}}\big), \quad t\to 0.
\]
In other words, for small $t$, one can explicitly approximate the positivity probability of $X_{t}+V_{t}$ by that of $X_{t}^{\text{stbl}}+V_{t}$, up to a term of order higher than $t^{(3-Y)/2}$.

\noindent{\rm (f)}
	One can find a more explicit expression for the constant $f$ by noting that 
	\begin{align*}
		\xi=\int_{0}^{\infty}\phi_{\sigma_0}(x)x^{1-Y}dx=\frac{(\sigma(y_0))^{1-Y}}{2}\bbe|W_{1}|^{1-Y}=\frac{(\sigma(y_0))^{1-Y}2^{-\frac{Y+1}{2}}}{\sqrt{\pi}}\Gamma\Big(1-\frac{Y}{2}\Big),
	\end{align*}
	using the well-known moment formula for centered Gaussian random variables.
	Moreover, 
	{we can further show (see \cite{Olafsson} for the details) that 
	the first two coefficients in (\ref{dk}) are given by} 
	\begin{align}
		d_1&:=\frac{C(1)-C(-1)}{(\sigma(y_0))^YY}\int_{0}^{\infty}(\phi(x)-\phi(0))x^{-Y}dx
		=-(C(1)-C(-1))\frac{(\sigma(y_0))^{-Y}2^{-\frac{Y}{2}}}{\sqrt{\pi}Y(Y-1)}\Gamma\Big(\frac{3-Y}{2}\Big),\label{d1c}\\
		d_2&:=  -\half\frac{C^2(1)-C^2(-1)}{(\sigma(y_0))^{2Y}Y^2}\int_{0}^{\infty}\int_{0}^{\infty}\Big((x+y)\phi(x+y)-x\phi(x)-y\phi(y)\Big)(xy)^{-Y}dxdy.\label{d2c}	
	\end{align}
\end{rem}

In the case when  
the asset price $S_t=S_0e^{X_t+V_t}$ is a well defined $\bbp$-martingale, 
the previous result can be viewed as an asymptotic expansion for ATM digital call prices.
Moreover, for models of this form, Theorem $4.2$ in \cite{LopOla:2014} supplies the following short-term expansion for the ATM implied volatility,
\begin{align}\label{sigImpXV}
\hat\sigma(t)=\sigma(y_0)+\bar\sigma_1 t^{\frac{2-Y}{2}} + o(t^{\frac{2-Y}{2}}),\quad t\to 0.
\end{align}
where 
\begin{align}\label{sigImpXV2}
\bar\sigma_1:=\frac{(C(1)+C(-1))2^{-\frac{Y}{2}}}{Y(Y-1)}\Gamma\left(1-\frac{Y}{2}\right)\sigma(y_0)^{1-Y}.
\end{align}
The above results can then be combined with (\ref{ATMslope}) to obtain an asymptotic expansion for the ATM implied volatility skew. Additionally, using (\ref{Delta}) and the ATM option price expansion in Theorem $4.2$ of \cite{LopOla:2014}, we {also} obtain an asymptotic expansion for the delta of ATM call options:
\begin{cor}\label{corXW}
Let $X$ and $V$ be as in Theorem \ref{thmXW}, with $b$ and $\nu$ satisfying the martingale condition (\ref{CndFrMrtX}),  $\mu=-\half\sigma^2$, 
and $\alpha$, $\gamma$, and $\sigma$ 
such that $(e^{V_t})_{t\geq 0}$ is a true martingale. Then,
\begin{enumerate}
\item
The ATM implied volatility skew satisfies
\begin{align}\label{volSlope1cont}
-\bigg.\frac{\partial\hat\sigma(\kappa,t)}{\partial\kappa}\bigg|_{\kappa=0} 
&=\sqrt{2\pi}\,\sum_{k=1}^{n}d_k\,t^{\left(1-\frac{Y}{2}\right)k-\half}
 + \frac{c}{\sigma_0}+\Big(\sqrt{2\pi} f+\half\bar\sigma_1\Big)\,t^{1-\frac{Y}{2}}+o(t^{1-\frac{Y}{2}}),\quad t\to 0,
\end{align}
where ${c}:=\tilde\gamma-\half\rho\sigma'(y_0)\gamma(y_0)$, and, if $C(1)=C(-1)$, the expansion becomes
\begin{align}\label{volSlope1contSymm}
-\bigg.\frac{\partial\hat\sigma(\kappa,t)}{\partial\kappa}\bigg|_{\kappa=0} 
= \frac{c}{\sigma_0}+\Big(\sqrt{2\pi}f+\half\bar\sigma_1\Big)\,t^{\frac{2-Y}{2}}+o(t^{\frac{2-Y}{2}}),\quad t\to 0.
\end{align}
\item
The delta of an ATM call option satisfies
\begin{align}
\Delta(t) &= \half + \sum_{k=1}^n d_{k}t^{k\left(1-\frac{Y}{2}\right)}  
+\Big(\frac{\sigma(y_0)}{\sqrt{2\pi}}+e\Big)\,t^{\half} + \Big(\frac{\bar{\sigma}_1}{\sqrt{2\pi}}+f\Big)\,t^{\frac{3-Y}{2}} + o(t^{\frac{3-Y}{2}}),\quad t\to 0.
\end{align}
\end{enumerate}
\end{cor}

\begin{rem}\label{RemSignCont}
The previous result shows that the order of convergence and the sign of the ATM implied volatility slope can easily be recovered from the model parameters. In the asymmetric case, i.e.\ when $C(1)\neq C(-1)$, it blows up like $t^{1/2-{Y}/{2}}$, and has the same sign as $C(1)-C(-1)$. However, when $C(1)=C(-1)$, the summation term in (\ref{volSlope1cont}) vanishes, and the slope converges to a nonzero value, $-{c}/\sigma_0$, as $t\to 0$, as in jump-diffusion models. In both cases it is observed that for a fixed value of the index of jump activity, $Y$, the short-term slope is less explosive than in the pure-jump case. 
\end{rem}

{\smallskip
{\noindent\textbf{Proof of Theorem \ref{thmXW}.}}}

\noindent
\textbf{Step 1:} 
We first show that (\ref{probXW}) is true when the functions $\mu$ and $\sigma$ are assumed to be bounded, by considering the stopped processes 
\begin{align}\label{stopped}
\bar\mu_t:=\mu(Y_{t\wedge\tau}),\quad  \bar\sigma_t:=\sigma(Y_{t\wedge\tau}),\quad \tau:=\inf\left\{t:Y_t\notin I\right\},
\end{align}
where $I$ is a {bounded} open interval containing $y_0$, such that $\sigma$ is $C^2$, $\mu$ and $\gamma$ are Lipschitz, and $\alpha$ is bounded on $I$. Note that due to the continuity of $\mu$ and $\sigma$ around $y_0$, we can find constants $0<m<M<\infty$ such that $|\bar\mu_t|<M$ and $m<\bar\sigma_t<M$. Throughout the proof, we set $\sigma_{0}=\sigma(y_{0})$, $\mu_{0}=\mu(y_{0})$, $\alpha_{0}=\alpha(y_{0})$, $\gamma_{0}=\gamma(y_{0})$, and  $\bar\rho:=\sqrt{1-\rho^2}$.
As in the pure-jump case, we also start by assuming that the $\bar\q$-function of $X$ satisfies (\ref{NewAssumEq}-ii) and (\ref{NewAssumEq}-iii). Then, the idea is to reduce the problem to the case where $\bar\mu_t$ and $\bar\sigma_t$ are deterministic, by conditioning the positivity probability on the realization of the process $(W^{1}_t)_{0\leq t\leq 1}$. To do that we follow similar steps as in the proof of Theorem $4.2$ in \cite{LopOla:2014}. On a filtered probability space $(\breve\Omega,\breve\calF,(\breve\calF_{t})_{t\geq{}0},\breve\bbp)$ satisfying the usual conditions, we define independent processes $\breve{X}$ and $\breve{W}^2$, such that the law of $(\breve{X}_t)_{0\leq{}t\leq{}1}$ under $\breve\bbp$ is the same as the law of $(X_{t})_{0\leq{}t\leq{}1}$ under $\bbp^{}$, and $(\breve{W}^2_t)_{0\leq{}t\leq{}1}$ is a standard Brownian motion. Also, for any deterministic functions $\breve\mu:=(\breve\mu_{s})_{s\in[0,1]}$, $\breve\sigma:=(\breve\sigma_{s})_{s\in[0,1]}$, and $\breve\q:=(\breve\q_{s})_{s\in[0,1]}$, belonging to $C([0,1])$, the set of continuous function on $(0,1)$,  we define the process $(\breve\V^{ \breve\mu,\breve\sigma,\breve\q}_{t})_{0\leq{}t\leq{}1}$ as follows:
\begin{align}\label{DtrmVolMod}
	\breve\V^{\breve\mu,\breve\sigma,\breve\q}_t & := \int_0^t\breve\mu_udu+\rho\breve\q_t + \bar\rho\int_{0}^{t}\breve\sigma_ud\breve{W}^2_u,\qquad 0\leq t\leq 1.
\end{align} 
With this notation at hand, we consider a functional $\Phi:[0,1]\times C([0,1])\times C([0,1])\times C([0,1])\to [0,1]$, defined as 
\begin{align}\label{OptPricDtrmVol}
\Phi\left(t,\breve\mu,\breve\sigma,\breve\q\right):=\breve{\bbp}\big(\breve{X}_{t}+\breve\V^{\breve\mu,\breve\sigma,\breve\q}_t\geq 0\big).
\end{align}
Then, for any $t\in[0,1]$, 
\begin{equation}\label{CndStep0}
\bbp\big(\left.X_t+V_t\geq 0\right|W_s^1,s\in[0,1]\big)=\Phi\big(t,(\bar\mu_{s})_{s\in[0,1]},(\bar\sigma_{s})_{s\in[0,1]}, (\bar\q_{s})_{s\in[0,1]}\big),
\end{equation}
where {$\bar\q_s:=\int_0^s\bar\sigma_udW_u^1$}.
For simplicity, we omit the superscripts in the process $\breve{V}^{\breve{\mu},\breve{\sigma},\breve\q}$, unless explicitly needed.
Throughout the proof, we assume that $\breve{\mu}$ and $\breve{\sigma}$ satisfy the same uniform boundedness conditions as $\bar{\mu}$ and $\bar\sigma$; namely, $m<\breve{\sigma}_{t}<M$ and $|\breve{\mu}_{t}|<M$ for any $t\in[0,1]$.

\noindent
As in the pure-jump case, we define {a} probability measure $\widetilde{\bbp}'$ on $(\breve\Omega,\breve\calF)$, {analogous to the probability measure $\widetilde{\bbp}$ described in Section \ref{TSP}, but} replacing the jump measure $N$ of the process $X$ by the jump measure of $\breve{X}$. We also define the strictly stable process $\breve{Z}_{t}:=\breve{X}_{t}-\tilde\gamma t$, where $\tilde{\gamma}:=\widetilde{\bbe}'\big(\breve{X}_{1}\big)$, and $\widetilde{\bbe}'$ denotes the expectation with respect to the probability measure $\widetilde{\bbp}'$. Note that the law of $(\breve{V}_{t})_{t\leq{}1}$ under $\widetilde\bbp'$ remains unchanged and, under both $\breve\bbp$ and $\widetilde{\bbp}'$, 
\begin{align}\label{DistV}
t^{-\half}\breve{V}_t \sim \mathcal{N}\big(t^{\half}\breve\mu^*_t+t^{-\half}\rho\breve\q_t,\bar\rho^2(\breve\sigma_t^*)^2\big)
\end{align}
where, for $t\in(0,1]$, 
\begin{align}\label{BndBarSigma}
\breve\mu^*_t:=\frac{1}{t}\int_0^t\breve\mu_sds\in[-M,M],\quad\quad
\breve\sigma^*_t:=\sqrt{\frac{1}{t}\int_0^t\breve\sigma^2_sds}\in[m,M].
\end{align}
Now, note that {$\bbp\left(X_t+V_t\geq 0\right) = \bbp\big(t^{-\half}X_t+t^{-\half}V_t\geq 0\big)$  converges to $1/2$, as $t\to{}0$, by} Slutsky's theorem, and the facts that $t^{-1/2}V_t\ld \Lambda\sim\mathcal{N}(0,\sigma_0^2)$ (cf. \cite{LopOla:2014}, Eq. (4.47)), and $t^{-{1}/{Y}}X_t\ld Z_1$, where $Z_1$ is a strictly $Y$-stable random variable (cf. \cite{rosenbaum.tankov.10}, Prop. $1$). {In} order to find higher order terms in the expansion, we investigate the limit of the process
\begin{align}\label{R}
R_t & := \bbp\left(X_t+V_t\geq 0\right)-\half,
\end{align}
as $t\to{}0$. In terms of the functional $\Phi$, $R_{t}$ can be expressed {as 
\begin{align}
R_t & = \bbe\Big(\bbp\left(\left.X_t+V_t\geq 0\right|W_s^1,s\in[0,1]\right)-\half\Big)
 = \bbe\big(\bar{\Phi}\big(t,(\bar\mu_{s})_{s\in[0,1]},(\bar\sigma_{s})_{s\in[0,1]},(\bar\q_{s})_{s\in[0,1]}\big)\big),
\end{align}
where $\bar{\Phi}\left(t,\breve\mu,\breve\sigma,\breve\q\right)=\Phi\left(t,\breve\mu,\breve\sigma,\breve\q\right)-1/2$. Note that 
\begin{align}\label{CGMYDecomp}
\bar{\Phi}\big(t,\breve\mu,\breve\sigma,{\breve\q}\big)
&= \widetilde\bbe'\big(e^{-U_t}{\bf 1}_{\{t^{-\half}\breve\V_t\geq -t^{-\half}\breve\Z_t-\tilde\gamma t^{\half}\}}-{\bf 1}_{\{\breve\W_1\geq 0\}}\big)\nonumber\\
&= \widetilde\bbe'\big({\bf 1}_{\{t^{-\half}\breve\V_t\geq -t^{-\half}\breve\Z_t-\tilde\gamma t^{\half}\}}-{\bf 1}_{\{\breve\W_1\geq 0\}}\big)
+ \widetilde\bbe'\big((e^{-U_t}-1){\bf 1}_{\{t^{-\half}\breve\V_t\geq -t^{-\half}\breve\Z_t-\tilde\gamma t^{\half}\}}\big)\nonumber\\
&=: I_1(t,\breve\mu,\breve\sigma,\breve\q)+I_2(t,\breve\mu,\breve\sigma,\breve\q),
\end{align} 
and we proceed to} analyze the two terms separately. For the first one, we use (\ref{DistV}) to show that
\begin{align}\label{CGMYT1}
I_1(t,\breve\mu,\breve\sigma,\breve\q)
&= \widetilde\bbe'\Big(\int_{{-t^{-\half}\breve\Z_t-\left(\tilde\gamma+\breve\mu^*_t\right) t^{\half}-t^{-\half}\rho \breve\q_t}}^0\phi_{\bar\rho\breve\sigma_t^*}(x)dx\Big)\nonumber\\
&= \widetilde\bbe'\big(\int_{{-t^{-\half}\breve\Z_t-\left(\tilde\gamma+\breve\mu^*_t\right) t^{\half}-t^{-\half}\rho \breve\q_t}}^{{-t^{-\half}\breve\Z_t-\left(\tilde\gamma+\mu_0\right) t^{\half}-{t^{-\half}\rho \breve\q_t}}}\phi_{\bar\rho\breve\sigma_t^*}(x)dx\big)
+ \widetilde\bbe'\big(\int_{{-t^{-\half}\breve\Z_t-\big(\tilde\gamma+\mu_0\big) t^{\half}-{t^{-\half}\rho \breve\q_t}}}^{{-t^{-\half}\breve\Z_t-{t^{-\half}\rho \breve\q_t}}}\left(\phi_{\bar\rho\breve\sigma_t^*}(x)-\phi_{\bar\rho\sigma_0}(x)\right)dx\big)\nonumber\\
&\quad+ \widetilde\bbe'\big(\int_{{-t^{-\half}\breve\Z_t-\left(\tilde\gamma+\mu_0\right) t^{\half}-{t^{-\half}\rho \breve\q_t}}}^{{-t^{-\half}\breve\Z_t-{t^{-\half}\rho \breve\q_t}}}\phi_{\bar\rho\sigma_0}(x)dx\big)
+\widetilde\bbe'\big(\int_{{-t^{-\half}\breve\Z_t-{t^{-\half}\rho \breve\q_t}}}^{0}\phi_{\bar\rho\breve\sigma^*_t}(x)dx\big)\nonumber\\
&=: I_1^1(t,\breve\mu,\breve\sigma,{\breve\q})+ I_1^2(t,\breve\sigma,{\breve\q}) + I_1^3(t,{\breve\q}) +  I_1^4(t,\breve\sigma,{\breve\q}),
\end{align}
where we recall that $\phi_{\sigma}$ denotes the density of a $\mathcal{N}(0,\sigma^2)$ random variable. For the first term of (\ref{CGMYT1}), we have
\begin{align}\label{CGMYT1p1Asy}
\bbe\left|I_1^1(t,\bar\mu,\bar\sigma,{\bar\q})\right| 
\leq t^{\half}{\phi_{\bar\rho m}(0)} \bbe\left|\bar\mu^*_t-\mu_0\right|
\leq t^{\half}\phi_{\bar\rho m}(0)\frac{1}{t}\int_0^t\bbe|\bar\mu_s-\mu_0|ds
=O(t),\quad t\to 0,
\end{align}
where the last step follows from Lemma \ref{SVlemma}-(i) below. Similarly, {by Lemma \ref{SVlemma}-(ii) and} the easily verifiable fact
$\displaystyle\sup_{x\in\bbr}|\phi_{\bar\rho\breve\sigma^*_t}(x)-\phi_{\bar\rho\sigma_0}(x)|= \left|\phi_{\breve\sigma^*_t}(0)-\phi_{\sigma_0}(0) \right|$,
\begin{align}\label{CGMYT1p2Asy}
\bbe\left|I_1^2(t,\bar\sigma,\bar\q)\right| 
\leq t^{\half}{|\tilde\gamma+\mu_0|}\bbe\left|\phi_{\bar\rho\bar\sigma^*_t}(0)-\phi_{\bar\rho\sigma_0}(0) \right|
\leq \frac{t^{\half}|\tilde\gamma+\mu_0|}{m^2\bar\rho\sqrt{2\pi}}\bbe\left|\bar\sigma_t-\sigma_0\right|
=O(t),\quad t\to 0.
\end{align}
{For the third term $I_1^3(t,\breve\q)$, assume} $\tilde\gamma':=\tilde\gamma+\mu_0>0$ (the analysis when $\tilde\gamma'<0$ is identical), and write 
\begin{align}\label{CGMYT1p1}
t^{-\half}I_1^3(t,\breve\q)-\tilde\gamma'\phi_{\sigma_0}(0)
 &= t^{-\half}\widetilde\bbe'\Big(\int_{{-t^{-\half}\breve\Z_t-\tilde\gamma' t^{\half}}}^{{-t^{-\half}\breve\Z_t}}\big(\phi_{\bar\rho\sigma_0}(x-t^{-\half}\rho\sigma_0\breve\w^1)-\phi_{\bar\rho\sigma_0}(-t^{-\half}\rho\sigma_0\breve\w^{1})\big)dx\Big)\nonumber\\
 &\quad + t^{-\half}\widetilde\bbe'\Big(\int_{{-t^{-\half}\breve\Z_t-\tilde\gamma' t^{\half}}}^{{-t^{-\half}\breve\Z_t}}\big(\phi_{\bar\rho\sigma_0}(x-t^{-\half}\rho\breve\q_t)-\phi_{\bar\rho\sigma_0}(x-t^{-\half}\rho\sigma_0\breve\w^1)\big)dx\Big)\nonumber\\
 &\quad + \tilde\gamma'\big(\phi_{\bar\rho\sigma_0}(-t^{-\half}\rho\sigma_0\breve\w^{1})-\phi_{\sigma_0}(0)\big)\nonumber\\
&=:I_1^{3,1}(t,\breve\w^1)+I_1^{3,2}(t,\breve\q,\breve\w^1)+I_1^{3,3}(t,\breve\w^1),
\end{align}
where $\breve\w^1\in\bbr$. First, for $I_1^{3,1}(t,\breve\w^1)$, use Fubini's theorem to write
\begin{align*}
I_1^{3,1}(t,\breve\w^1)
&=t^{-\half}\int_{-\infty}^{\infty}\big(\phi_{\bar\rho\sigma_0}(x-\rho\sigma_0t^{-\half}\breve\w^1)-\phi_{\bar\rho\sigma_0}(-\rho\sigma_0t^{-\half}\breve\w^1)\big)J_t(x)dx,
\end{align*}
where $J_{t}(x):=\widetilde\bbp'(-t^{\half-\frac{1}{Y}}x-\tilde\gamma' t^{1-\frac{1}{Y}}\leq \breve\Z_1\leq -t^{\half-\frac{1}{Y}}x)\leq\tilde\kappa t^{\frac{3-Y}{2}}|x|^{-Y-1}$, {with the inequality being a} special case of (\ref{JboundNew}) below, and holds for some constant $\tilde\kappa$, all $x\neq 0$ and $t<1$. Moreover, 
$J_{t}(x)\sim \tilde\gamma' C({-x}/{|x|}) |x|^{-Y-1}t^{{(3-Y)}/{2}}$, as $t\to 0$, which follows from $f_Z(x)\sim C({x}/{|x|})|x|^{-Y-1}$, $|x|\to\infty$ (cf. \cite{Sato}, 14.37).
On the other hand, it is easy to see {that
$\bbe\left(\phi_{\bar\rho\sigma_0}(x-\rho\sigma_0W_1^1)-\phi_{\bar\rho\sigma_0}(-\rho\sigma_0W_1^1)\right)=\phi_{\sigma_0}(x)-\phi_{\sigma_0}(0)=O(x^2)$, as $x\to 0$, 
so,} in light of the above relations, the dominated convergence theorem can be applied to $\bbe\big(I_1^{3,1}(t,W_t^1)\big)$, to obtain
\begin{align}\label{I13first}
\lim_{t\to 0}t^{-\frac{2-Y}{2}}\bbe\big(I_1^{3,1}\big(t,W_t^1\big)\big)
&=\tilde\gamma'\left(C(1)+C(-1)\right)\int_{0}^{\infty}\left(\phi_{\sigma_0}(x)-\phi_{\sigma_0}(0)\right) x^{-Y-1}dx.
\end{align}
For the second part of (\ref{CGMYT1p1}), we can find a constant $\tilde\kappa$ such that 
\begin{align*} 
\big|\phi_{\sigma_0\bar\rho}(x-t^{-\half}\rho\breve\q_t)-\phi_{\sigma_0\bar\rho}(x-t^{-\half}\rho\sigma_0\breve\w^1)\big|\leq \tilde\kappa\big|t^{-\half}\breve\q_t-t^{-\half}\sigma_0\breve\w\big|,
\end{align*}
for all $x\in\bbr$, and $\bbe\big|t^{-\half}\bar\q_t-\sigma_0t^{-\half}\W^1_t\big|=O(t^{\half})$ as $t\to 0$, by Lemma \ref{SVlemma}-(v), so 
\begin{align}\label{I13second}
\bbe\big(I_1^{3,2}\big(t,\bar\q,W_t^1\big)\big) = O(t^{\half}) = o(t^{\frac{2-Y}{2}}),\quad t\to 0.
\end{align}
Finally, $\bbe\big(I_1^{3,3}\big(t,W_t^1\big)\big)=0$ since $\bbe\left(\phi_{\bar\rho\sigma_0}\left(-{\rho}\sigma_0W_1^1\right)\right)= \phi_{\sigma_0}(0)$, which together with (\ref{CGMYT1p1})-(\ref{I13second}) gives
\begin{equation}\label{CGMYT1p3Asy}
	\bbe\left(I_1^3(t,\bar\q)\right) =
	\tilde\gamma'\phi_{\sigma_0}(0)\,t^{\half}+\tilde\gamma'\left(C(1)+C(-1)\right)\int_{0}^{\infty}(\phi_{\sigma_0}(x)-\phi_{\sigma_0}(0))x^{-Y-1}dx\,t^{\frac{3-Y}{2}}+o(t^{\frac{3-Y}{2}}),\quad t\to 0.  
\end{equation}
Before handling the fourth part of (\ref{CGMYT1}), let us introduce some further notation. First, note that by It\^o's formula,
\begin{align}\label{xi12}
\bar\q_t = \int_0^t\bar\sigma_sdW_s^1 
&=\sigma_0W_t^1+\int_0^t\int_0^s\bar\sigma'_u\bar\gamma_udW_u^1dW_s^1+\int_0^t\int_0^s\big( \bar\sigma'_u\bar\alpha_u+\half\bar\sigma''_u\bar\gamma^2_u\big)du dW_s^1
=: \sigma_0W_t^1 + \xi_t^1 + \xi_t^2,
\end{align}
where {$\bar\sigma'_{u}=\sigma'(Y_{u}){\bf 1}_{\{u<\tau\}}$, $\bar\sigma''_{u}=\sigma''(Y_{u}){\bf 1}_{\{u<\tau\}}$, $\bar\alpha_u:=\alpha(Y_{u\wedge\tau})$, and $\bar\gamma_u:=\gamma(Y_{u\wedge\tau})$.}
Also, define 
\begin{align}\label{xi10}
\xi_t^{1,0} := \int_0^t\int_0^s\bar\sigma'_0\bar\gamma_0dW_u^1 dW_s^1  
 = \half \bar\sigma'_0\bar\gamma_0\left( (W_t^1)^2 - t\right) \ed \half \bar\sigma'_0\bar\gamma_0 t\left( (W_1^1)^2 -1\right),
\end{align}
and, for reals $\breve{w}^{1}$ and $\breve\xi$, let
\begin{align}\label{J3decomp}
	I_1^4(t,\breve\sigma,\breve\q)
&=  \widetilde\bbe'\Big(\int_{{-t^{-\half}\breve\Z_t- t^{-\half}\rho\breve\q_t}}^{{-t^{-\half}\breve\Z_t-t^{-\half}\rho \big(\sigma_0 \breve{w}^{1} + \breve\xi\big)}}\phi_{\bar\rho\breve\sigma^*_t}(x)dx\Big)
+ \widetilde\bbe'\Big(\int^{{-t^{-\half}\breve\Z_t-t^{-\half}\rho\sigma_0\breve{w}^1}}_{{-t^{-\half}\breve\Z_t-t^{-\half}\rho \big(\sigma_0\breve{w}^{1} + \breve\xi\big)}}\phi_{\bar\rho\breve\sigma^*_t}(x)dx\Big) \nonumber\\
&\quad + \widetilde\bbe'\Big(\int^{0}_{{-t^{-\half}\breve\Z_t-t^{-\half}\rho \sigma_0\breve{w}^1}}\phi_{\bar\rho\sigma_0}(x)dx\Big)
+\widetilde\bbe'\Big(\int^{{0}}_{{-t^{-\half}\breve\Z_t-t^{-\half}\rho \sigma_0 \breve{w}^1}}\big(\phi_{\bar\rho\breve\sigma^*_t}(x)-\phi_{\bar\rho\sigma_0}(x)\big)dx\Big)\nonumber\\
&= J^1(t,\breve\sigma,\breve\q,\breve{w}^{1},\breve\xi) + J^2(t,\breve\sigma,\breve{w}^{1},\breve\xi) + J^3(t,\breve{w}^{1}) + J^4(t,\breve\sigma,\breve{w}^{1}).
\end{align}

For the first term, {by Lemma \ref{SVlemma}-(v),}
\begin{align}\label{J31}
\bbe\big|J^1\big(t,\bar\sigma,\bar\q,W_{t}^{1},\xi_{t}^{1,0}\big)\big| \leq \phi_{\bar\rho m}(0){\rho}t^{-\half}\big(\bbe\big|\xi_t^2\big|+\bbe\big|\xi_t^1 -\xi_t^{1,0}\big|\big) = O(t),\quad t\to 0, 
\end{align}
For the second term, Cauchy's inequality, (\ref{xi10}), and Lemma \ref{SVlemma}-(iii) can be used to show that {$\bbe\big|\xi_t^{1,0}\big(\phi_{\bar\rho\bar\sigma^*_t}(0)-\phi_{\bar\rho\bar\sigma_0}(0)\big)\big|
= O(t^{\frac{3}{2}})$, as $t\to{}0$,
and, thus, using again that $\big|\phi_{\bar\rho\breve\sigma^*_t}(x)-\phi_{\bar\rho\sigma_0}(x) \big|$ attains its maximum at $x=0$,}
\begin{align*}
\bbe\big(J^2\big(t,\bar\sigma,W_{t}^{1},\xi_{t}^{1,0}\big)\big) 
&=\bbe\Big(\Big.\widetilde\bbe'\Big(\int^{{-t^{-\half}\breve\Z_t-\rho\sigma_0\breve{w}}}_{{-t^{-\half}\breve\Z_t-\rho \sigma_0\breve{w} - t^{\half}\frac{\rho}{2}\sigma'_0\gamma_0(\breve{w}^2-1)}}
\phi_{\bar\rho\sigma_0}(x)dx\Big)\Big|_{\breve{w}=W_{1}^{1}}\Big) + O(t),\quad t\to 0,
\end{align*}
where above we also used that $(\xi_{t}^{1,0},W_{t}^{1})\ed{}(\half \bar\sigma'_0\bar\gamma_0 t\left( (W_1^1)^2 -1\right),t^{1/2}W_{1}^{1})$. Thus, by the dominated convergence theorem,
\begin{align}\label{J320}
\lim_{t\to 0} t^{-\half}\bbe\big(J^2\big(t,\bar\sigma,W_{t}^{1},\xi_{t}^{1,0}\big)\big) 
&=\frac{\rho}{2}{\sigma'_0\gamma_0}\bbe\big(\big((W_1^1)^2-1\big)\phi_{\bar\rho\sigma_0}\big({\rho\sigma_0}W_1^1\big)\big)
=-\frac{\rho^3}{2}\phi_{\sigma_0}(0)\sigma'_0\gamma_0,
\end{align}
where the last step is the result of elementary calculations. To find the second order term of $J^2$,  define
\begin{align*}
 \tilde\J^2(t,\breve\w)&:=t^{-\half}\widetilde\bbe'\Big(\int^{{-t^{-\half}\breve\Z_t-\rho\sigma_0\breve{w}}}_{{-t^{-\half}\breve\Z_t-\rho\sigma_0\breve{w}-t^{\half}\frac{\rho}{2}\sigma'_0\gamma_0\big(\breve{w}^2-1\big)}}\phi_{\bar\rho\sigma_0}(x)dx\Big)-\frac{\rho}{2}{\sigma'_0\gamma_0}\left(\breve\w^2-1\right)\phi_{\bar\rho\sigma_0}\left({\rho\sigma_0}\breve\w\right)\\ 
&=t^{-\half}\int_{-\infty}^{\infty}\left(\phi_{\sigma_0\bar\rho}(x+\rho\sigma_0\breve{w})-\phi_{\sigma_0\bar\rho}\left({\rho\sigma_0}\breve\w\right)\right)J_t(x,\breve\w)dx, 
\end{align*}
where 
\begin{align*}
J_t(x,\breve\w) &:= \widetilde\bbp'\big({{t^{-\half}\breve\Z_t}\leq x\leq t^{-\half}\breve\Z_t + t^{\half}\frac{\rho}{2}\sigma'_0\gamma_0\left(\breve{w}^2-1\right)}\big)
-\widetilde\bbp'\big({{t^{-\half}\breve\Z_t + t^{\half}\frac{\rho}{2}\sigma'_0\gamma_0\left(\breve{w}^2-1\right)\leq x\leq t^{-\half}\breve\Z_t}}\big)\\
&=\int_{t^{\half-\frac{1}{Y}}x - t^{1-\frac{1}{Y}}\frac{\rho}{2}\sigma'_0\gamma_0\left(\breve{w}^2-1\right)}^{t^{\half-\frac{1}{Y}}x}f_Z(z)dz,
\end{align*}
for which, we can {use the estimate $f_Z(z)\leq R |z|^{-Y-1}$ (see, e.g.\ \cite{Sato}, (14.37)) to find a function $f$ such that $\bbe\left(f(W_1^1)\right)<\infty$ and, for all $x\neq 0$ and $t\leq 1$,
\begin{align}\label{JboundNew}
|J_{t}(x,\breve\w)| \leq f(\breve\w)t^{\frac{3-Y}{2}}|x|^{-Y-1};
\end{align}
see \cite{Olafsson} for the details. Next,} by applying the dominated convergence theorem twice, we get
\begin{align}\label{J321}
\lim_{t\to 0}t^{-\frac{2-Y}{2}}\bbe\big(\tilde\J^2\big(t,W_1^1\big)\big)
&=\int_{-\infty}^{\infty}\bbe\left(\left(\phi_{\sigma_0\bar\rho}(x+\rho\sigma_0W_1^1)-\phi_{\sigma_0\bar\rho}\left({\rho\sigma_0}W_1^1\right)\right)\lim_{t\to 0}t^{-\frac{3-Y}{2}}J_t(x,W_1^1)\right)dx\nonumber\\
&=\frac{\rho}{2}\sigma'_0\gamma_0\int_{-\infty}^{\infty}\bbe\left(\left(\phi_{\sigma_0\bar\rho}(x+\rho\sigma_0W_1^1)-\phi_{\sigma_0\bar\rho}\left({\rho\sigma_0}W_1^1\right)\right)\left((W_1^1)^2-1\right)\right)C\Big(\frac{x}{|x|}\Big)|x|^{-Y-1}dx\nonumber\\
&=(C(1)+C(-1))\frac{\rho^3}{2}\sigma'_0\gamma_0\int_{0}^{\infty}\big(\phi_{\sigma_0}(x)\big(\frac{x^2}{\sigma_0^2}-1\big)+\phi_{\sigma_0}(0)\big)x^{-Y-1}dx,
\end{align}
where} the final two equalities follow from the tail estimate $f_Z(x)\sim C({x}/{|x|})|x|^{-Y-1}$, $|x|\to\infty$ (cf. \cite{Sato}, 14.37), and standard calculations. {Note the second} application of the dominated convergence theorem above follows from (\ref{JboundNew}), and the boundedness of $\phi_{\bar\rho\sigma_0}$. 
The first application of it can also be justified using (\ref{JboundNew}) for $|x|\geq 1$, but for $|x|\leq 1$ we use Taylor's theorem to switch the order of limit and integration. More precisely, we can write
\begin{align}\label{Taylor}
\phi_{\sigma_0\bar\rho}(x+\rho\sigma_0\breve\w) = \phi_{\sigma_0\bar\rho}(\rho\sigma_0\breve\w)
+x\left.\phi_{\sigma_0\bar\rho}'(x+\rho\sigma_0\breve\w)\right|_{x=0}
+{\half\left.x^2\phi_{\sigma_0\bar\rho}''(x+\rho\sigma_0\breve\w)\right.|_{x=\xi_x}}
\end{align}
where $0\leq |\xi_x|\leq |x|$, so
\begin{align*}
\bbe\Big(\int_{-1}^{1}\left(\left(\phi_{\sigma_0\bar\rho}(x+\rho\sigma_0W_1^1)-\phi_{\sigma_0\bar\rho}\left({\rho\sigma_0}W_1^1\right)\right)J_t(x,W_1^1)\right)dx\Big)
&=\int_{-1}^{1}\bbe\Big(\half \left.x^2\phi_{\sigma_0\bar\rho}''(x+\rho\sigma_0{W_1^1})\right|_{x=\xi_x}J_t(x,W_1^1)\Big)dx,
\end{align*}
because $\left.\phi_{\sigma_0\bar\rho}'(x+\rho\sigma_0\breve\w)\right|_{x=0}=-{(\rho \breve\w)}/{(\bar\rho\sigma_0)}\phi_{\sigma_0\bar\rho}(\rho\sigma_0\breve{w})$ and  $\bbe\left(W_1^1\phi_{\sigma_0\bar\rho}(\rho\sigma_0W_1^1)J_t(x,W_1^1)\right)=0$, due to symmetry.
Then, (\ref{JboundNew}), and the fact that $\phi''$ is a bounded function, allows us to apply the dominated convergence {theorem,
\begin{align*}
\lim_{t\to 0}t^{-\frac{3-Y}{2}}\int_{-1}^{1}\bbe\Big(\frac{x^2}{2}\left. \phi_{\sigma_0\bar\rho}''(x+\rho\sigma_0\breve\w)\right|_{x=\xi_x}J_t(x,W_1^1)\Big)dx
&=\int_{-1}^{1}\lim_{t\to 0}t^{-\frac{3-Y}{2}}\bbe\big(\left(\phi_{\sigma_0\bar\rho}(x+\rho\sigma_0W_1^1)-\phi_{\sigma_0\bar\rho}\left({\rho\sigma_0}W_1^1\right)\right)J_t(x,W_1^1)\big)dx,
\end{align*}
where} we have again used (\ref{Taylor}). Finally, from (\ref{J320}) and (\ref{J321}) we get
\begin{align}\label{J32}
\frac{2}{\rho^3\sigma'_0\gamma_0}\bbe\left(J^2\left(t,\bar\sigma,W_{t}^{1},\xi_{t}^{1,0}\right)\right) 
&=-\phi_{\sigma_0}(0)t^{\half}
-(C(1)+C(-1))\int_{0}^{\infty}\left(\phi_{\sigma_0}(x)-\phi_{\sigma_0}(0)\right)x^{-Y-1}dx\,t^{\frac{3-Y}{2}}\nonumber\\
&\quad+(C(1)+C(-1))\frac{1}{\sigma_0^2}\int_{0}^{\infty}\phi_{\sigma_0}(x)x^{-Y+1}dx\,t^{\frac{3-Y}{2}} + o(t^{\frac{3-Y}{2}}),\quad t\to 0.
\end{align}
Next, for the third term in (\ref{J3decomp}), let $\widetilde{\bbp}$ denote the probability measure on $(\Omega,\calF,(\calF_t)_{t\geq 0})$ defined in Section \ref{TSP} and let $Z:=(Z_{t})_{t\geq{}0}$ be the process defined in (\ref{ZDecomp}). Note that by the independence of $Z$ and $W^{1}$, and the fact that the law of $\breve{Z}$ under $\widetilde{\bbp}'$ is the same as that of $Z$ under {$\widetilde{\bbp}$, 
\begin{align}\label{Ptilde}
\bbe\left(J^3\left(t,W_{t}^{1}\right)\right) 
=\widetilde\bbe\Big(\int^{0}_{{-t^{-\half}\Z_t}}\phi_{\sigma_0}\left(y\right)dy\Big)
=\widetilde\bbe\Big(\int_{0}^{\Z\left(\sigma_{0}^{-Y}t^{1-Y/2}\right)}\phi\left(y\right)dy\Big),
\end{align}
where 
the} last equality we used the self-similarity relationship $s^{{1}/{Y}}\Z_t\ed \Z_{st}$. Therefore, it is sufficient to find the asymptotic behavior, as $t\to{}0$, of $\widetilde\bbe\big(\Psi(\Z_{t})\big)$, with $\Psi$ as in (\ref{functional}). 
But, since $\Psi(z)$ has continuous and bounded derivatives of all orders,
an iterated Dynkin-type formula (see \cite{LopHou:2009}, Eq.~(1.6)) can be applied to obtain 
\begin{align*}
	\widetilde\bbe\left(\Psi(\Z_{t})\right)=\Psi(0)+\sum_{k=1}^{n}\frac{t^{k}}{k!}L_{Z}^{k}\Psi(0)+
	\frac{t^{n+1}}{n!}\int_{0}^{1}(1-\alpha)^{n}\widetilde\bbe\left(L^{n+1}\Psi(Z_{\alpha t})\right)d\alpha,
\end{align*}
for any $n\in\bbn$, where $L_{Z}$ is the infinitesimal generator of the strictly stable process $Z$, defined in (\ref{DfnLZ}). Therefore,
\begin{align}\label{CGMYT1p41Asy}         
	\bbe\left(\J^3\left(t,W_{t}^{1}\right)\right)
	&=\sum_{k=1}^{n} \frac{ \sigma_{0}^{-kY}L_{Z}^{k}\Psi(0)}{k!}\,t^{{k}(1-\frac{Y}{2})} +O\big(t^{\frac{({n}+1)(2-Y)}{2}}\big),\quad t\to 0.
\end{align}
Finally, for the fourth part of (\ref{J3decomp}), use Taylor's theorem to write
\begin{align*}
\phi_{\bar\rho\breve\sigma_t^*}(x) = \phi_{\bar\rho\sigma_0}(x) - \frac{\phi_{\bar\rho\sigma_0}(x)}{\sigma_0}\big(1-\frac{x^2}{\bar\rho^2\sigma_0^2}\big)\big(\breve\sigma_t^*-\sigma_0\big) + h_x(\breve\sigma_t^*)\big(\breve\sigma_t^*-\sigma_0\big)^2,
\end{align*}
where $h_x(\breve\sigma_t^*)\to 0$, as $\breve\sigma_t^*\to\sigma_0$, and the boundedness of $\breve\sigma_t^*$ away from $0$ and $\infty$ allows us to find a constant $K$ such that $0\leq|h_x(\breve\sigma_t^*)|<K$, for all $t\leq 1$ and all $x\in\bbr$. 
From the {latter, and Lemma \ref{SVlemma}, parts (iii) -(iv)}, it follows that 
\begin{align*}
\bbe\left(J^{4}(t,\bar\sigma,W_t^1)\right) &= -\bbe\Big(\left(\bar\sigma_t^*-\sigma_0\right)\Big. \widetilde\bbe'\Big(\int^{{-t^{-\half}\breve\Z_t}}_{{-t^{-\half}\breve\Z_t-t^{-\half}\rho \sigma_0{\breve\w^1}}}\frac{\phi_{\bar\rho\sigma_0}(x)}{\sigma_0}\big(1-\frac{x^2}{\bar\rho^2\sigma_0^2}\big)dx\Big)\Big|_{\breve\w^1=W_t^1}\Big)+O(t)\\
&=-\frac{\sigma'_0\gamma_0}{\sigma_0}\bbe\Big(\frac{1}{t}\int_0^tW_s^1ds\, \widetilde\bbe'\Big.\Big(\int^{{-t^{-\half}\breve\Z_t}}_{{-t^{-\half}\breve\Z_t-t^{-\half}\rho \sigma_0{ \breve\w^1}}}{\phi_{\bar\rho\sigma_0}(x)}\big(1-\frac{x^2}{\bar\rho^2\sigma_0^2}\big)dx\Big)\Big|_{\breve\w^1=W_t^1}\Big)
+O(t),\quad t\to 0,
\end{align*}
where the second equality follows from Lemma \ref{SVlemma}-(vi). To handle the last expression, let us first note that, {conditionally on $W_t^1$, $\int_0^tW_s^1ds$ is normally distributed with mean and variance, $tW_t^1/2$ and $t^{3}/12$, respectively.} 
Therefore, again using the probability measure $\widetilde{\bbp}$ and the process $Z:=(Z_{t})_{t\geq{}0}$, as in (\ref{Ptilde}), we can {write
\begin{align*}
-t^{-\half}\frac{\sigma_0}{\sigma'_0\gamma_0}\bbe\left(J^{4}(t,\bar\sigma,W_t^1)\right) 
&=\widetilde\bbe\Big(\frac{1}{2} t^{-\half}W_t^1\int^{{-t^{-\half}\Z_t}}_{{-t^{-\half}\Z_t-\rho\sigma_0{t^{-\half}W_t^1}}}{\phi_{\bar\rho\sigma_0}(x)}\big(1-\frac{x^2}{\bar\rho^2\sigma_0^2}\big)dx\Big)
=\half\widetilde\bbe\Big(\hat\Psi\big(Z\big(t^{1-\frac{Y}{2}}\big)\big)\Big),
\end{align*}
where} we have used the self-similarity relationships $s^{\half}W_t^1\ed W_{st}^1$ and $s^{\frac{1}{Y}}\Z_t\ed \Z_{st}$, and the notation
\[
	\hat\Psi(z):=\widetilde\bbe\Big(W_1^1\int_{{z}}^{{z+\rho\sigma_0{{W_1}^1}}}{\phi_{\bar\rho\sigma_0}(x)}\big(1-\frac{x^2}{\bar\rho^2\sigma_0^2}\big)dx\Big) = \frac{1}{\sigma_0}\phi_{\sigma_0}(z)(\sigma_0^2-z^2)\bar\rho^2\rho.
\]
Since $\hat\Psi(z)$ has continuous and bounded derivatives of all orders, we proceed as in (\ref{CGMYT1p41Asy}) and obtain
\begin{align}\label{J4}      
	-2\frac{\sigma_0}{\sigma_0'\gamma_0}\bbe\left(\J^4\left(t,\bar\mu,\bar\sigma,\bar\q,{W_{t}^{1}}\right)\right)
	&=\hat\Psi(0)t^{\half} + {L_{Z}\hat\Psi(0)}\,t^{\frac{3-Y}{2}} +o(t^{\frac{3-Y}{2}}),\quad t\to 0,
\end{align}
where $\hat\Psi(0)=\rho\bar\rho^2\phi(0)$ and
\begin{align*} L_{Z}\hat\Psi(0)=(C(1)+C(-1))\rho\bar\rho^2\int_{0}^{\infty}\big(\frac{1}{\sigma_0}\phi_{\sigma_0}(u)(\sigma_0^2-u^2)-\sigma_0\phi_{\sigma_0}(0)\big)u^{-Y-1}du,
\end{align*} 
which trivially follows from (\ref{DfnLZ}).
Thus, combining (\ref{J3decomp})-(\ref{J31}), and (\ref{J32})-(\ref{J4}), gives an asymptotic
expansion for $\bbe\left(I_1^4(t,\bar\sigma,\bar\q)\right)$, which, together with (\ref{CGMYT1})-(\ref{CGMYT1p2Asy}) and (\ref{CGMYT1p3Asy}), finally gives 
\begin{align}\label{CGMYT1Asy}
\bbe\left(I_1(t,\bar\mu,\bar\sigma,{\bar\q})\right)
&=\sum_{k=1}^{n} \frac{ \sigma_{0}^{-kY}L_{Z}^{k}\Psi(0)}{k!}\,t^{{k}(1-\frac{Y}{2})} 
+\big(\tilde\gamma'-\frac{\rho}{2}\sigma'_0\gamma_0\big)\phi_{\sigma_0}(0)\,t^{\half}  \nonumber\\
&\quad-\frac{C(1)+C(-1)}{\sigma_0^2Y}\big(\tilde\gamma'-\frac{\rho}{2}\sigma'_0\gamma_0\left(1+Y\right)\big)\int_{0}^{\infty}\phi_{\sigma_0}(x)x^{-Y+1}dx\,t^{\frac{3-Y}{2}}+o(t^{\frac{3-Y}{2}}), 
\end{align}
as $t\to 0$, where $n:=\max\{k\geq 3:k(1-Y/2)\leq (3-Y)/2\}$, and we have used integration by parts to write
\begin{align*}
\int_{0}^{\infty}(\phi_{\sigma_0}(x)-\phi_{\sigma_0}(0))x^{-Y-1}dx=-\frac{1}{\sigma_0^2Y}\int_{0}^{\infty}\phi_{\sigma_0}(x)x^{-Y+1}dx.
\end{align*}

\noindent Now consider the second part of (\ref{CGMYDecomp}). By using (\ref{Uplusminuseta}) and similar steps as in (\ref{Decomp121}), followed by the decomposition (\ref{UDecomp}), we can {write
\begin{align}\label{CGMYT2}
I_2(t,\breve\mu,\breve\sigma,{\breve\q})
&=-\widetilde\bbe'\big(\widetilde\U^{(1)}_t{\bf 1}_{\{t^{-\half}\breve\V_t\geq -t^{-\half}\breve\Z_t-\tilde\gamma t^{\half}\}}\big)
+\widetilde\bbe'\big(\widetilde\U^{(2)}_t{\bf 1}_{\{t^{-\half}\breve\V_t\geq -t^{-\half}\breve\Z_t-\tilde\gamma t^{\half}\}}\big) +R(t,\breve{\mu},\breve{\sigma},\breve{q})\nonumber\\
&=-\tilde\I_1(t,\breve\mu,\breve\sigma,{\breve\q})
+\tilde\I_2(t,\breve\mu,\breve\sigma,{\breve\q}) +R(t,\breve{\mu},\breve{\sigma},\breve{q})
\end{align}
where} $\bbe(R(t,\bar\mu,\bar\sigma,\bar{q}))=O(t)$, as $t\to 0$, and $\bbe(\tilde\I_1(t,\bar\mu,\bar\sigma,\bar{q}))=O(t)$ since $\widetilde\U_t^{(1)}$ is a finite variation process. We further decompose $\tilde\I_2(t,\breve\mu,\breve\sigma,{\breve\q})$ as
\begin{align}\label{CGMYT2p2}
\tilde\I_2(t,\breve\mu,\breve\sigma,\breve\q) 
&=\alpha(1)\widetilde\bbe'\big(\breve\Z_t^{(p)}{\bf 1}_{\{t^{-\half}\breve\V_t\geq -t^{-\half}\left(\breve\Z_t^{(p)} + \breve\Z_t^{(n)}\right)-\tilde\gamma t^{\half}\}}\big)
+\alpha(-1)\widetilde\bbe'\big(\breve\Z_t^{(n)}{\bf 1}_{\{t^{-\half}\breve\V_t\geq -t^{-\half}\left(\breve\Z_t^{(p)} + \breve\Z_t^{(n)}\right)-\tilde\gamma t^{\half}\}}\big) 
\nonumber\\
&=: \alpha(1)I_2^1(t,\breve\mu,\breve\sigma,\breve\q)+\alpha(-1)I_2^2(t,\breve\mu,\breve\sigma,\breve\q), 
\end{align}
and we look at the two terms separately. For the first one we have
\begin{align}\label{I21decomp}
I_2^1(t,\breve\mu,\breve\sigma,\breve\q)
&= {\widetilde\bbe'}\Big(\breve\Z_t^{(p)}\int_{-t^{-\half}\left(\breve\Z_t^{(p)} + \breve\Z_t^{(n)}\right)-(\tilde\gamma+\breve\mu^*_t) t^{\half}{-t^{-\half}\rho\breve\q_t}}^{\infty}\phi_{{\bar\rho}\breve\sigma^*_t}(x)dx\Big)\nonumber\\
&={\widetilde\bbe'}\Big(\breve\Z_t^{(p)}\Big(\int_{-t^{-\half}\left(\breve\Z_t^{(p)} + \breve\Z_t^{(n)}\right)-(\tilde\gamma+\breve\mu^*_t) t^{\half}{-t^{-\half}\rho\breve\q_t}}^{-t^{-\half}\breve\Z_t^{(p)}{-t^{-\half}\rho\breve\q_t}}
+\int_{-t^{-\half}\breve\Z_t^{(p)}{-t^{-\half}\rho\breve\q_t}}^{{-t^{-\half}\rho\breve\q_t}}
+\int_{{-t^{-\half}\rho\breve\q_t}}^{\infty}\Big)
\phi_{{\bar\rho}\breve\sigma^*_t}(x)dx\Big),
\end{align}
and the third integral is zero since $\widetilde\bbe\big(\breve\Z_t^{(p)}\big)=0$. For the first one, use the independence of $\breve\Z_t^{(p)}$ and $\breve\Z_t^{(n)}$ to write
\begin{align}
\widetilde\bbe'\Big|\breve\Z_t^{(p)}\int_{-t^{-\half}\left(\breve\Z_t^{(p)} + \breve\Z_t^{(n)}\right)-(\tilde\gamma+\bar\mu^*_t) t^{\half}{-t^{-\half}\rho\bar\q_t}}^{-t^{-\half}\breve\Z_t^{(p)}{-t^{-\half}\rho\bar\q_t}}\phi_{{\bar\rho}\bar\sigma^*_t}(x)dx\Big|
&\leq t^{\frac{1}{Y}}\widetilde\bbe'\big|\breve\Z_1^{(p)}\big|\phi_{{\bar\rho}m}(0)\big(t^{\frac{1}{Y}-\half}\widetilde\bbe'\big|\breve\Z_1^{(n)}\big|+(|\tilde\gamma|+M)t^{\half}\big)\nonumber\\
&= O(t^{\frac{2}{Y}-\half}) = o(t^{\frac{3-Y}{2}}),\quad t\to 0.
\end{align}
Finally, for the second integral, let $\breve\w^1\in\bbr$, and write
\begin{align}\label{CGMYT2UpUm1}
{\widetilde\bbe'}\Big(\breve\Z_t^{(p)}\int_{t^{-\half}\rho\bar\q_t}^{t^{-\half}\breve\Z_t^{(p)}+ t^{-\half}\rho\bar\q_t}\phi_{\bar\rho\breve\sigma^*_t}(x)dx\Big)
&={\widetilde\bbe'}\Big(\breve\Z_t^{(p)}\int_{t^{-\half}\rho\bar\q_t}^{ t^{-\half}\rho\sigma_0\breve\w^1}\phi_{\bar\rho\breve\sigma^*_t}(x)dx\Big)
+{\widetilde\bbe'}\Big(\breve\Z_t^{(p)}\int_{t^{-\half}\rho\sigma_0\breve\w^1}^{t^{-\half}\breve\Z_t^{(p)}+ t^{-\half}\rho\sigma_0\breve\w^1}\phi_{\bar\rho\breve\sigma^*_t}(x)dx\Big)\nonumber\\
&\quad +{\widetilde\bbe'}\Big(\breve\Z_t^{(p)}\int^{t^{-\half}\breve\Z_t^{(p)}+ t^{-\half}\rho\bar\q_t}_{t^{-\half}\breve\Z_t^{(p)}+ t^{-\half}\rho\sigma_0\breve\w^1}\phi_{\bar\rho\breve\sigma^*_t}(x)dx\Big)\nonumber\\
&=:J_1(t,\breve\sigma,\breve\q,\breve\w^1)+J_2(t,\breve\sigma,\breve\w^1)+J_3(t,\breve\sigma,\breve\q,\breve\w^1),
\end{align}
and observe that 
\begin{align}
\bbe\left|J_1(t,\bar\sigma,\bar\q,W_t^1)+J_3(t,\bar\sigma,\bar\q,W_t^1)\right|
\leq 2\rho\phi_{\bar\rho m}(0) t^{\frac{1}{Y}-\half} \widetilde\bbe'\left|\breve\Z_1^{(p)}\right|\bbe\big|\bar\q_t-\sigma_0W_t^1\big|=O(t),\quad t\to 0,
\end{align}
by Lemma \ref{SVlemma}-(v), which implies that $\bbe\left|\bar\q_t-\sigma_0W_t^1\right|=O(t^{3/2})$. 
Next, write
\begin{align}\label{CGMYT2UpUm}
J_2(t,\breve\sigma,\breve\w^1)&=t^{\frac{1}{Y}}\int_{0}^{\infty}\widetilde\bbe\big(\breve\Z_1^{(p)}{\bf 1}_{\{0\leq {y}\leq t^{\frac{1}{Y}-\half}\breve\Z_1^{(p)} \}}\big)\phi_{\bar\rho\breve\sigma_t^*}({y}-t^{-\half}\rho\sigma_0\breve\w^1)d{y}\nonumber\\
&\quad-t^{\frac{1}{Y}}\int_{-\infty}^{0}\widetilde\bbe\big(\breve\Z_1^{(p)}{\bf 1}_{\{ t^{\frac{1}{Y}-\half}\breve\Z_1^{(p)}\leq y\leq 0 \}}\big)\phi_{\bar\rho\breve\sigma_t^*}({y}-t^{-\half}\rho\sigma_0\breve\w^1)d{y}\nonumber\\
&=:J_2^1(t,\breve\sigma,\breve\w^1)-J_2^2(t,\breve\sigma,\breve\w^1).
\end{align}
{It then follows that 
\begin{align}\label{CGMYT2Asy1}
\lim_{t\to 0}t^{-\frac{3-Y}{2}}\bbe\left(J_2^1(t,\bar\sigma,W_t^1)\right)
&=\frac{C(1)}{Y-1}\int_{0}^{\infty}{y}^{1-Y}\phi_{\sigma_0}(x)d{x},\qquad 
\bbe\left(J_2^2(t,\bar\sigma,W_t^1)\right)=o(t^{\frac{3-Y}{2}}),\quad t\to 0.
\end{align}
The first relation above follows from a similar procedure as in the proof of (4.26) in \cite{LopOla:2014}, while the second holds because} the jump support of $\breve\Z_1^{(p)}$ is concentrated on the positive axis.
Combining {(\ref{I21decomp})-(\ref{CGMYT2Asy1})} gives
\begin{align}\label{CGMYT2p1Asy}
\bbe\left(I_2^1(t,\bar\mu,\bar\sigma,\bar\rho)\right)=\frac{C(1)}{Y-1}\int_{0}^{\infty}x^{1-Y}\phi_{\sigma_0}(x)dx\,t^{\frac{3-Y}{2}}+o(t^{\frac{3-Y}{2}}),\quad t\to 0.
\end{align}
Finally, for the term $I_2^2(t)$ in (\ref{CGMYT2p2}), the same procedure can be used to obtain 
\begin{align}\label{CGMYT2p2Asy}
\bbe\left(I_2^2(t,\bar\mu,\bar\sigma,\bar\rho)\right)
= \frac{C(-1)}{Y-1}\int_{0}^{\infty}x^{1-Y}\phi_{\sigma_0}(x)dx\,t^{\frac{3-Y}{2}}+o(t^{\frac{3-Y}{2}}),\quad t\to 0,
\end{align}
which, together with (\ref{CGMYT2})-(\ref{CGMYT2p2}), yields 
\begin{equation}\label{CGMYT2Asy}
	\bbe\left(I_2(t,\bar\mu,\bar\sigma,\bar\q)\right) = 
	\frac{\alpha(1)C(1)+\alpha(-1)C(-1)}{Y-1}\int_{0}^{\infty}\phi_{\sigma_0}(x)x^{1-Y}dx\,t^{\frac{3-Y}{2}} + o(t^{\frac{3-Y}{2}}),\quad t\to 0. 
\end{equation}
Combining (\ref{R})-(\ref{CGMYDecomp}), (\ref{CGMYT1Asy}), and (\ref{CGMYT2Asy}), then gives (\ref{probXW}).

{\noindent\textbf{Step 2:} 
The} next step is to show that the expansion (\ref{probXW}) extends to the case when the $\bar\q$-function of $X$ does not necessarily satisfy conditions (\ref{NewAssumEq}-ii) and (\ref{NewAssumEq}-iii). That can be done exactly as in Step $2$ of the pure-jump case, by defining a process $\breve\X$ on an extended probability space $(\breve\Omega,\breve\calF,\breve\bbp)$, satisfying those conditions, and using the triangle inequality to show that the terms of order lower than $t$ in the asymptotic expansion of $\bbp(\breve\X_t+V_t\geq 0)$ extend to $\bbp(X_t+V_t\geq 0)$.

{\noindent\textbf{Step 3:} 
Lastly,} we will show that the expansion (\ref{probXW}) extends to the general case when $\sigma$ and $\mu$ are not necessarily bounded functions. To that end, define a process $(\bar\V_t)_{t\leq{}1}$ as in (\ref{modelVY})), but replacing $\sigma(Y_t)$ and $\mu(Y_t)$ with the stopped processes $\bar\sigma_t:=\sigma(Y_{t\wedge\tau})$ and $\bar\mu_t:=\mu(Y_{t\wedge\tau})$, introduced in (\ref{stopped}). By Steps 1-2 above, the asymptotic expansion (\ref{probXW}) holds for the process $X+\bar\V$. For it to extend to the process $X+V$, it is then sufficient to show that
\begin{align*}
\left|\bbp\left(X_t+V_t\geq 0\right)-\bbp\left(X_t+\bar\V_t\geq 0\right)\right|=O(t),\quad t\to 0,
\end{align*}
because $(3-Y)/{2}<1$ for $Y\in(1,2)$. 
But since $\bar\V_t=V_t$ for $t<\tau$, we have
\begin{align*}
\left|\bbp\left(X_t+V_t\geq 0\right)-\bbp\left(X_t+\bar\V_t\geq 0\right)\right|
&= \left|\bbp\left(X_t+V_t\geq 0,\tau<t\right)-\bbp\left(X_t+\bar\V_t\geq 0,\tau<t\right)\right|
\leq 2\bbp\left(\tau<t\right)
=O(t),
\end{align*}
as $t\to 0$, where the last step follows form Lemma 4.1 in \cite{LopOla:2014}. 
\hfill\qed

\begin{rem}\label{OTMSkewH}
We finish this section by briefly considering the OTM
implied volatility skew, 
under an exponential L\'evy model $S_t:=S_0e^{X_t+\sigma\W_t}$, where $(X_t)_{t\geq 0}$ is a L\'evy process with generating triplet $(0,b,\nu)$, and $(W_t)_{t\geq 0}$ is an independent standard Brownian motion. 
To this end, fix a log-moneyness level $\kappa\neq 0$, and recall  
the following well known asymptotic relationship (see, e.g., \cite{Sato}, Corollary 8.9),
\begin{align}
\bbp\left(X_t+\sigma\W_t\geq \kappa\right)=t\nu([\kappa,\infty)){\bf 1}_{\{\kappa>0\}}+(1-t\nu((-\infty,\kappa])){\bf 1}_{\{\kappa<0\}}+o(t),\quad t\to 0,
\end{align}
and that for the implied volatility of an option with log-moneyness $\kappa$, Theorem 2.3 in \cite{FigForde:2012} states that
\begin{align}\label{OTMvol}
\hat\sigma^2(\kappa,t)t=\frac{\kappa^2}{2\ln\frac{1}{t}}\Big(1+V_1(t,\kappa)+o\Big(\frac{1}{{\ln\frac{1}{t}}}\Big)\Big),\quad t\to 0,
\end{align}
under some very mild conditions on the L\'evy measure, where 
\begin{align}\label{V1}
V_1(t,\kappa):=\frac{1}{\ln\frac{1}{t}}\ln\Big(\frac{4\sqrt{\pi}a_0(\kappa)e^{-\kappa/2}}{|\kappa|}\Big(\ln\frac{1}{t}\Big)^{\frac{3}{2}}\Big),
\end{align} 
and $a_0(\kappa):=\int_{\bbr_0}(e^x-e^{\kappa})^+\nu(dx){\bf 1}_{\{\kappa>0\}}+\int_{\bbr_0}(e^{\kappa}-e^x)^+\nu(dx){\bf 1}_{\{\kappa<0\}}$. 
Plugging the above relations into (\ref{slope1}-i) then yields 
the following higher order expansion for the OTM implied volatility {skew:
\begin{align}\label{OTMslope2}
\frac{\partial \hat\sigma(\kappa,t)}{\partial \kappa}\sqrt{2t\ln\frac{1}{t}}
&=\frac{\kappa}{|\kappa|}\Big(1 + \frac{V_1(t,\kappa)}{2}\Big)
-\frac{\kappa}{|\kappa|}\frac{1}{2\ln\frac{1}{t}}\Big(1+\frac{\kappa}{2}-\kappa\frac{ b_0(\kappa)}{a_0(\kappa)}\Big)+o\Big(\frac{1}{\ln\frac{1}{t}}\Big),\quad t\to 0,
\end{align}
where} $b_0(\kappa):=-e^{\kappa}\big(\nu([\kappa,\infty)){\bf 1}_{\{\kappa>0\}}+\nu((-\infty,\kappa]){\bf 1}_{\{\kappa<0\}}\big)$ (see Chapter 3 in \cite{Olafsson} for the details). It is noteworthy that this expression can be obtained in much more generality than the ATM skew expressions in Corollaries \ref{corX} and \ref{corXW}. For instance, the above result does not depend on the presence of a nonzero Brownian component, or whether $X$ has finite or infinite jump activity. 
Furthermore, it is interesting that this expression for the limiting skew can also be deduced (at least formally) by differentiating the expression (\ref{OTMvol}) for the implied volatility - something that is not possible {for the} ATM skew.
\end{rem}
 
\section{Numerical examples}\label{Examples}

In the first part of this section we use Monte Carlo simulation to assess the accuracy of the asymptotic expansions presented in Sections \ref{SectionPureJump} and \ref{SectionContComp}. In the second part we investigate the implied volatility skew in S\&P500 option prices and how it compares to the short-term skew of the models in Sections \ref{SectionPureJump} and \ref{SectionContComp}.

\subsection{Accuracy of the asymptotic expansions}

In this section we carry out a numerical analysis for the popular class of the tempered stable L\'evy processes, with and without an independent continuous component (see, e.g., \cite{Andersen}, \cite{CT04}, {\cite{Koponen:1995}}). They are an extension of the CGMY model of \cite{CGMY}, and characterized by a L\'evy measure of the form 
\begin{align}
\nu(dx)=C\Big(\frac{x}{|x|}\Big)|x|^{-Y-1}\big(e^{-Mx}{\bf 1}_{\{x>0\}}+e^{-G|x|}{\bf 1}_{\{x<0\}}\big)dx,
\end{align}
where $C(1)$ and $C(-1)$ are nonnegative such that $C(1)+C(-1)>0$, $G$ and $M$ are strictly positive constants, and $Y\in(1,2)$. The martingale condition (\ref{CndFrMrtX}) also implies that $M>1$. Note that in terms of the notation of Section \ref{TSP}, we have $\alpha(1)=-M$ and $\alpha(-1)=G$, and the constants $\tilde\gamma$ and $\eta$ defined in Eqs.~(\ref{tildegamma}) and (\ref{Uplusminuseta}) are given by
\begin{align*}
\tilde\gamma 
&= -\Gamma(-Y)\big(C(1)\big((M-1)^Y-M^Y\big)+C(-1)\big((G+1)^Y-G^Y\big)\big),\\
\eta 
&= \Gamma(-Y)\big(C(1)M^Y+C(-1)G^Y\big).
\end{align*}
In the pure-jump case, Theorem \ref{thmX} and Corollary \ref{corX} present short-term expansions for digital call option prices and the implied volatility skew. 
For tempered stable processes, the $d_k$-coefficients are as in (\ref{d}), while
\begin{align}
e & = -M\widetilde\bbe\big(Z_1^{(p)}{\bf 1}_{\{Z_1^{(p)} + Z_1^{(n)}\geq 0\}}\big)
+ G\widetilde\bbe\big(Z_1^{(n)}{\bf 1}_{\{Z_1^{(p)} + Z_1^{(n)}\geq 0\}}\big),\\ 
f & = -\tilde\gamma(M+G)\widetilde\bbe\big(Z_1^{(p)}f_{Z_1^{(n)}}(-Z_1^{(p)})\big)
+  \Gamma(-Y)(\widetilde\bbp\big(Z_1\leq 0\big)C(1)M^Y-\widetilde\bbp(Z_1> 0)C(-1)G^Y).
\end{align}
It is informative to note that the terms can be further simplified in the CGMY-case, i.e.\ when $C:=C(1)=C(-1)$. In that case, $Z_1^{(p)}\ed-Z_1^{(n)}$, 
which implies $\widetilde\bbe(Z_1^{(p)}{\bf 1}{\{Z_1^{(p)} + Z_1^{(n)}\geq 0\}})=\widetilde\bbe(Z_1^{(n)}{\bf 1}{\{Z_1^{(p)} + Z_1^{(n)}\geq 0\}})$, and thus,
\begin{align}\label{d0}
e 
= \frac{G-M}{2\pi}\Gamma\Big(1-\frac{1}{Y}\Big)\Big(2C\Gamma(-Y)\Big|\cos\Big(\frac{\pi Y}{2}\Big)\Big|\Big)^{\frac{1}{Y}},
\end{align}
where we have also used the expression for $\widetilde\bbe(Z_1^+)$ given in Remark \ref{remark1}, and 
\begin{align}\label{d1}
f = -\tilde\gamma(M+G)\widetilde\bbe\big(Z_1^{(p)}f_{Z_1^{(p)}}(Z_1^{(p)})\big)
+  \frac{C\Gamma(-Y)}{2}(M^Y-G^Y).
\end{align}
In the presence of a continuous component, Theorem \ref{thmXW} and Corollary \ref{corXW} supply short-term expansions for ATM digital call prices and the implied volatility skew. 
The coefficients $d_k$, $e$, and $f$, are as in (\ref{dk})-(\ref{dn1c}), but in the zero-correlation case, the last of these reduces to 
\begin{align}
f&=\frac{\sigma^{1-Y}2^{-\frac{Y+1}{2}}}{\sqrt{\pi}}\Gamma\Big(1-\frac{Y}{2}\Big)\Big(\frac{-MC(1)+GC(-1)}{Y-1}-\frac{C(1)+C(-1)}{\sigma^2Y}\big(\tilde\gamma-\half\sigma^2\big)\Big).
\end{align}
We also note that the first two $d_k$-coefficients are given in Remark \ref{remark2}-(f), whereas in the CGMY-case, $C(1)=C(-1)$, and all the $d_k$'s vanish.

To assess the accuracy of the approximations, we compare them to the {``true"} values of the quantities, which {are estimated} using Monte Carlo simulation. First, for ATM digital call options, 
we use the measure transformation introduced in Section \ref{TSP} to write 
\begin{align}\label{MC}
\bbp(X_t+V_t\geq 0) 
= \widetilde\bbe\big(e^{-U_t}{\bf 1}_{\{X_t+V_t\geq 0\}}\big)
=\widetilde\bbe\big(e^{-MZ_t^{(p)}+GZ_t^{(n)}-\eta t}{\bf 1}_{\{Z_t^{(p)} + Z_t^{(n)}+\tilde\gamma t+V_t \geq 0\}}\big),
\end{align}
where 
$Z_t^{(p)}$ and $Z_t^{(n)}$ are strictly $Y$-stable random variables with L\'evy measures of the form (\ref{nutildePM})\footnote{Equivalently, $Z_t^{(p)}$ and $Z_t^{(n)}$ are $Y$-stable random variables with location parameter $0$, skewness parameters $1$ and $-1$, and scale parameters $\big(tC(1)|\cos(\pi Y/2)|\Gamma(-Y)\big)^{1/Y}$ and $\big(tC(-1)|\cos(\pi Y/2)|\Gamma(-Y)\big)^{1/Y}$, respectively.}. Such variables can be simulated efficiently, and standard discretization schemes can be used for the continuous component $V$, so (\ref{MC}) can be used to obtain an unbiased estimate of ATM digital call prices. For the implied volatility slope, we first estimate near-the-money option prices and implied volatility, using a Monte Carlo procedure similar to the one based on (\ref{MC}) for digital options, and then use those values to numerically estimate the ATM volatility slope.

As mentioned in the introduction, a common drawback of short-term expressions  
is that their convergence 
may be slow, and only satisfactory at extremely small time scales. However, the performance is highly parameter-dependent, and in this section we consider parameter values that are of relevance in financial applications, based on results in \cite[Tables 1 and 5]{Andersen}, \cite{Kawai} and \cite[p.82]{Schoutens}, where the tempered stable model is calibrated to observed option prices.

Figure \ref{PureJumpProb} compares the asymptotic expansions for ATM digital call prices, under a pure-jump tempered stable model, to the true price estimated by Monte Carlo simulation. The axes are on a $\log_{10}$-scale, with time-to-maturity in years. The first and second order approximations are of order $t^{{1}/{Y}}$, and $t$, respectively, and it is clear that the second order approximation significantly improves the first order approximation, and gives good estimates for maturities up to at least one month. 
Figure \ref{PureJumpSmile} then shows the implied volatility smile and the ATM slope approximation. 
The maturity is $0.1$ years (over one month), and the slope approximation captures the sign of the slope, and even gives a good estimate of its magnitude
: In Panel (a), the slope approximation is $0.320$ compared to a Monte Carlo estimate of $0.328$; 
in Panel (b) the approximation gives $-0.456$ while the Monte Carlo estimate is $-0.521$. 

Figure \ref{ContCompAll} carries out the same analysis for a tempered stable model with a nonzero Brownian component. Panel (a) compares the Monte Carlo estimates of ATM digital call prices to the first and second order approximations, which are of order $t^{1/2}$ and $t^{{(3-Y)}/{2}}$, respectively. As in the pure-jump case, the second order approximation gives a good estimate for maturities up to at least a month. 
Panel (b) then displays (in red) the volatility smile and the ATM slope approximation for maturity $t=0.1$ years (over 1 month). {For comparison, it also shows (in blue) the volatility smile and slope approximation when the constant volatility Brownian component is replaced by a Heston stochastic volatility process with the same spot volatility, and leverage parameter $\rho=-0.3$}. In both cases the approximations clearly capture the sign of the slope, and are in fact very close to the Monte Carlo estimates: 0.112 (resp.\ 0.305) compared to Monte Carlo estimates of 0.116 (resp.\ 0.289) in the Brownian case (resp.\ Heston case). 

Finally, for completeness, we include in Figure \ref{PriceAll} a comparison of the first and second order approximations for ATM call option prices from \cite{LopOla:2014}, and the true values estimated by Monte Carlo simulation.

\begin{figure}
	\centering
	\includegraphics[width=7cm,height=5.0cm]{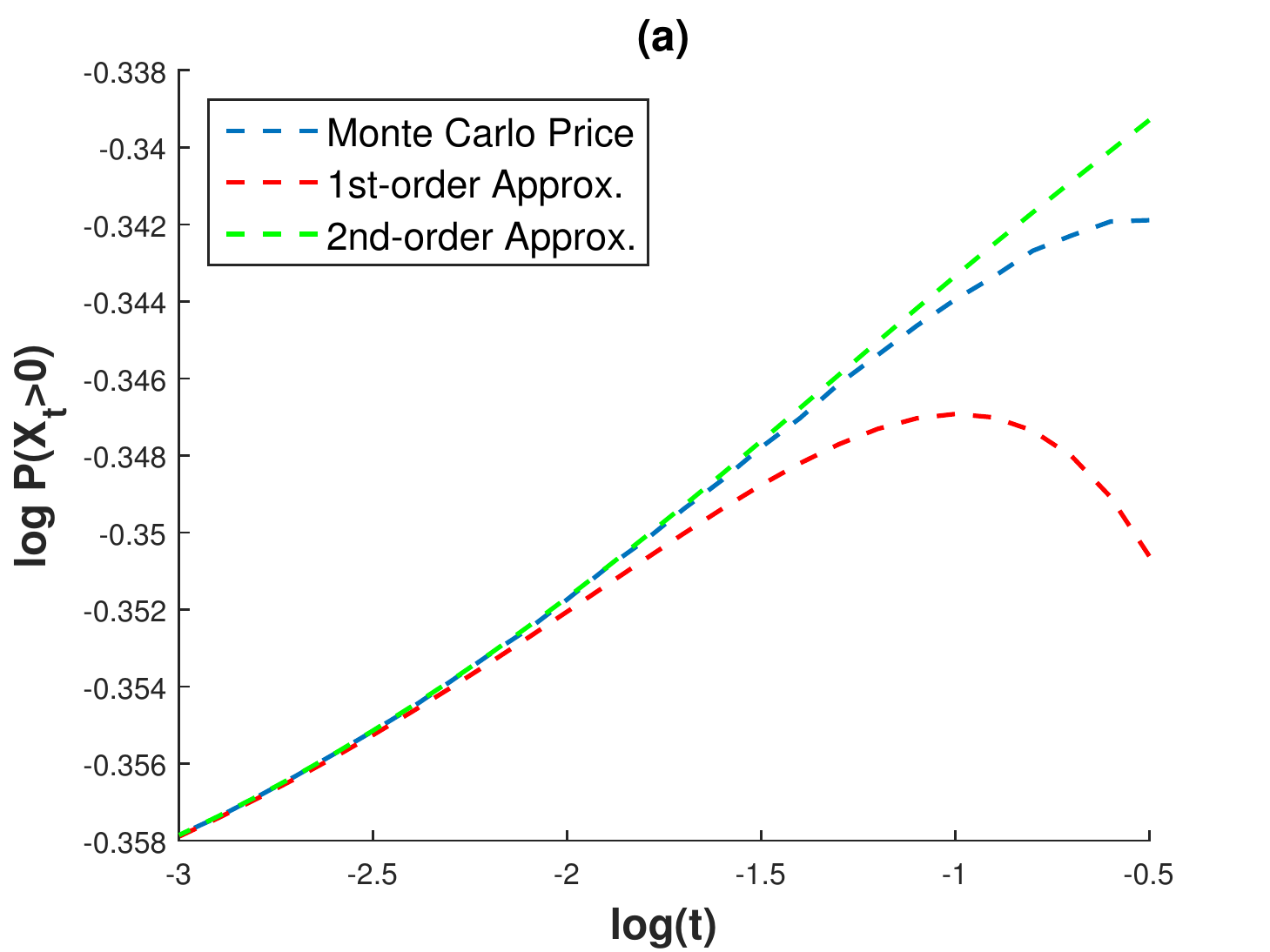}\hspace{1cm}
    \includegraphics[width=7cm,height=5.0cm]{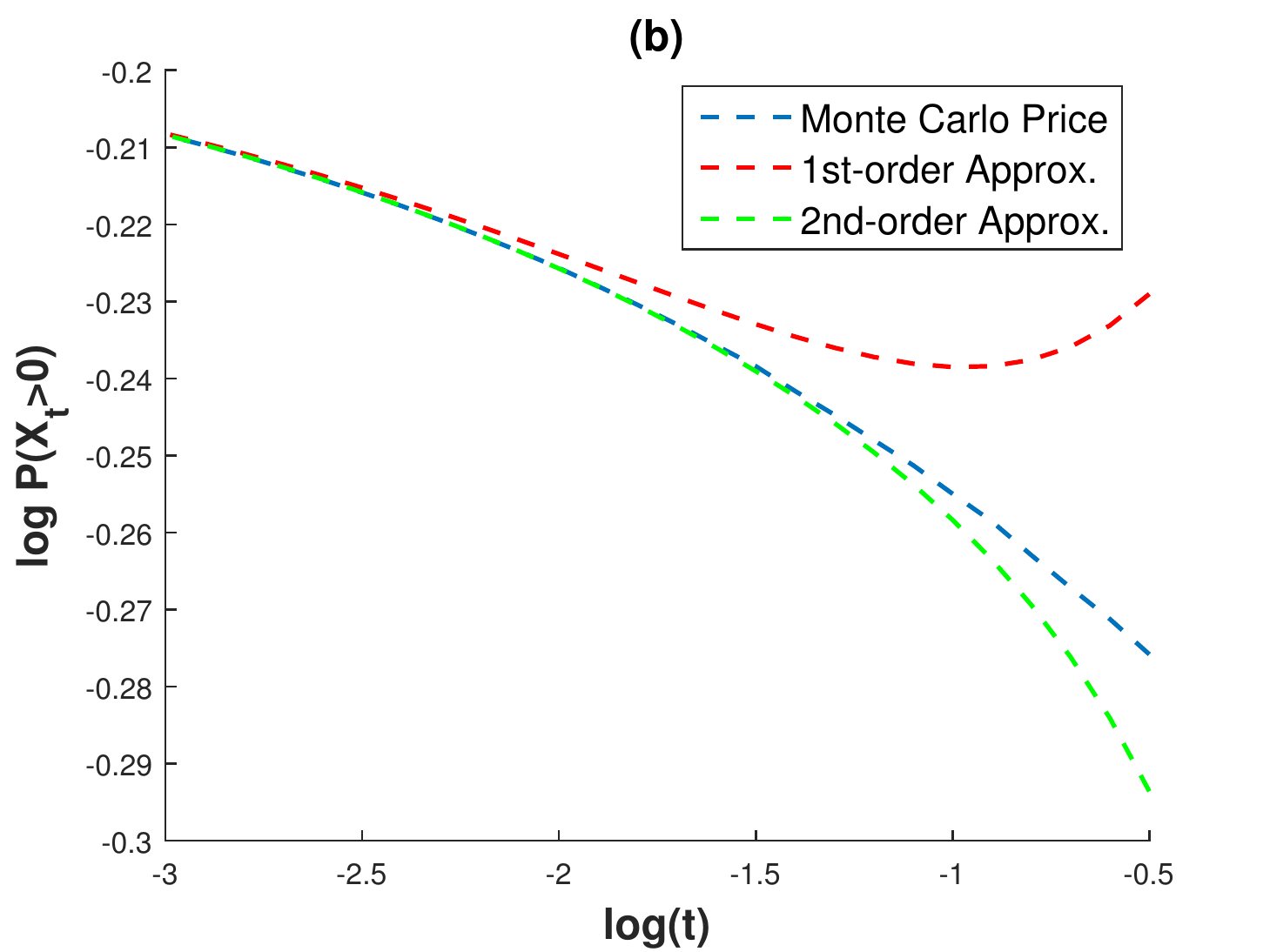}
    \caption{{\small ATM digital call option prices computed by Monte Carlo, and the first- and second-order approximations, under a pure-jump tempered stable model. Time is in years and both axes on a $\text{log}_{10}$-scale. {\bf(a)} $(C(1),C(-1),G,M,Y)=(0.0088,0.0044,0.41,1.93,1.5)$ as suggested in \cite{Andersen}. {\bf(b)} $(C(1),C(-1),G,M,Y)=(0.015,0.041,2.318,4.025,1.35)$ as suggested in \cite{Kawai}}.}\label{PureJumpProb}
\end{figure}

\begin{figure}
	\centering
	\includegraphics[width=7cm,height=5.0cm]{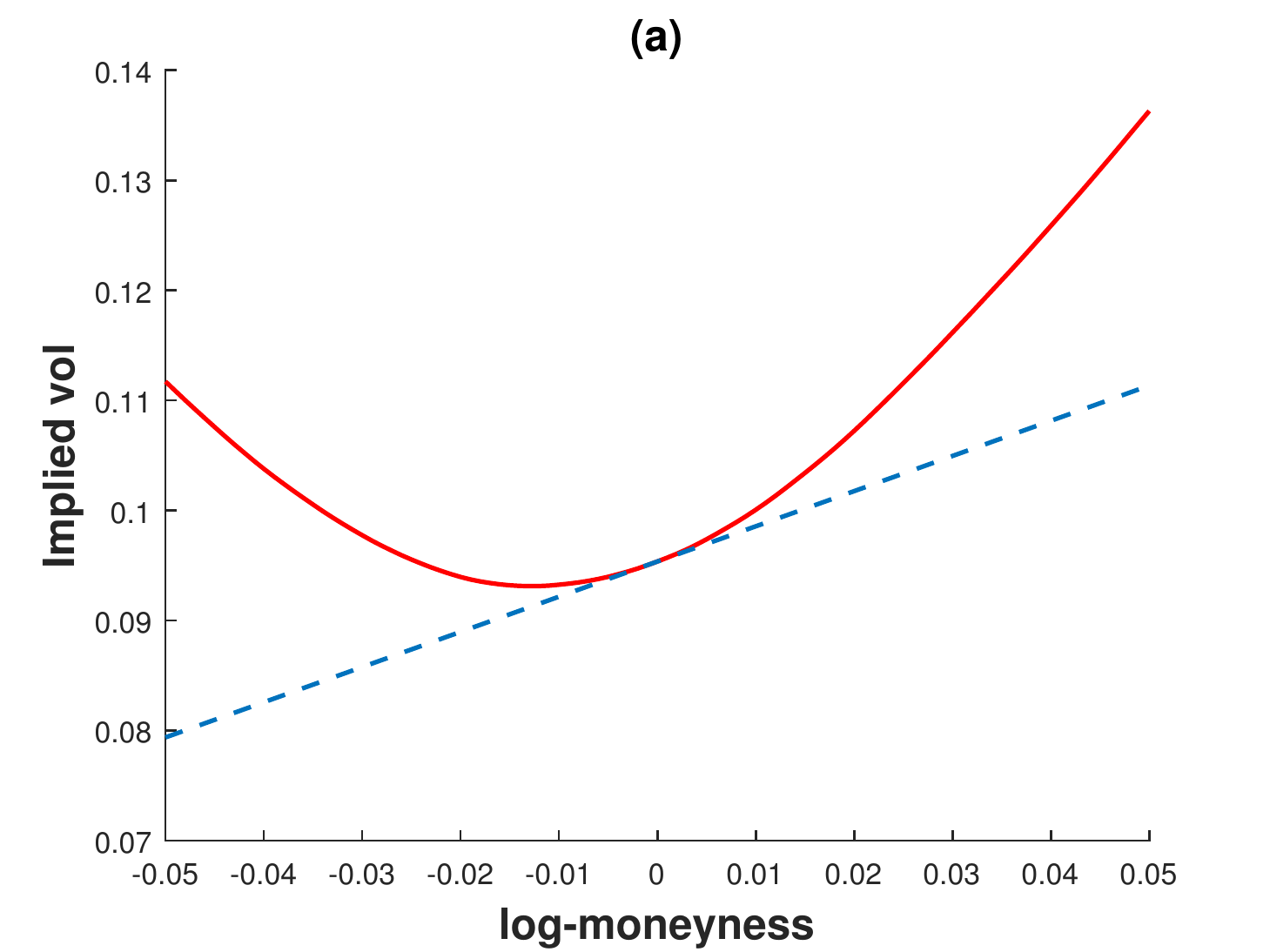}\hspace{1cm}
    \includegraphics[width=7cm,height=5.0cm]{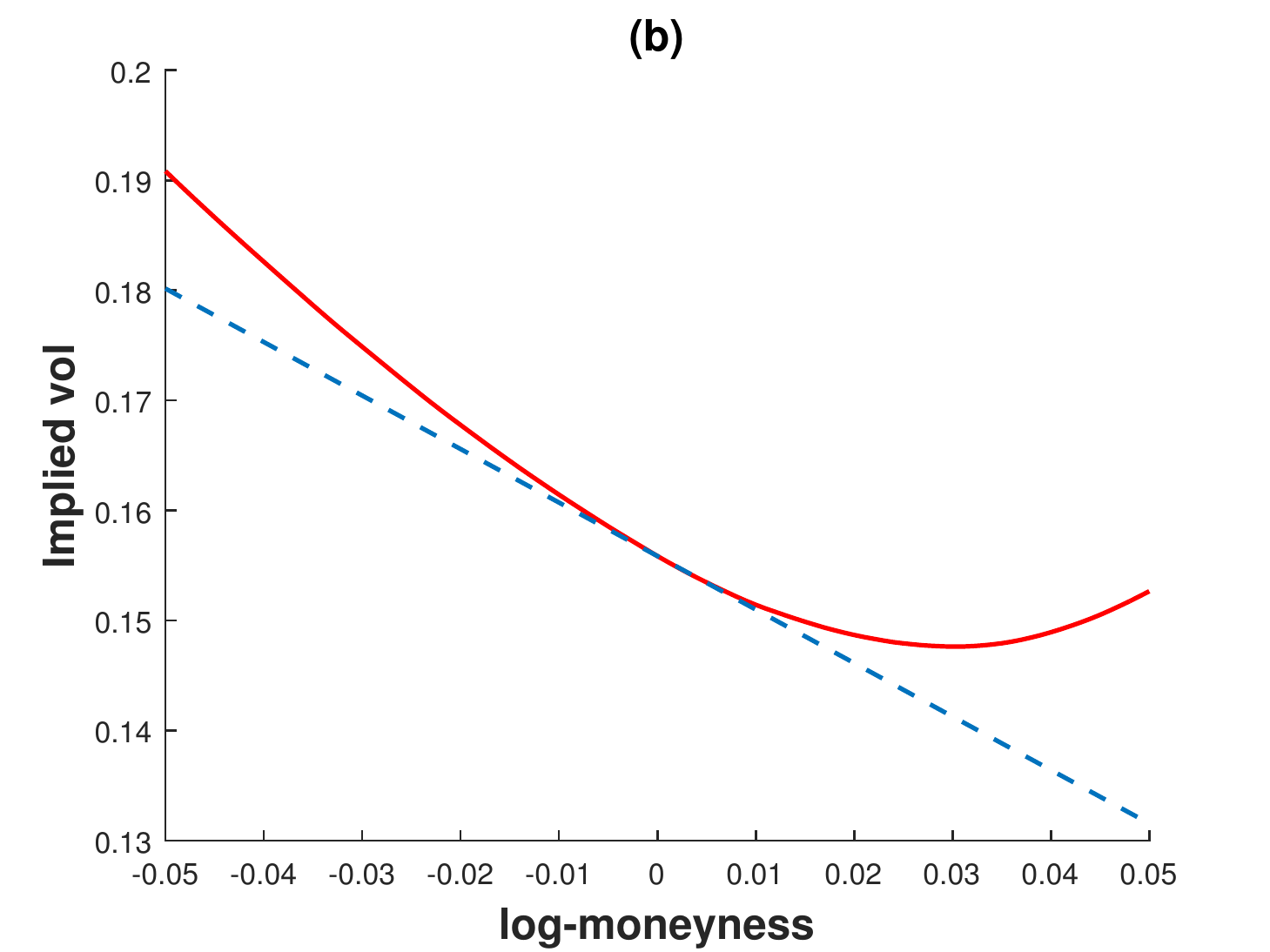}
    \caption{{\small The volatility smile (red) as a function of $\text{log}$-moneyness, and the second-order slope approximation (blue). Time-to-maturity is $0.1$ years, and the models are the same as in Figure \ref{PureJumpProb}.}}\label{PureJumpSmile}
\end{figure}

\begin{figure}
	\centering
	\includegraphics[width=7cm,height=5.0cm]{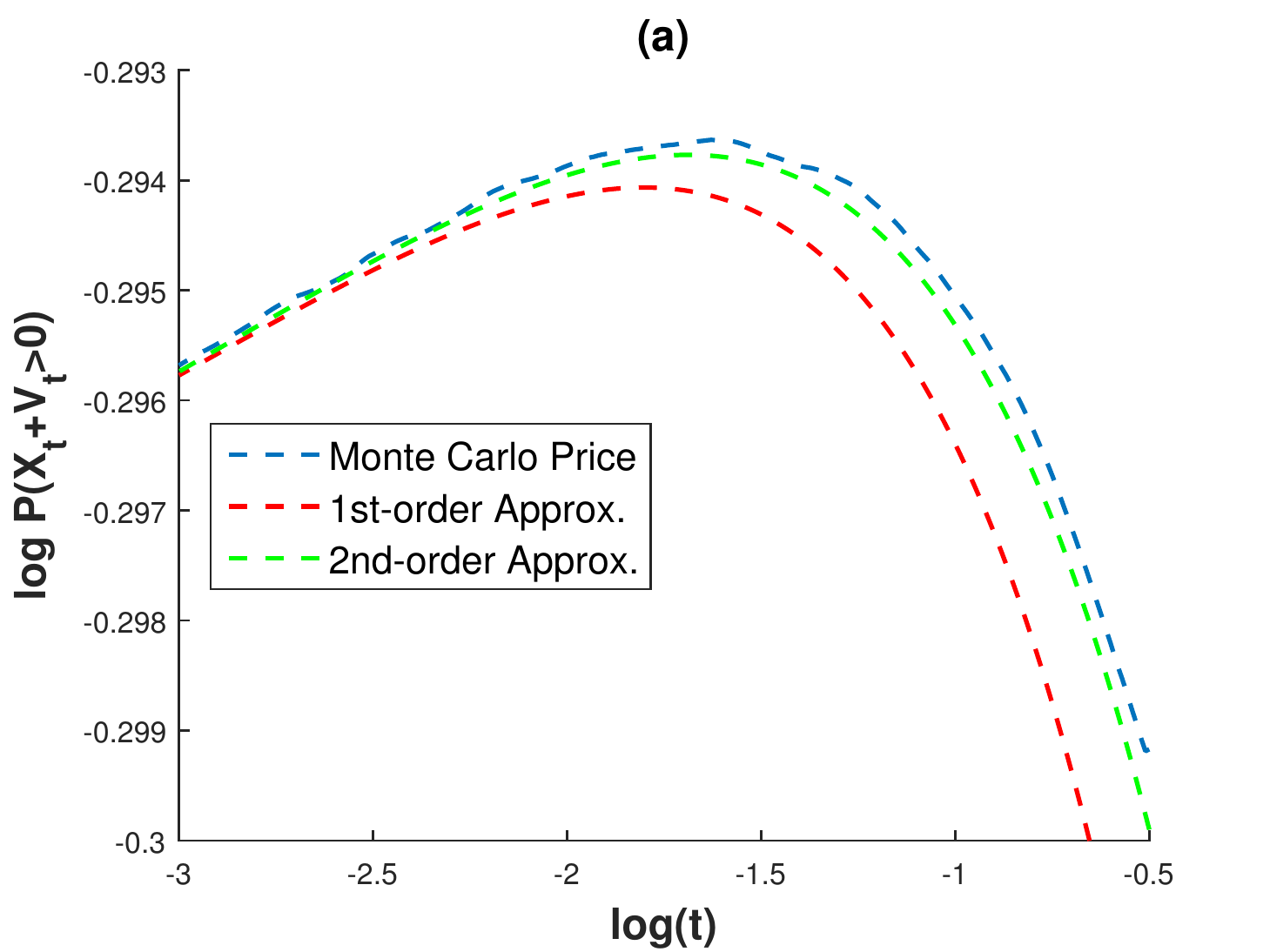}\hspace{1cm}
	\includegraphics[width=7cm,height=5.0cm]{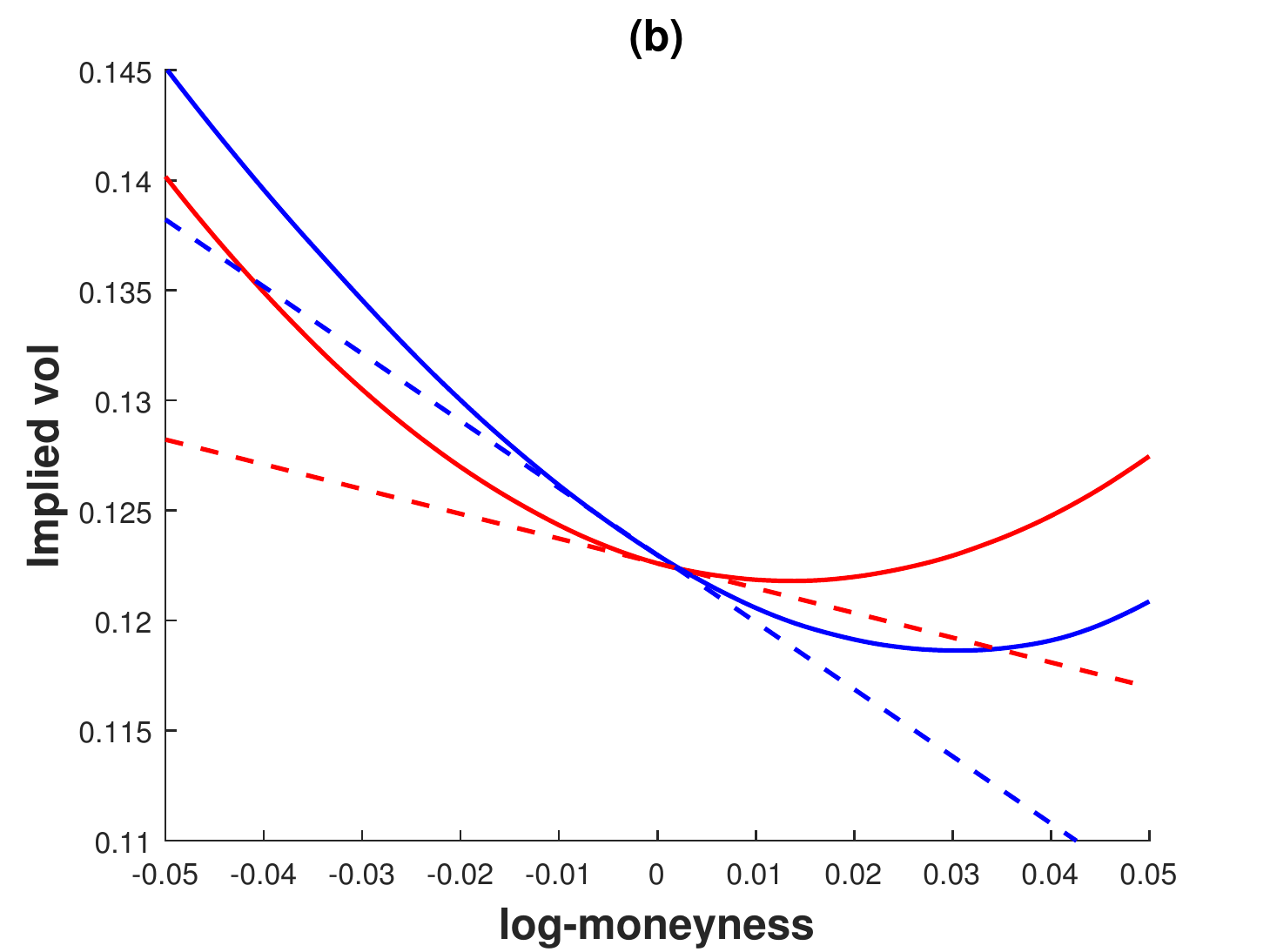}
    \caption{{\small {\bf(a)} ATM digital call option prices computed by Monte Carlo, and the first- and second-order approximations. Time is in years and both axes on a $\text{log}_{10}$-scale. The model is tempered stable with a Brownian component and parameters $(C(1),C(-1),G,M,Y,\sigma)=(0.0040,0.0013,0.41,1.93,1.5,0.1)$ as suggested in \cite{Andersen}. {\bf(b)} The red curves show the volatility smile and the second-order slope approximation for maturity $0.1$ years. The blue curves show the same quantities after replacing the Brownian component by a Heston process with the same spot volatility and $\rho=-0.3$.}}\label{ContCompAll}
\end{figure}

\begin{figure}
	\centering
	\includegraphics[width=19.5cm,height=5cm]{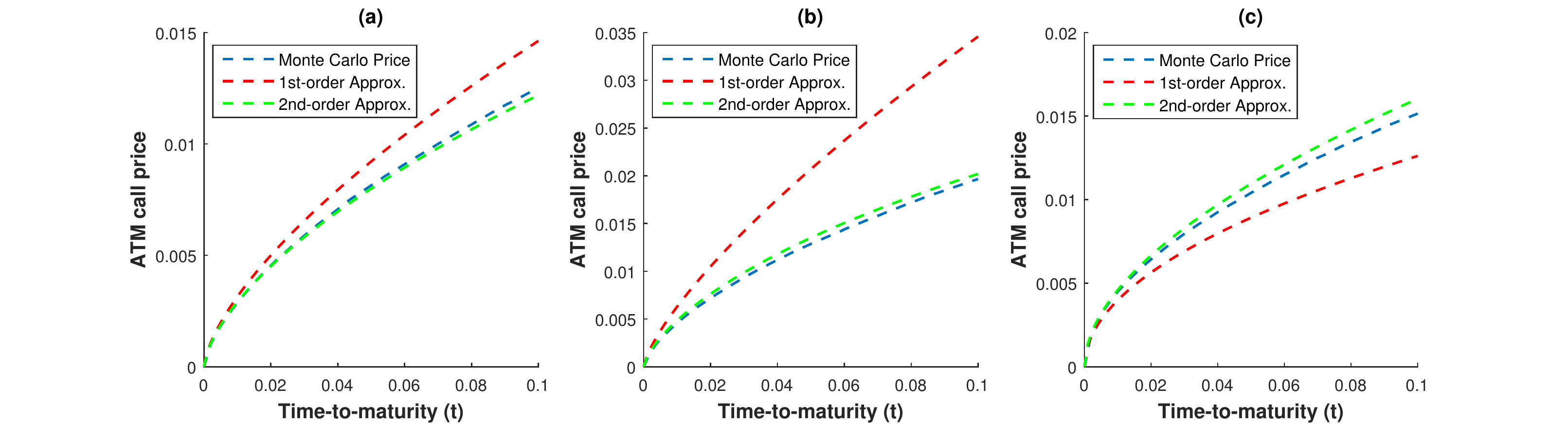}
    \caption{\small ATM call option prices computed by Monte Carlo, and the first- and second-order approximations from \cite{LopOla:2014}. Panels (a) and (b) correspond to the pure-jump cases in Figures \ref{PureJumpProb}-\ref{PureJumpSmile}, while panel (c) corresponds to the mixed case in Figure \ref{ContCompAll}.}\label{PriceAll}
\end{figure}

\subsection{Empirical application: S\&P500 implied volatility skew}

In this section we analyze the implied volatility skew in S\&P500 options, and compare it to the results of Sections \ref{SectionPureJump} and \ref{SectionContComp}. Our dataset consists of daily closing bid and ask prices for S\&P500 index options, across all strikes, $K$, and maturities, $t$, from January 2, 2014, to January 31, 2014 (21 business days). We take the mid-values of bid and ask prices as our raw data, and observations with time-to-maturity of less than five days are dropped to minimize the impact of microstructure effects. For each day and maturity we also visualize the quoted prices to check for obvious outliers. 

{We base the construction of the implied volatility curve on out-of-the-money (OTM) options since in-the-money (ITM) options are infrequently traded compared to OTM options and, thus, their prices are typically less reliable}.  More specifically, we follow a standard procedure, described in \cite{SahaliaLo} and \cite{Zhang}, {in which the put-call parity and liquid at-the-money (ATM) options are first} used to compute the \emph{implied} forward price of the underlying\footnote{The ATM strike {is taken to be the strike price at which the call and put options prices} are closest in value. We also set the risk-free interest rate to zero, but using a nonzero rate based on U.S.\ treasury yields did not change the results of our analysis {since the rate is close to zero over the sample period and the time-to-maturity is small}.}. Then, OTM options are used to compute the implied volatility for different strike prices. That is, put options (resp.\ call options) are used for strike prices that are below (resp.\ above) the forward price.

Figure \ref{NearTheMoney} shows some stylized features of our data. Panel (a) shows how the strike prices of {the typically} liquid 25-delta options\footnote{The 25-delta put (resp.,\ call) is the put (resp.,\ call) whose strike price has been chosen such that the option's delta is -25\% (resp.,\ 25\%).} become increasingly concentrated around the ATM strike, as time-to-maturity decreases, which is one reason for the importance of considering a small-moneyness regime in short-time. Panel (b) then shows how the implied volatility smile becomes increasingly skewed as time-to-maturity decreases. 
It also clearly shows that the left wing ({corresponding to} OTM put options) is steeper than the right wing ({corresponding to} OTM call options), which has {consistently} been observed in S\&P option prices since the market crash of 1987, and reflects the negative skewness in the underlying distribution of risk-neutral returns, or, equivalently, the {high} demand for protective put options against downward index movements.

As explained in previous sections, the short-term behavior of the skew differs significantly from one model setting to the next: In purely continuous models the skew is bounded, but in jump-models the relationship between the skew and time-to-maturity is a power-law, where the exponent depends on properties of the jump-component. This observation invites a model selection and calibration procedure based on a comparing the skew observed in real markets, to the model-skew under different {model assumptions}. This is similar in spirit to the approach in \cite{CarrWu}, where the short-term decay of option prices is used to infer properties of the underlying asset price process.

First, we need to decide on a measure of the implied volatility skew. Various skew-measures have been proposed in the literature ({see, e.g.,} \cite{Mixon}). {In this work, we simply estimate the ATM skew by taking the slope of the volatility smile between two near-the-money options}. For consistency we always choose OTM 25-delta options, since they are actively traded\footnote{Repeating the analysis using 10-delta options did not have a qualitative effect on the outcome.}, and are indeed near-the-money options (see Figure \ref{NTM}). Also, in order to fairly compare the skew across different dates and overall volatility levels, all volatilities for a given date are normalized by 
the CBOE Volatility Index (VIX), which is a measure of the average volatility of the S\&P index.

Figure \ref{Skew}(a) shows the estimated skew for options with maturities up to one year, for each day in our dataset. The power-law behavior is {evident}, not only for short-term options, but it seems to continue to hold for longer maturities as well. {Consequently, in Figure \ref{Skew}(b) the log-skew seems to be linear in log-maturity. Thus, 
to} estimate the exponent of the power law, we run a linear regression of log-skew on log-maturity, and take the slope coefficient as an estimate of the exponent of the power law. Our results can then be used for model selection and calibration: A nonnegative slope coefficient implies a purely continuous model or a jump-diffusion model, while a negative slope coefficient is consistent with a jump component of infinite activity. In the latter case, the magnitude of the slope coefficient can also be used to distinguish between jump-components of finite and infinite variation.

We carry out this regression for each of the 21 days in our dataset, using options with time-to-maturity less than 0.25 years (3 months), and days with at least four maturities less than 0.25 years {(which discards only} one day). We find that the linear model fits the data {extremely} well, with the average $R$-squared being 0.98. Moreover, the slope coefficient is consistently observed to be between $-0.3$ and $-0.4$, with an average of -0.36, which, as explained above, contradicts a {jump component with finite activity}, as well as an infinite activity jump-component of finite variation, in which case the order of the skew is -0.5 (cf.\ \cite{GerGul:2014}). On the other hand, this slope coefficient is in line with the results for the infinite variation jump-models studied in Sections \ref{SectionPureJump}-\ref{SectionContComp}, and can be used to calibrate important model parameters. 

In particular, this yields a simple procedure to calibrate the index of jump activity of the process, $Y$, which can be viewed as a new forward-looking tool to assess this fundamental parameter, complementing the popular rear-facing estimation methods based on high-frequency observations of the underlying asset's returns {(cf.\ \cite{AitJacod09})}. First, in the pure-jump case of Section \ref{SectionPureJump}, the order of the skew is $-1/2$ if $C(1)\neq C(-1)$, but $1/2-1/Y\in(-1/2,0)$ otherwise, so our regression results suggests that $C(1)=C(-1)$ and $Y\in(1.11,1.25)$. {However}, for the mixed model of Section \ref{SectionContComp}, the skew is bounded if $C(1)=C(-1)$, but of order $1/2-Y/2\in(-1/2,0)$ otherwise, so the regression results, together with the sign of the skew (see Remark \ref{RemSignCont}), {suggest} that $C(1)< C(-1)$ and $Y\in(1.6,1.8)$. These results show that both the pure-jump model and the mixed model are able to capture the short-term order of the skew, but the latter value of $Y$ is perhaps of greater interest, as several studies point to the presence of a continuous component in the returns process, and $Y\geq 1.5$ for actively traded stocks (cf.\ \cite{AitJacod09}, \cite{AitJacod10}, \cite{CarrWu})}.  

In {the analysis above} we used options with time-to-maturity less than 3 {months. 
However, one of the} main conclusion of the first part of this section was that the asymptotic expansions seem to give good approximations for options with maturities up to 0.1 years (just over one month). We therefore repeat the analysis using only options with maturities below or around the one month mark, excluding days with fewer than three maturities in that range, which eliminates 3 days out of 21. In this case, the average $R$-squared is 0.97, and the average slope is {-0.31, which, 
in the pure-jump case, indicates} $C(1)=C(-1)$ and $Y=1.23$, {while, in the mixed-case, it {again} suggests} $C(1)< C(-1)$ and $Y=1.62$. Those $Y$-values are of similar magnitude as in the previous analysis, but slightly more moderate in the sense that they {are closer to $1.5$}. We also note that {the analysis 
based on options with} maturities shorter than 0.1 years resulted in {more stable} regression results between days, compared to when including maturities shorter than 0.25, 
i.e.\ most of the slope coefficients were quite close to the average -0.31.

\begin{figure}
	\centering
	\includegraphics[width=7cm,height=5.0cm]{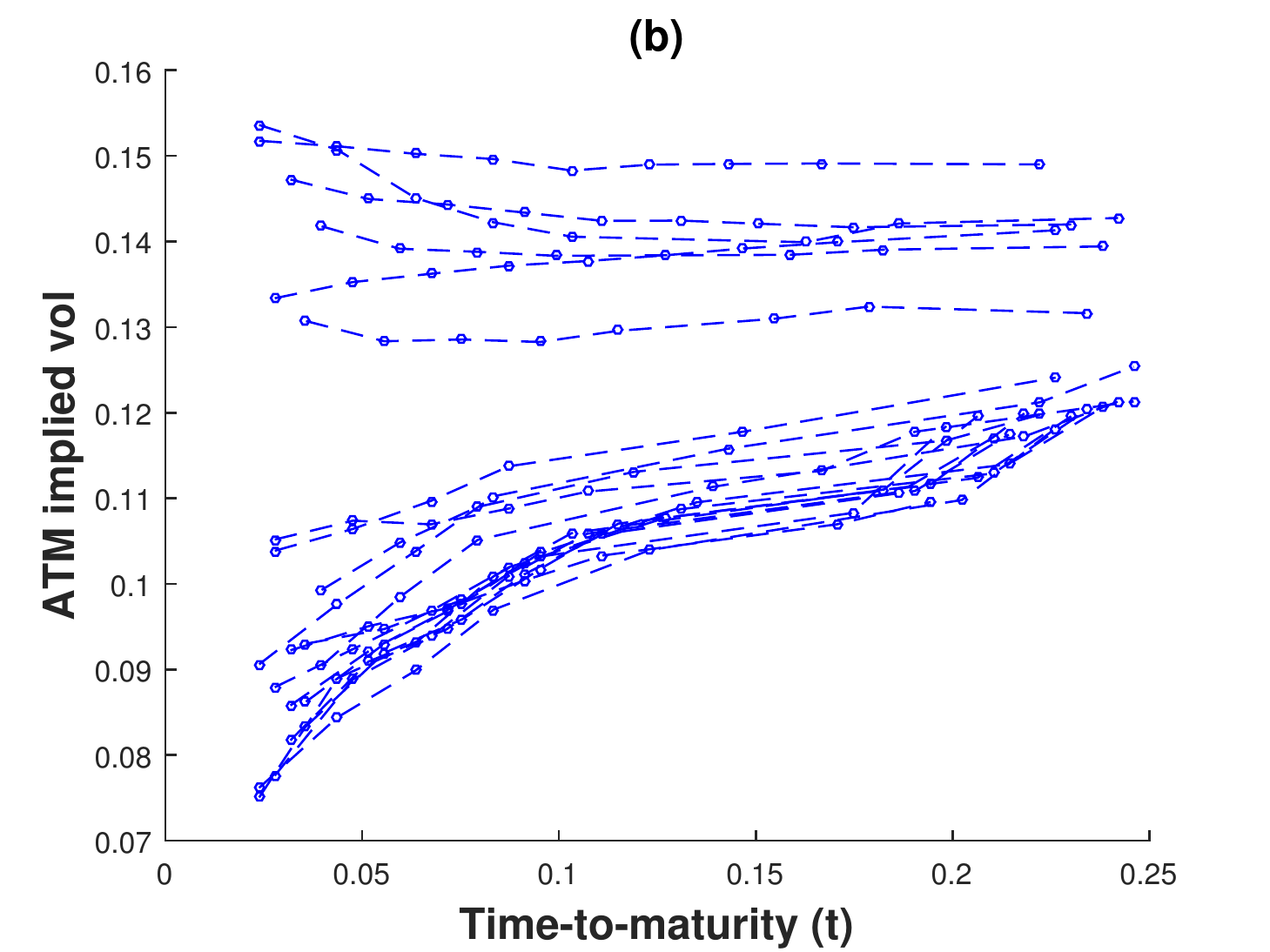}\hspace{1cm}
	\includegraphics[width=7cm,height=5.0cm]{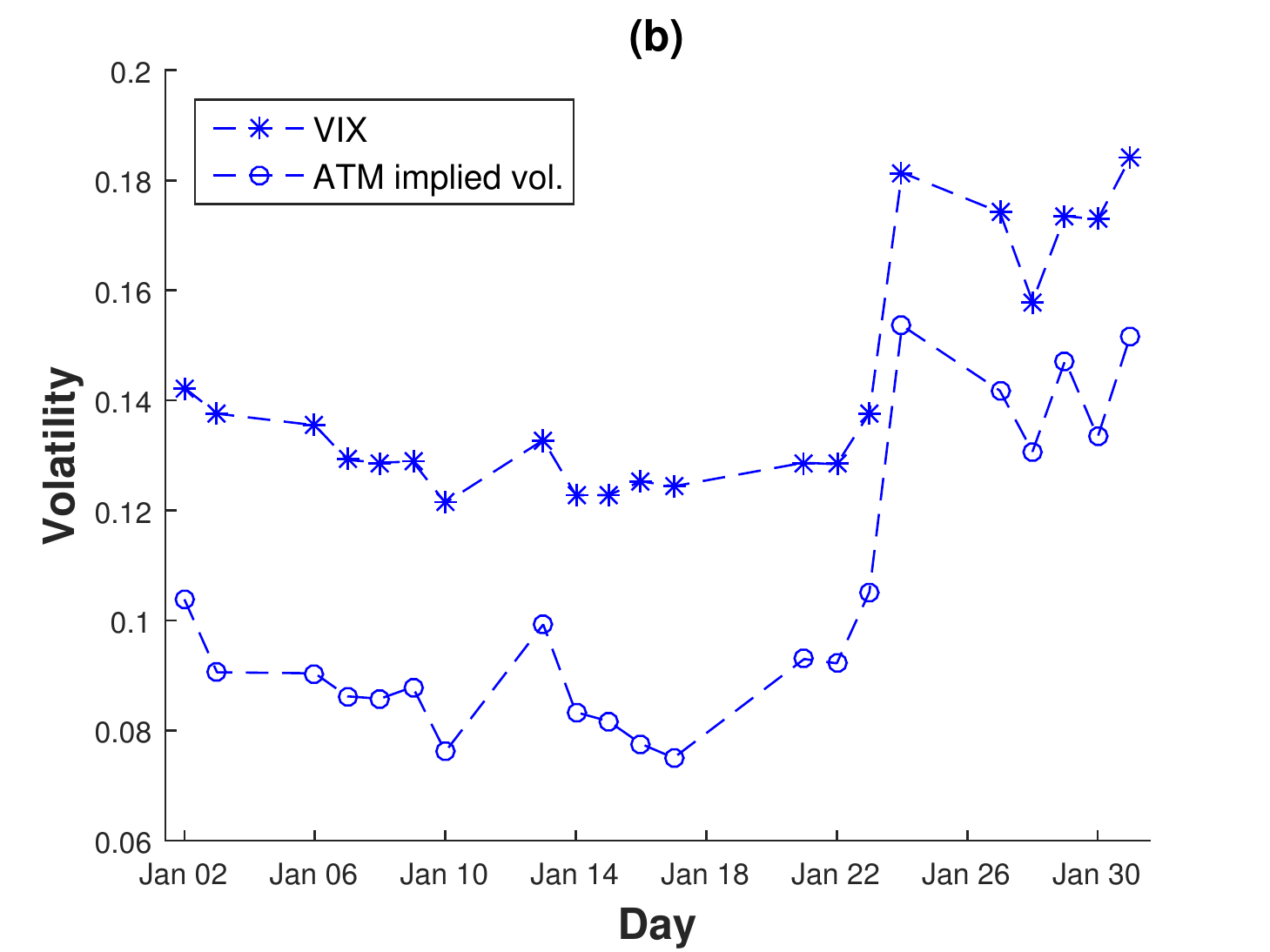}
    \caption{\small {\bf(a)} ATM implied volatility as a function of time-to-maturity for each business day in Jan 2014. {\bf(b)} ATM implied volatility for the shortest outstanding maturity compared to the VIX index.}\label{ATMvol}
\end{figure}

Finally, let us remark that even though one cannot reject one of our models in favor of the other based solely on the behavior of the skew, they can be distinguished, in principle, by the fact that the ATM volatility converges to zero in the pure-jump case, but to the nonzero spot volatility in the mixed case. As already described in \cite{LopGonHou:2014}, this property of models with a continuous component can also be exploited when using short-term expansions to calibrate their parameters, to circumvent the fact that the spot volatility is not directly observable (see, e.g., \cite{MedSca07}).\ Figure \ref{ATMvol}(a) displays for each day the ATM volatility as a function of time-to-maturity, and a quick inspection {indicates that} extrapolating to the zero-maturity does not seem to result in zero volatility, as would be the case in a pure-jump model. Furthermore, Figure \ref{ATMvol}(b) shows how the ATM implied volatility for the shortest outstanding maturity moves in tandem with the corresponding VIX measurement.

\label{NTM}

\begin{figure}
	\centering
	\includegraphics[width=7cm,height=5.0cm]{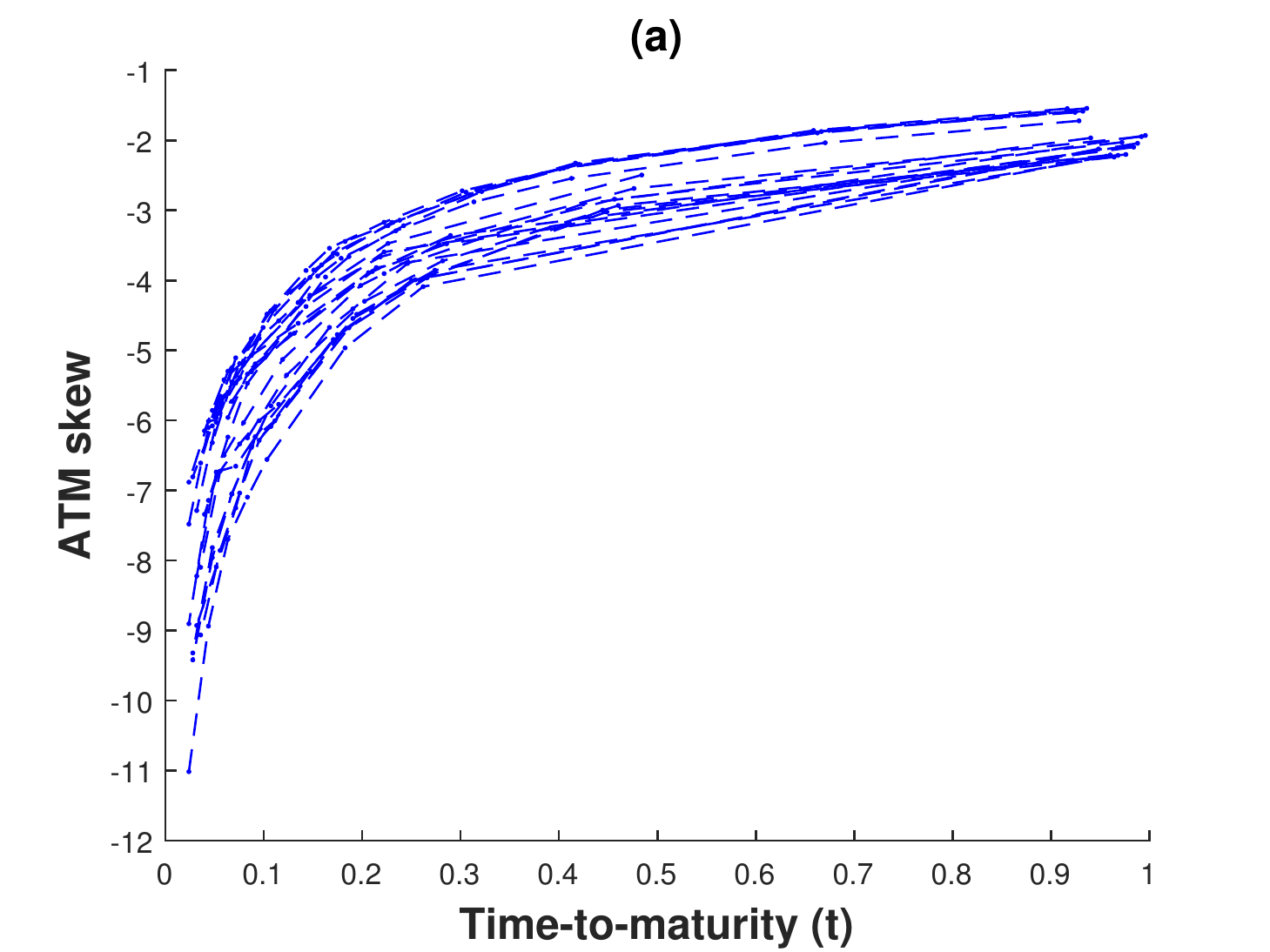}\hspace{1cm}
	\includegraphics[width=7cm,height=5.0cm]{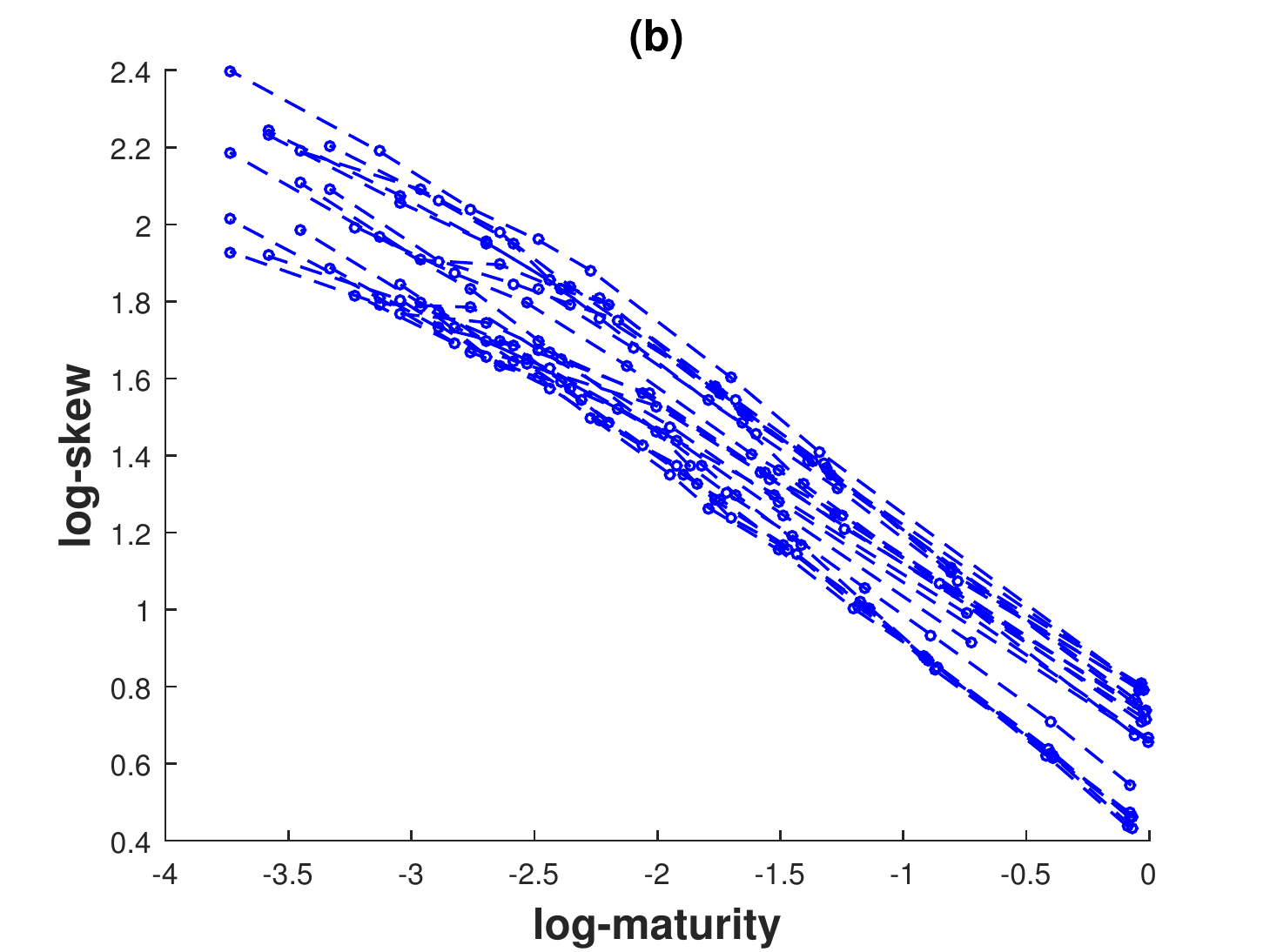}
    \caption{\small {\bf(a)} The ATM implied volatility skew as a function of time-to-maturity, for each business day in Jan 2014. {\bf(b)} The natural logarithm of the skew in panel (a) as a function of the natural logarithm of time-to-maturity.}\label{Skew}
\end{figure}

\section{Conclusions and future work}\label{conclusions}

As time-to-maturity becomes short, it is observed empirically that the liquid strike prices become increasingly concentrated around the ATM strike. The short-term volatility smile is therefore sometimes analyzed in terms of three quantities: the ATM implied volatility level, skew, and convexity (see, e.g., \cite{Bouchaud} and \cite{Zhang}).

The present work focuses on the ATM implied volatility skew (i.e.\ the strike-derivative). We obtain high-order short-term expansions for the skew under models with a L\'evy jump-component of infinite variation, and in the presence of a nonhomogeneous continuous component we quantify explicitly the short-term skew-effects of both jumps and stochastic volatility. 
Our proofs utilize a relationship between the skew and transition probabilities of the form $\bbp(S_t\geq S_0)$, i.e.\ prices of ATM digital call options, and as auxiliary results we {also} obtain short-term approximations for the delta of ATM options. Simulation results indicate the validity of our results for options with maturities up to at least one month, and we show that the volatility skew in recent S\&P500 index options is in accordance with the infinite variation jump-component of our models. 

{It is also natural to wonder about the short-time behavior of the ATM smile convexity, which, just like the skew, is of great importance in financial markets (see Section \ref{motivation}). Specifically, 
the ATM convexity is defined as the second order strike-derivative of the implied volatility,
\begin{align}
\left.\frac{\partial^2\hat\sigma\left(\kappa,t\right)}{\partial\kappa^2}\right|_{\kappa=0},
\end{align}
and, following similar steps as those used to derive (\ref{ATMslope}),
it is possible to show that the key quantity needed to analyze the convexity is 
$f_{S_t}(S_0)$,  where $f_{S_{t}}$ is the risk-neutral probability density of $S_{t}$. 
Moreover, just like the transition probability $\bbp(S_t\geq S_0)$ could be linked to the delta of ATM options, the probability density $f_{S_t}(S_0)$ is connected to the gamma of ATM options. 
In order to analyze the behavior of $f_{S_t}(S_0)$,  it seems necessary to also analyze the characteristic function of the log-returns process and, more specifically, its decay properties. This is in sharp contrast {\Blue to} the methods used in the present work that {\Blue do not} rely on inverse Fourier representations of the option prices and transition probabilities. The latter approach is left for future work.}

\appendix

\section{{Additional Proofs}}\label{appendixA}

\begin{lem}\label{SVlemma}
Let $V$ be as in (\ref{modelVY})), with $\mu(Y_t)$ and $\sigma(Y_t)$ replaced by $\bar\mu_t$ and $\bar\sigma_t$, defined in (\ref{stopped}). Also let $\bar\sigma'_t$, $\bar\sigma''_t$, $\bar\alpha_t$, and $\bar\gamma_t$, be {the stopped processes in (\ref{xi12})}, and $\bar\sigma_t^*:=\sqrt{\frac{1}{t}\int_0^t\bar\sigma_s^2ds}$. Then the following relations hold for any $p\geq 1$:
\begin{enumerate}
\item[(i)]
$\bbe\left|\bar\mu_t-\mu_0\right|^p = O(t^{\frac{p}{2}}),\quad t\to 0$.\label{SVlemMu}
\item[(ii)]
$\bbe\left|\bar\sigma_t-\sigma_0\right|^p = O(t^{\frac{p}{2}}),\quad t\to 0$.\label{SVlemSigAbs}
\item[(iii)]
$\bbe\left|\bar\sigma^*_t-\sigma_0\right|^p = O(t^{\frac{p}{2}}),\quad t\to 0$.\label{SVlemSigBar}
\item[(iv)]
$\bbe\left(\bar\sigma_t^*-\sigma_0\right) = O(t), \quad t\to 0$.\label{SVlemSig}
\item[(v)]
For $\xi_t^1,\xi_t^2$, and $\xi_t^{1,0}$, as in (\ref{xi12})-(\ref{xi10}), we have $\bbe\left|\xi_t^1\right|=O(t)$ and $\bbe\left|\xi_t^2\right|+\bbe\big|\xi_t^1-\xi_t^{1,0}\big| = O(t^{\frac{3}{2}}),\quad t\to 0$.\label{SVlemXi}
\item[(vi)]
$\bbe\big|\bar\sigma_t^*-\sigma_0-\sigma_0'\gamma_0\frac{1}{t}\int_0^tW_s^1ds\big|=O(t),\quad t\to 0$.\label{SVlemInt}
\end{enumerate}
\end{lem}

\noindent{\textbf{Proof.}} Let $L$ be a common Lipschitz constant for $\bar\mu_t$, $\bar\sigma_t$, and $\bar\gamma_t$.

\noindent {(i)} 
By the Lipschitz continuity of $\bar\mu_t$, and the Burkholder-Davis-Gundy (BDG) inequality, we can find a constant $C_p$ such that
\begin{align*}
\bbe\left|\bar\mu_t-\mu_0\right|^p \leq L^p\bbe\left|Y_{t\wedge\tau}-y_0\right|^p
\leq L^pC_p\Big(\bbe\Big(\int_0^t\bar\alpha_sds\Big)^p+\bbe\Big(\int_0^t\bar\gamma^2_sds\Big)^{\frac{p}{2}}\Big)
=O(t^{\frac{p}{2}}),\quad t\to 0, 
\end{align*}
since $\bar\alpha_s$ and $\bar\gamma_s$ are bounded.

\noindent {(ii)} is proved in a similar way, and for {(iii)} we use the boundedness of $\bar\sigma_t$, Jensen's inequality, and (ii) to write
\begin{align*}
\bbe\left|\bar\sigma^*_t-\sigma_0\right|^p 
\leq\frac{1}{(2m)^p}\bbe\Big(\frac{1}{t}\int_0^t\left(\bar\sigma^2_s-\sigma^2_0\right)ds\Big)^p
\leq\Big(\frac{M}{m}\Big)^p\frac{1}{t}\int_0^t\bbe\big(\bar\sigma_s-\sigma_0\big)^pds
=O(t^{\frac{p}{2}}),\quad t\to 0.
\end{align*}

\noindent{(iv)}
We can write
$\bbe(\bar\sigma_t^*-\sigma_0)
=\bbe\big((\bar\sigma_t^*)^2-\sigma_0^2\big)/(2\sigma_0)
+\bbe\big(\big(\big(\bar\sigma_t^*\big)^2-\sigma_0^2\big)\big({1}/{(\bar\sigma_t^*+\sigma_0)}-{1}/{(2\sigma_0)}\big)\big)$,
where the second term is $O(t)$ by (iii), while for the first term we have by It\^o's lemma
\begin{align*}
\bbe\big(\big(\bar\sigma_t^*\big)^2-\sigma_0^2\big) 
=\bbe\Big(\frac{1}{t}\int_0^t\Big(\int_0^s2\bar\sigma_u\bar\sigma'_u\bar\gamma_udW_u^1+\int_0^s\Big(2\bar\sigma_u\bar\sigma'_u\bar\alpha_u+\big(\bar\sigma'_u\big)^2+\bar\sigma_u\bar\sigma''_u\Big)du\Big)ds\Big)=O(t),\quad t\to 0, 
\end{align*}
due to the fact that the expected value of the stochastic integral is zero.


\noindent {(v)}
By Cauchy's inequality and It\^o's isometry we have
\begin{align*}
\bbe\left|\xi_t^2\right| \leq \sqrt{\int_0^t\bbe\Big(\int_0^s\Big(\bar\sigma'_u\bar\alpha_u+\half\bar\sigma''_u \bar\gamma^2_u\Big)du\Big)^2 ds}=O(t^{\frac{3}{2}}),\quad t\to 0.
\end{align*}
Similarly,
\begin{align*}
\bbe\big|\xi_t^1-\xi_t^{1,0}\big| \leq \sqrt{\int_0^t\int_0^s\bbe\big(\bar\sigma'_u\bar\gamma_u-\sigma'_0 \gamma_0\big)^2duds}=O(t^{\frac{3}{2}}),\quad t\to 0,
\end{align*}
because by the boundedness of $\bar\sigma'_u$ and $\bar\gamma_u$, we can find a constant K such that
\begin{align*}
\bbe\left(\bar\sigma'_u\bar\gamma_u-\sigma'_0\gamma_0\right)^2 
\leq  K\bbe\left(\bar\gamma_u-\gamma_0\right)^2 + K\bbe\left(\bar\sigma'_u-\sigma'_0\right)^2
\leq 2LK{\bbe\left(Y_{u\wedge{}\tau}-y_0\right)^2} = O(u),\quad u\to 0, 
\end{align*}
where in the last step we again used the BDG inequality. Similarly, Cauchy's inequality and It\^o's isometry yield $\bbe\left|\xi_t^1\right|=O(t)$.

\noindent{(vi)}
follows from the triangle inequality and the following three identities. First, by (iii) above, we have
\begin{align*}
\bbe\Big|\bar\sigma_t^*-\sigma_0-\frac{(\bar\sigma_t^*)^2-\sigma_0^2}{2\sigma_0}\Big|
=\frac{1}{2\sigma_0}\bbe\left(\bar\sigma_t^*-\sigma_0\right)^2=O(t),\quad t\to 0.
\end{align*}
Second, by It\^o's Lemma, 
\begin{align*}
\bbe\Big|\left(\bar\sigma_t^*\right)^2-\sigma_0^2-\frac{1}{t}\int_0^t\int_0^s2\bar\sigma_u\bar\sigma'_u\bar\gamma_udW_u^1ds\Big|
&\leq \frac{1}{t} \int_0^t\int_0^s\bbe\Big|2\bar\sigma_u\bar\sigma'_u\bar\alpha_u+\half\left((\bar\sigma'_u)^2+\bar\sigma_u\bar\sigma''_u\right)\bar\gamma^2_u\Big|duds
=O(t),\quad t\to 0,
\end{align*}
since the integrand in the last integral is bounded. Third, Cauchy's inequality and It\^o's isometry can be used to show
\begin{align*}
\bbe\Big|\frac{1}{t}\int_0^t\int_0^s\left(\bar\sigma_u\bar\sigma'_u\bar\gamma_u-\sigma_0\sigma'_0\gamma_0\right)dW_u^1ds\Big|
=O(t),\quad t\to 0,
\end{align*}
by following similar steps as in the proof of (v).
\hfill\qed

\bibliographystyle{plain}

\vspace{0.2 cm}

\end{document}